\documentclass[10pt,a4paper]{article}
 
\usepackage[latin1]{inputenc}
\usepackage{float}
\usepackage{bm}
\usepackage{amsfonts}
\usepackage{amsthm}
\usepackage{amsmath}
\usepackage{mathtools}
\usepackage{mathrsfs}
\usepackage{axodraw4j}
 
\renewcommand{\[}{\begin{equation}\begin{aligned}}
\renewcommand{\]}{\end{aligned}\end{equation}}

\usepackage{slashed}
\usepackage{cite}
\usepackage{bbold}
\usepackage{todonotes}
\usepackage{graphics}
\usepackage{graphicx}
\usepackage{tikz}
\usetikzlibrary{patterns}
\usetikzlibrary{calc,decorations.markings,arrows.meta,shapes.misc,decorations.pathmorphing,bending}
\tikzset{
  branch point/.style={cross out,draw=black,fill=none,minimum size=2*(#1-\pgflinewidth),inner sep=0pt,outer sep=0pt}, 
  branch point/.default=5
}
\tikzset{
  branch cut/.style={
    decorate,decoration=snake,
    to path={
      (\tikztostart) -- (\tikztotarget) \tikztonodes
    },
    execute at begin to={{
      \coordinate (A) at ($(\tikztostart)!.8!-10:(\tikztotarget)$);
      \coordinate (B) at ($(\tikztostart)!.8!10:(\tikztotarget)$);
      \coordinate (AB/3) at ($(A)!1/3!(B)$);
      \coordinate (2AB/3) at ($(A)!2/3!(B)$);
      \coordinate (C) at ($(AB/3)!2/(3*sqrt(3))!-90:(B)$);
      \coordinate (D) at ($(2AB/3)!4/(3*sqrt(3))!-90:(B)$);
    }}
  }
}

\newcommand{\bq}{\bar{q}}
\newcommand{\bl}{\bar{\ell}}

\newcommand{\h}[3]{{{#1}\cdot \tilde{h}(#2)\cdot {#3}}}

\renewcommand{\vec}[1]{\mathbf{#1}}

\newcommand{\beq}{\begin{equation}}
\newcommand{\eeq}{\end{equation}}
\newcommand{\bea}{\begin{eqnarray}}
\newcommand{\eea}{\end{eqnarray}}
\newcommand{\nn}{\nonumber}

\def\lsi{\raise0.3ex\hbox{$<$\kern-0.75em\raise-1.1ex\hbox{$\sim$}}}
\def\gsi{\raise0.3ex\hbox{$>$\kern-0.75em\raise-1.1ex\hbox{$\sim$}}}

\newcommand\brwrap[3]{%
  \setbox0=\hbox{$#2$}
  \left#1\vbox to \the\ht0{\hbox to 0pt{}}\right.\kern-.2em
  \begingroup #2\endgroup\kern-.15em
  \left.\vbox to \the\ht0{\hbox to 0pt{}}\right#3
}
\def\Lexp{\biggl\langle\!\!\!\biggl\langle}
\def\Rexp{\biggr\rangle\!\!\!\biggr\rangle}
\newcommand{\KMOCav}[1]{\Lexp #1 \Rexp}

\usepackage{setspace}
\usepackage{fancyhdr}
\usepackage{listings}
\usepackage{changepage}
\usepackage[colorlinks=true, urlcolor=black, plainpages=false, pdfpagelabels]{hyperref}
\usepackage{etoolbox}
\usepackage{amssymb}
\usepackage[parfill]{parskip}
\usepackage{tableaux}
\def\next{\hfill\break} 

\title{\textbf{NLO deflections for spinning particles and Kerr black holes}}

\author{Gabriel~Menezes$^{a}$\footnote{\href{mailto:: gabrielmenezes@ufrrj.br}{gabrielmenezes@ufrrj.br}} ~and Matteo~Sergola$^{b, c}$\footnote{\href{mailto:matteo.sergola@ed.ac.uk}{matteo.sergola@ed.ac.uk}}
\vspace{14pt} 
\\
\it $^{a}$Departamento de F\'isica, Universidade Federal Rural do Rio de Janeiro, 
\\\vspace{8pt} 
23897-000, Seropédica, RJ, Brazil.\\
\it $^{b}$Higgs Centre for Theoretical Physics, School of Physics and Astronomy,
\\\vspace{8pt} 
\it  The University of Edinburgh, EH9 3FD, Scotland.
\\
\it $^{c}$Kavli Institute for Theoretical Physics,
\\ \it University of California, Santa Barbara, CA, 93106-4030, USA.}

\begin{document}
\maketitle

\begin{abstract}
We employ the ``KMOC" formalism of \cite{Kosower:19} to compute classical momentum deflections of spinning bodies with arbitrary spin orientations {up to next-to-leading order (one loop)}. We do this in electrodynamics and gravity. {The final result, valid for generic masses, is true  for all spins at tree level and up to second (fourth) spin order  for the electromagnetic (gravity) case at one loop. Furthermore, emphasis is given to the probe limit scenario where our results extend to all spin orders in the heavy source, even at next-to-leading order.}  We carry out our computations both using a unitarity based framework and Feynman diagrammatic approach which relies on scattering amplitudes computed on fixed backgrounds.
\end{abstract}



\section{Introduction}

Revolutionary observations by the LIGO/VIRGO collaboration \cite{LIGO1,LIGO2,LIGO3,LIGO4,LIGO5} are paving the way towards a new era of ground breaking astrophysical discoveries. Furthermore, these were instrumental in triggering the ongoing surge of research activities dedicated to modern techniques for treating the two-body problem in gravity. The idea is that such techniques would complement several methods presented in the literature~\cite{13,14,15,16,17,18,19}. Here we intend to explore the application of the modern amplitudes program to fulfill this current trend. Amplitudes and QFT methods were applied successfully to understand in an alternative perspective the general relativistic two-body problem~\cite{18,69,Bjerrum-Bohr:14,71,72,Bjerrum-Bohr:16,76,Bjerrum-Bohr:2021vuf,Bjerrum-Bohr:2021din, Herrmann:2021tct,Cristofoli:2021vyo, Bern:2020buy, Bern:2021dqo,Herrmann:2021lqe, DiVecchia:2021bdo, Bautista:2021wfy,Bern:2020gjj,Moynihan:2020gxj,Cristofoli:2020uzm,Parra-Martinez:2020dzs,Haddad:2020tvs,AccettulliHuber:2020oou,Moynihan:2020ejh,Manu:2020zxl,Sahoo:2020ryf,delaCruz:2020bbn,Bonocore:2020xuj,Mogull:2020sak,Emond:2020lwi,Cheung:2020gbf,Mougiakakos:2020laz,Carrasco:2020ywq,Kim:2020cvf,Bjerrum-Bohr:2020syg,Gonzo:2020xza,delaCruz:2020cpc,Cristofoli:2021jas,Bautista:2021llr, Aoude:2021oqj, Cho:2022syn, Bern:22, yutinspin, Alessio:2022kwv, Brandhuber:2021eyq,Brandhuber:2021bsf,Brandhuber:2021kpo, FebresCordero:2022jts}. On the other hand, the relevance of a loop amplitude to the classical potential was already perceived in the context of the quantum effective field theory of gravity~\cite{77,78,79,80,81,82} and later emphasized in Ref.~\cite{83}.

Gravitational amplitudes are intimately connected to the amplitudes of Yang-Mills theory due to the so-called double copy, which says that gravitational amplitudes can be obtained as a product of two Yang-Mills amplitudes. The original incarnation by Kawai, Lewellen and Tye states that a tree-level gravitational amplitude is given by a sum over the products of two tree-level color-ordered Yang-Mills amplitudes (weighted by Mandelstam invariants)~\cite{Kawai:1985xq}. In a celebrated result, Bern, Carrasco and Johansson demonstrated that gravitational numerators are, in a given sense, the square of the kinematic numerators in Yang-Mills theory~\cite{Bern:2008qj,Bern:2010ue,zvirev}. One may expect these insights to be of utmost importance to the physics of gravitational waves. The double-copy approach also connects classical solutions of Yang-Mills theory and gravity, where point charges in Yang-Mills theory map to point sources in gravity~\cite{Monteiro:2014cda,Luna:2016due,Goldberger:2016iau,Luna:2016hge}. Furthermore, there are also indications that the double copy can be useful in the study of bound states and spinning particles~\cite{Goldberger:2017vcg,Plefka:2018dpa,Goldberger:2017ogt,Li:2018qap, Kalin:2019rwq, Kalin:2019inp}. These results suggest that one can greatly benefit from the knowledge of scattering-amplitudes methods in the investigation of classical gravitational physics.

In this paper we will study scattering of spinning particles in electrodynamics and black holes in Einstein general relativity. The no-hair theorem asserts that they are characterized by only three observable classical parameters: mass, electric charge and angular momentum. This suggests that a black hole can be viewed pretty much like a point particle. Indeed, recently it has been shown that black holes can be envisaged as particularly rigid compact objects -- this point of view has been pivotal in the derivation of the spin-independent part of the two-body classical potential for inspiralling bodies, from the scattering amplitudes of gravitationally coupled scalars~\cite{Cheung:2018wkq,Bern:2019nnu,Foffa:2019yfl,Cristofoli:2019neg,Maybee:19}. In turn, from the perspective of on-shell scattering amplitudes, the first step is the determination of the $3$-particle amplitudes. For the Kerr   case, this three-particle amplitude is given by the notion of a minimal coupling, as expected from the no-hair theorem. This makes the massive particle look as elementary as possible to the graviton probe. Moreover, it was shown recently that the potential for Kerr black holes can be indeed recovered to all orders from minimal coupling~\cite{Guevara:19a,Chung:19}.

Furthermore, Newman and Janis (NJ) found that one can obtain the Kerr metric by suitably complexifying the Schwarzschild solution in null polar coordinates and performing a shift~\cite{Newman:1965tw}. Modern methods of amplitudes also allow us to comprehend the origin of the complex shift. The shift is simply a consequence of the spin effects generated when one goes from a minimally coupled scalar to a spinning particle -- in the classical limit, the minimal coupling exponentiates~\cite{Guevara:19a}. When applied to the computation of classical observables, this exponentiation precisely induces the relevant shift at tree level~\cite{Arkani-Hamed:20}. This refines the origin of the equivalence between black holes and particles. 

Here we  report on some progress concerning the scattering of two Kerr black holes in a similar spirit as in Ref.~\cite{Arkani-Hamed:20}. Namely, we wish to study the impulse up to next-to-leading order communicated to a spinning black hole in the collision to another spinning black hole in accordance with the on-shell perspective, in which minimally coupled higher-spin particles at large spin correspond to Kerr black holes. We mainly follow the method designed in Ref.~\cite{Kosower:19}, where a direct mapping between amplitudes and observables was disclosed. We will see that the Newman-Janis shift is implemented in a rather non-trivial way. As a warm-up exercise, we consider an Abelian gauge theory and construct the relevant scattering amplitudes for the classical impulse associated with the collision of two massive spinning particles due to photon exchange. This will offer a simple illustration of the calculation that one should implement in the gravity case, for which we  will rely on the existence of the double-copy procedure. 

An important part of our paper will also be dedicated to the probe limit which is obtained by sending $M_1/M_2\to 0$ and possibly by setting the light particle's spin to zero: $a_1\to 0$. This limit turns out to simplify the calculation of observables quite a bit, and allows us to write down new results  such as the non-aligned classical momentum deflection to all spin orders in the heavy source for both EM and gravity. We will also investigate this limit with, perhaps, a more traditional approach in section \eqref{probesss} by dealing with Feynman rules of sQED/Gravity in presence of a fixed classical source. This framework, although maybe computationally less powerful than the unitarity method, allows us to establish another connection  between amplitudes and the classical formulation of the  double copy of \cite{Monteiro:2014cda}. The importance of the probe limit scenario has also been recently underlined in works such as \cite{Siemonsen:2019dsu, Gonzo:2021drq, Adamo:2022rmp}.

The plan of the paper is the following: in  Sec.~\ref{linimp} we recap the basic formulae needed for the computation of classical deflections. We then proceed first by computing the EM results in Sec.~\ref{emmgg} and the gravity ones in Sec.~\ref{GRRRR} using unitarity-based techniques and finally we re-examine the probe limit with source amplitudes in Sec.~\ref{probesss}. Conclusions and further directions are outlined in Sec.~\ref{fin}.

\section{Linear impulse}\label{linimp}

In this section we will review the relevant formulae to calculate specific classical observables from scattering amplitudes. We are particularly interested in calculating the leading-order and next-to-leading order impulse in the scattering of spinning particles.

Following Ref.~\cite{Kosower:19}, we can study classically measurable quantities directly from on-shell quantum scattering amplitudes. In particular, we will calculate the momentum transfer up to next-to-leading order in the scattering of two massive spinning particle, thereby extending the results of Ref.~\cite{Arkani-Hamed:20}. The collision of such spinning particles to be discussed can thus be envisaged as the collision of two Kerr black holes, as explained in the previous section. In other words, minimally coupled higher-spin particles at large spin correspond to Kerr black holes.

The classical impulse on the scalar particle during the scattering event reads (up to next-to-leading order)~\cite{Kosower:19} 
\bea\label{dp}
\Delta p^{\mu}_{1} &=&\Delta p^\mu_{\text{LO}, 1} +\Delta p^\mu_{\text{NLO}, 1}  
\eea
with
\bea
\nn\\
\Delta p^\mu_{\text{LO}, 1} &=& i \frac{g^2}{4} \Biggl\langle \!\!\! \Biggl\langle
\hbar^2 \int {\hat{d} ^4 \bar{q}} \,\hat{\delta}( \bar{q} \cdot p_1 ) \hat{\delta}( \bar{q} \cdot p_2 )
e^{- i b \cdot {\bar q}} \bar{q}^{\mu} {\cal A}^{(0)}(p_1, p_2 \to p_1 + \hbar {\bar q}, p_2 - \hbar {\bar q})
\Biggr\rangle\!\!\! \Biggr\rangle\eea
and
\bea
\nn\\
\Delta p^\mu_{\text{NLO}, 1} &=& i \frac{g^4}{4} \Biggl\langle \!\!\! \Biggl\langle
 \int {\hat{d} ^4 \bar{q}} \,\hat{\delta}( \bar{q} \cdot p_1 ) \hat{\delta}( \bar{q} \cdot p_2 )
e^{- i b \cdot {\bar q}} {\cal I}^{\mu (1)}
\Biggr\rangle \!\!\! \Biggr\rangle
\nn\\
{\cal I}^{\mu (1)} &=& \hbar  \bar{q}^{\mu} 
{\cal A}^{(1)}(p_1, p_2 \to p_1 + \hbar {\bar q}, p_2 - \hbar {\bar q})
\nn\\
&-& i \hbar^3 \int \frac{d^4 \bar{w}}{(2\pi)^4} \hat{\delta}( 2 \bar{w} \cdot p_1 + \hbar \bar{w}^2) 
\hat{\delta}( 2 \bar{w} \cdot p_2 - \hbar \bar{w}^2) \bar{w}^{\mu}
\nn\\
&\times& \sum_{\textrm{states}} {\cal A}^{(0)}(p_1, p_2 \to p_1 + \hbar {\bar w}, p_2 - \hbar {\bar w})
{\cal A}^{(0)\, *}(p_1+ \hbar {\bar q}, p_2 - \hbar {\bar q} \to p_1 + \hbar {\bar w}, p_2 - \hbar {\bar w})
\label{impulse}
\eea
here $b$ is the impact parameter, we have removed overall factors of powers of the coupling constant $g$ and accompanying $\hbar$s from the vertices in the amplitudes and
\bea
\Biggl\langle\!\!\! \Biggl\langle f(p_1, p_2, \ldots) \Biggr\rangle\!\!\! \Biggr\rangle
& \equiv& \int d\Phi(p_1) \int d\Phi(p_2) \, |\phi_1(p_1)|^2 |\phi_2(p_2)|^2  f(p_1, p_2, \ldots)
\nn\\
d\Phi(p_i) &=& \frac{d^4 p_i}{(2 \pi)^4} (2\pi) \theta(p^{0}_{i}) \delta(p^2_{i} - m^2_{i}),
\eea
$\phi_{i}(p_i)$ is a suitable relativistic wave function that obeys certain classical constraints \cite{Kosower:19}. In the non-relativistic limit one obtains minimum-uncertainty wave functions in momentum space. Moreover, within the large angle brackets we have approximated $\phi(p+\hbar{\bar q}) \sim \phi(p)$ and when evaluating the integrals we should set $p_{i} \sim M_i u_i$, $u_i$ being the classical four-velocity normalized to $u_{i}^2 = 1$. We finally note that the definition of this double angle brackets has to be able to deal with amplitudes with spin, which is the subject of the present study. Associated discussions will be presented in due course.

Observe the presence of tree-level $\mathcal{A}^{(0)}$ and one-loop $\mathcal{A}^{(1)}$  scattering amplitudes in the above expressions. Below we shall evaluate such amplitudes and extract the contribution to the classical processes. In addition, notice that the term quadratic in the tree amplitude appearing in ${\cal I}^{\mu (1)}$ can be envisaged as the two-particle cut of a one-loop box, weighted by the loop momentum $\bar{w}^{\mu}$; that explains the sum over all possible on-shell states that can cross the cut that appears in such an expression.

 We again remark that the amplitudes appearing in the above expressions have factors of $g/\sqrt{\hbar}$ removed for every interaction. Here $g=e$ for the electromagnetic case and $g = \kappa = \sqrt{32 \pi G}$ in the gravitational case. {In addition, observe that, as in Ref.~\cite{Kosower:19}, we are employing relativistically natural units, with $c=1$.}

\section{Electromagnetic linear impulse from quantum scattering amplitudes}\label{emmgg}

Let us begin our study by working in the context of electrodynamics. Here we consider two spin-$S_k$ fields $\Phi_{k}$, $k=1,2$, with masses $M_k$, interacting only through the minimal coupling with an electromagnetic field. A possible Lagrangian density  describing the system is given by
\beq
{\cal L} = - \frac{1}{4} F_{\mu\nu} F^{\mu\nu} 
+ {\cal L}[\Phi_{k}]
+ {\cal L}_{\textrm{int}}[\Phi_{k}, A_{\mu}]
\eeq
where ${\cal L}[\Phi_{k}]$ describes the free massive fields and ${\cal L}_{\textrm{int}}$ is the interaction Lagrangian. We will not display any particular contribution in the Lagrangian coming from the spinning particles, we will only assume that it has the expected symmetries. We exclude local interactions between matter fields because they violate the classical assumption that the inter-particle separation must be larger than their de Broglie wavelength. On the other hand, even if we do not have access to a explicit Lagrangian, we can still investigate the relativistic scattering of the associated spinning particles by considering modern on-shell amplitudes. As discussed above, these can serve as an efficient tool to calculate classical observables, such as the classical linear impulse imparted on a particle during a scattering event. Classical physics tells us this is simply the total change in the momentum of one of the particles during the collision.

So now let us discuss with some detail the amplitudes involved in this calculation. We will be mainly interested in the pieces required for the extraction of the classical limit of the amplitude.

\subsection{Calculation of the classical contribution from scattering amplitudes}

  $3$-particle tree-level amplitudes involving a photon and two massive spin-$S$ states with the same mass $M_2$ and minimal coupling are given by~\cite{Arkani-Hamed:17,Johansson:19}
\bea
A_{S}({\bf 1}, {\bf 2}, 3^{+}) &=&  \sqrt{2} e 
\frac{\langle {\bf 1} {\bf 2} \rangle^{2S}}{M_2^{2S-1}} x
\nn\\
A_{S}({\bf 1}, {\bf 2}, 3^{-}) &=& - \sqrt{2} e 
\frac{\bigl[ {\bf 1} {\bf 2} \bigr]^{2S}}{M_2^{2S-1}} \frac{1}{x}
\eea
where
\bea
x &=& \frac{\langle \zeta| {\bf 1} |3\bigr]}{M_2 \langle 3\zeta \rangle} 
\nn\\
\frac{1}{x} &=& \frac{\langle 3| {\bf 1} |\zeta\bigr]}{M_2 \bigl[ 3 \zeta \bigr]} .
\eea
$\zeta$ is a gauge choice and even though the $x$ factor has an explicit $\zeta$ dependence, the amplitudes are in fact  gauge invariant.

\begin{figure}[H]
\begin{center}

\tikzset{every picture/.style={line width=0.75pt}} 

\begin{tikzpicture}[x=0.85pt,y=0.85pt,yscale=-1,xscale=1]

\draw    (452.67,1799.44) -- (408.33,1758) ;
\draw [shift={(430.5,1778.72)}, rotate = 43.07] [fill={rgb, 255:red, 0; green, 0; blue, 0 }  ][line width=0.08]  [draw opacity=0] (5.36,-2.57) -- (0,0) -- (5.36,2.57) -- cycle    ;
\draw    (408.33,1758) -- (457.67,1718.67) ;
\draw [shift={(433,1738.33)}, rotate = 141.43] [fill={rgb, 255:red, 0; green, 0; blue, 0 }  ][line width=0.08]  [draw opacity=0] (5.36,-2.57) -- (0,0) -- (5.36,2.57) -- cycle    ;
\draw    (408.33,1758) .. controls (406.7,1759.69) and (405.03,1759.72) .. (403.33,1758.09) .. controls (401.64,1756.46) and (399.97,1756.49) .. (398.33,1758.18) .. controls (396.7,1759.87) and (395.03,1759.9) .. (393.34,1758.27) .. controls (391.65,1756.64) and (389.98,1756.67) .. (388.34,1758.36) .. controls (386.7,1760.05) and (385.03,1760.08) .. (383.34,1758.45) .. controls (381.65,1756.82) and (379.98,1756.85) .. (378.34,1758.54) .. controls (376.7,1760.23) and (375.03,1760.26) .. (373.34,1758.62) .. controls (371.65,1756.99) and (369.98,1757.02) .. (368.34,1758.71) .. controls (366.7,1760.4) and (365.03,1760.43) .. (363.34,1758.8) .. controls (361.65,1757.17) and (359.98,1757.2) .. (358.34,1758.89) .. controls (356.7,1760.58) and (355.03,1760.61) .. (353.34,1758.98) -- (351.08,1759.02) -- (351.08,1759.02) ;
\draw    (366.44,1766.8) -- (383,1767.2) ;
\draw [shift={(386,1767.27)}, rotate = 181.37] [fill={rgb, 255:red, 0; green, 0; blue, 0 }  ][line width=0.08]  [draw opacity=0] (5.36,-2.57) -- (0,0) -- (5.36,2.57) -- cycle    ;
\draw [line width=1.5]    (306.75,1800.46) -- (351.08,1759.02) ;
\draw [shift={(328.92,1779.74)}, rotate = 136.93] [fill={rgb, 255:red, 0; green, 0; blue, 0 }  ][line width=0.08]  [draw opacity=0] (6.97,-3.35) -- (0,0) -- (6.97,3.35) -- cycle    ;
\draw [line width=1.5]    (351.08,1759.02) -- (301.75,1719.69) ;
\draw [shift={(326.42,1739.35)}, rotate = 38.57] [fill={rgb, 255:red, 0; green, 0; blue, 0 }  ][line width=0.08]  [draw opacity=0] (6.97,-3.35) -- (0,0) -- (6.97,3.35) -- cycle    ;

\draw (450,1723.33) node [anchor=north west][inner sep=0.75pt]    {$p'_{1}$};
\draw (445,1769.33) node [anchor=north west][inner sep=0.75pt]    {$p_{1}$};
\draw (369.17,1773.17) node [anchor=north west][inner sep=0.75pt]    {$q$};
\draw (290,1774.33) node [anchor=north west][inner sep=0.75pt]    {$p_{2}$};
\draw (288,1719.33) node [anchor=north west][inner sep=0.75pt]    {$p'_{2}$};

\end{tikzpicture}
\caption{Tree-level scattering amplitude $A^{(0)}$ between the spin-$S$, labeled by $2$, and the scalar particle, labeled by $1$. The wiggly line is either a photon or a graviton.}
\label{treelevel}
\end{center}
\end{figure}
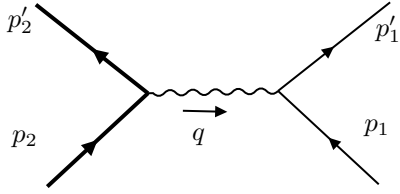

The four-point amplitude corresponding to the scattering of a scalar particle with mass $m$ with the minimally coupled spin-$S$ particle with exchange of photons can be calculated as follows. The residue associated with this amplitude can be calculated by gluing appropriate $3$-particle amplitudes. One finds
\beq
{\textrm{Res} [A^{(0)}] \Big |_{q^2 = 0}} = 2 e^2  M_1M_2
\left( \frac{\langle {\bf 2} {\bf 2}^{\prime} \rangle^{2S}}{M_2^{2S}} 
\frac{x_{22^{\prime}}}{x_{11^{\prime}}}
+ \frac{\bigl[ {\bf 2} {\bf 2}^{\prime} \bigr]^{2S}}{M_2^{2S}} \frac{x_{11^{\prime}}}{x_{22^{\prime}}}
\right) .
\eeq
Despite the presence of the $x$ factors, the above expression can be brought to a local form~\cite{Arkani-Hamed:17}. In fact, we can also use a different representation for the $x$-ratios, which are little-group invariant. Using that $u_{1} \cdot u_2 = \gamma $,  we can write
\beq
\frac{x_{11^{\prime}}}{x_{22^{\prime}}} = u_{1} \cdot u_2 + \sinh\phi = e^{\phi}
\eeq
where in the last step we used the standard definition of the rapidity, $\cosh\phi = \gamma$\footnote{It may be also useful to recall that $\cosh 2\phi=2\gamma^2-1, \,\,\,\sinh 2\phi=2\gamma\sqrt{\gamma^2-1},\,\,\, \sinh \phi=\sqrt{\gamma^2-1} .$}. 
The tree-level amplitude is then given by (see Fig.~\ref{treelevel})
\bea
A^{(0)}(1,2 \to 1^{\prime}, 2^{\prime}) &=& M_1M_2 \frac{2 e^2}{q^2} 
\left( \frac{\langle {\bf 2} {\bf 2}^{\prime} \rangle^{2S}}{M_2^{2S}} 
\frac{x_{22^{\prime}}}{x_{11^{\prime}}}
+ \frac{\bigl[ {\bf 2} {\bf 2}^{\prime} \bigr]^{2S}}{M_2^{2S}} \frac{x_{11^{\prime}}}{x_{22^{\prime}}}
\right) 
\nn\\
&=& M_1M_2 \frac{2 e^2}{q^2} 
\left( \frac{\langle {\bf 2} {\bf 2}^{\prime} \rangle^{2S}}{M_2^{2S}} e^{- \phi}
+ \frac{\bigl[ {\bf 2} {\bf 2}^{\prime} \bigr]^{2S}}{M_2^{2S}} e^{\phi} \right) .
\eea
Note we are using the incoming momenta convention here.  From the expression of the classical impulse, we know that $p_1^{\prime} = p_1 + \hbar {\bar q}$ and 
$p_2^{\prime} = p_2 - \hbar {\bar q}$. For the calculation of classical observables, we know that $\hbar{\bar q}$ should be a small quantity. This implies that the spinor $| {\bf 2}^{\prime} \rangle$ is only a small boost of the spinor $| {\bf 2} \rangle$. We may therefore write~\cite{Arkani-Hamed:20}
\beq
| {\bf 2}^{\prime} \rangle = | {\bf 2} \rangle + \frac{1}{8} \omega_{\mu\nu} ( \sigma^{\mu} \bar{\sigma}^{\nu} 
- \sigma^{\nu} \bar{\sigma}^{\mu} ) | {\bf 2} \rangle
\eeq
where the boost parameters $\omega_{\mu\nu}$ are small and given by
\beq
\omega_{\mu\nu} = - \frac{1}{M^2_2} ( p_{2 \mu} q_{\nu} - p_{2 \nu} q_{\mu} ) 
\eeq
and we have  also taken into account the on-shell relation $2 p_2 \cdot q = q^2 \approx 0$. Hence
\beq
| {\bf 2}^{\prime} \rangle 
= | {\bf 2} \rangle + \frac{1}{2    M_2} q  | {\bf 2} \bigr]
\eeq
where we used that $p | {\bf p} \rangle = M_2 | {\bf p} \bigr]$ and the on-shell relation just mentioned above. Likewise, with $p | {\bf p} \bigr] = M_2 | {\bf p} \rangle$, one finds that
\beq
| {\bf 2}^{\prime} \bigr] = | {\bf 2} \bigr] - \frac{1}{2    M_2} q  | {\bf 2} \rangle .
\eeq
The sign difference is due to the fact that $| {\bf 2} \rangle$ ($| {\bf 2} ] $) transforms in the $(1/2,0)$ ($(0,1/2)$) representation of the Lorentz group and these are complex-conjugate representations. Hence 
when taking the limits\footnote{One of the first instances where the $S\to \infty$ limit of a massive particle was related to black hole spin multipoles is \cite{Vaidyafirstspin}.} $S \to \infty$ and $\hbar \to 0$, $\hbar S$ fixed, we find that
\bea
\frac{\langle {\bf 2} {\bf 2}^{\prime} \rangle^{2S}}{M_2^{2S}} \Bigg|_{\hbar \to 0,S \to \infty} 
&=& e^{\bar{q} \cdot a}
\nn\\
\frac{\bigl[ {\bf 2} {\bf 2}^{\prime} \bigr]^{2S}}{M_2^{2S}} \Bigg|_{\hbar \to 0,S \to \infty}  
&=& e^{- \bar{q} \cdot a}
\eea
where $a^{\mu} = s^{\mu}/M_2$, and $s^{\mu}$ is the Pauli-Lubanski pseudovector associated with a spin $S$ particle:
\beq
s^{\mu} = \frac{\hbar S}{M_2} \langle {\bf 2} | \sigma^{\mu}  | {\bf 2} \bigr] .
\eeq
Now we are ready to display the classical limit for the above tree-level amplitude:
\beq
A^{(0)}(1,2 \to 1^{\prime}, 2^{\prime}) \bigg|_{\textrm{classical}}
= M_1 M_2 \frac{2 e^2}{ \hbar^3 \bar{q}^2 } 
\left( e^{\bar{q} \cdot a} e^{- \phi} + e^{ - \bar{q} \cdot a} e^{\phi} \right).
\eeq
Defining 
\begin{equation}
A^{(0)}=\frac{e^2}{\hbar} \mathcal{A} ^{(0)}
\end{equation}
we finally arrive at
\beq
{\cal A}^{(0)}(1,2 \to 1^{\prime}, 2^{\prime}) \bigg|_{\textrm{classical}}
=   M_1 M_2 \frac{2}{ \hbar^2 \bar{q}^2 } 
\left( e^{\bar{q} \cdot a} e^{- \phi} + e^{ - \bar{q} \cdot a} e^{\phi} \right) .
\eeq
It is easy to generalize the above result to the case of two particles with spin. The four-point tree-level amplitude corresponding to the scattering of a spin-$S_1$ particle with mass $M_1$ with a spin-$S_2$ particle with mass $M_2$ with exchange of photons is given by
\begin{equation}
\begin{split}
 {\cal A}^{\textrm{tree}}&(1,2 \to 1^{\prime}, 2^{\prime}) = 
M_1 M_2 \frac{2}{\hbar^2 \bar{q}^2} 
\left[ e^{ \bar{q} \cdot a_1} \, e^{ \bar{q} \cdot a_2}  e^{- \phi} 
 + e^{ -\bar{q} \cdot a_1} \, e^{ -\bar{q} \cdot a_2}  e^{ \phi} \right]
= M_1 M_2 \frac{4}{\hbar^2 \bar{q}^2} 
\cosh[\bar{q} \cdot (a_1+a_2) - \phi]
\nn\\&\,\,\,\,\,\,\,\,
=  
 \frac{2}{\hbar^2 \bar{q}^2} 
 \left[ (s - M_1^2 - M_2^2) \cosh[\bar{q} \cdot (a_1+a_2)]  
 - \sqrt{( s - M_1^2 - M_2^2 )^2 - 4 M_1^2 M_2^2} \sinh[\bar{q} \cdot (a_1+a_2)]
\right] 
\end{split}
\end{equation}
where $s = (p_1 + p_2)^2$. We can express this amplitude in a more convenient way exploring the fact that, on the support of the delta functions, the Gram determinant constraint reads
$$
( \epsilon^{\alpha\beta\rho\sigma} u_{1 \alpha} u_{2 \beta} a_{\rho} \bar{q}_{\sigma} )^2
= - \sinh^2 \phi (a \cdot \bar{q})^2 
$$
where $a=a_1, a_2$. Hence we can write that
\beq
\cosh[ \bar{q} \cdot (a_1+a_2)]  \cosh \phi = 
u_{1 \mu} u_{2 \nu} \left[ \eta^{\mu\nu} + \frac{1}{2!} \eta^{\mu\nu} \Bigl( \bar{q} \cdot (a_1+a_2) \Bigr)^2
+ \frac{1}{4!} \eta^{\mu\nu} \Bigl( \bar{q} \cdot (a_1+a_2) \Bigr)^4 + \cdots \right]
\eeq
and
\bea
\sinh[ \bar{q} \cdot (a_1+a_2)]  \sinh \phi &=& 
\sinh \phi \left[ \bar{q} \cdot (a_1+a_2) + \frac{1}{3!} \Bigl( \bar{q} \cdot (a_1+a_2) \Bigr)^3 + \cdots  \right]
\nn\\
&=& \left[  i \epsilon^{\alpha\beta\rho\sigma} u_{1 \alpha} u_{2 \beta} (a_1 + a_2)_{\rho} \bar{q}_{\sigma} +  \cdots  \right]
\eea
and hence
\bea\label{firstcosh}
&& \cosh[ \bar{q} \cdot (a_1+a_2)]  \cosh \phi - \sinh[ \bar{q} \cdot (a_1+a_2)]  \sinh \phi 
\nn\\
&& =
u_{1 \mu} u_{2 \nu} \left[ \eta^{\mu\nu} 
-  i \epsilon^{\mu\nu\rho\sigma}  (a_1 + a_2)_{\rho} \bar{q}_{\sigma}
+ \frac{1}{2!} \eta^{\mu\nu} \Bigl( \bar{q} \cdot (a_1+a_2) \Bigr)^2
+ \cdots \right]
\nn\\
&& =  u_{1}^{\mu} \exp(- i (a_1+a_2)* \bar{q})_{\mu}^{\ \nu} u_{2 \nu}
\eea
where $(a*q)^{\mu}_{\ \nu} = \epsilon^{\mu}_{\ \nu\rho\sigma} a^{\rho} \bar{q}^{\sigma}$ and we have used the support of the delta functions. Hence we obtain that
\beq
 {\cal A}^{\textrm{tree}}(1,2 \to 1^{\prime}, 2^{\prime}) =  \frac{4 M_1 M_2}{\hbar^2 \bar{q}^2}
u_{1}^{\mu} \exp(- i (a_1+a_2)* \bar{q})_{\mu}^{\ \nu} u_{2 \nu}.
\eeq
As will be clear in what follows, this form of the tree-level amplitude will be more convenient to our purposes.

Now let us present the tree-level Compton amplitudes which will be relevant when constructing the one-loop amplitudes. For minimally coupled general spin-$S$ amplitude one finds (using BCFW recursion relations)~\cite{Arkani-Hamed:17,Johansson:19}
\bea
A_{S}({\bf 1},2^{+}, 3^{-}, {\bf 4}) &=& - 2 e^2 
\left( \frac{1}{(s - \mu^2) t} + \frac{1}{(u - \mu^2) t} \right)
\langle 3| {\bf 1} |2 \bigl]^{2-2S}
\Bigl( \langle {\bf 4} 3 \rangle \bigl[ {\bf 1} 2 \bigr] + \langle {\bf 1} 3 \rangle \bigl[ {\bf 4} 2 \bigr] \Bigr)^{2S}
\,\,\,\,\,\,\,\,
(S \leq 1)
\nn\\
A_{S}({\bf 1},2^{+}, 3^{+}, {\bf 4}) &=& 2 \mu^{2-2S} e^2 \bigl[ 23 \bigr]^2
\frac{\langle {\bf 1} {\bf 4} \rangle^{2S}}{t} 
\left( \frac{1}{s - \mu^2} + \frac{1}{u - \mu^2} \right) 
\eea
where the mandelstams are $s = (p_1 + p_2)^2$, $u = (p_1 + p_3)^2$, $t = (p_2 + p_3)^2$, and $s +t+u = 2\mu^2$. For scalars there is also an extra contribution (which is a contact term) derived from a four-point vertex. One can also compute the associated Compton amplitudes from the helicity amplitudes just presented by using suitable monodromy relations~\cite{Bjerrum-Bohr:14,Bjerrum-Bohr:16} or simply by summing up the contribution calculated from the BCFW recursion with respect to the $u$-channel. Finally, observe that the amplitude with opposite helicities is not valid for all $S$. Indeed, such Compton amplitudes can be obtained from exponentiated soft factors~\cite{Guevara:19a}. A soft expansion then produces spurious poles due to the fact that one is extrapolating the universal limit -- for the electromagnetic case, the universal terms are precisely the ones associated with $S \leq 1$ in the above Compton amplitude. For $S > 1$ one finds spurious poles which can be eliminated by adding contact terms in order to take into account the non-universality feature of such contributions~\cite{paolocompton,Aoude:22}.

We are now in position to derive the one-loop amplitude associated with the scattering of the massive spinning particles. In order to extract the classical contribution  we will consider the formalism of the leading singularities to obtain the classical pieces associated with triangle diagrams, as described in Refs.~\cite{Forde,Cachazo:17,Guevara:17,Chung:19, nathanls2,Menezes:2022jow}. The reason behind this choice of technique is convenience -- in the course of the calculation, we will see that the triangle contribution is not valid for all $S$ \cite{Arkani-Hamed:20}. This is a known and important aspect of the Compton amplitude; as discussed above, such a representation is indeed not valid for all $S$, unlike the three-point.

For simplicity we will consider only integer spin so that factors of $(-1)^{2S}$ can be ignored. For convenience of the reader, we outline here the main aspects of this calculation. Details can be found in Refs.~\cite{Guevara:17,Guevara:19a}. We will employ the parametrization
\bea
p_1 &=& | \eta\bigr] \langle \lambda | + | \lambda\bigr] \langle \eta |
\nn\\
p_1^{\prime} &=& x_1 | \eta\bigr] \langle \lambda | + \frac{1}{x_1} | \lambda\bigr] \langle \eta |
+ | \lambda\bigr] \langle \lambda |
\nn\\
-\frac{t}{M_1^2} &=& \frac{(x_1-1)^2}{x_1},
\,\,\,
4-\frac{t}{M_1^2} = \frac{(x_1+1)^2}{x_1}
\nn\\
\langle \lambda \eta \rangle &=& \bigl[ \lambda \eta \bigr] = M_1.
\eea
Similar expressions hold for $p_2, p_2^{\prime}$. Eventually we will identify the complex null vector 
$| \lambda\bigr] \langle \lambda |$ with the momentum mismatch $q$ when $x_1 \to 1$. We obtain~\cite{Guevara:17}
\beq
\textrm{LS}^{(a+-)}_{\gamma} = \frac{x_1}{4 M_1^2 \left(x_1^2-1\right)} 
\frac{1}{2 \pi i} \oint_{\Gamma_a} \frac{dy}{y}
\Bigl\langle
A_{S_1}(-{\bf 1}^{\prime}, \boldsymbol\ell, - \ell_1^{+}) 
A_{S_1}({\bf 1}, -\boldsymbol\ell, -\ell_3^{-}) A_{S_2}({\bf 2},\ell_3^{+}, \ell_1^{-}, -{\bf 2}^{\prime})
\Bigr\rangle
\eeq
and
\beq
\textrm{LS}^{(b+-)}_{\gamma} =  \frac{x_2}{4 M_2^2 \left(x_2^2-1\right)} 
\frac{1}{2 \pi i} \oint_{\Gamma_b} \frac{dy}{y}
\Bigl\langle
A_{S_2}(-{\bf 2}^{\prime},\boldsymbol\ell, \ell_1^{+})
A_{S_2}({\bf 2}, -\boldsymbol\ell, \ell_3^{-}) 
A_{S_1}({\bf 1},-\ell_3^{+}, -\ell_1^{-}, -{\bf 1}^{\prime})
\Bigr\rangle
\eeq
where a change of variables was performed. The generalized expectation value $\langle (\cdots) \rangle$ was defined in~\cite{Guevara:19a}: it amounts to a proper normalization given by tensor products of the spin-$1$ polarization vectors:
$$
m^{2 S} \epsilon_{2,\mu_1 \ldots \mu_S} \epsilon_{1}^{\mu_1 \ldots \mu_S}
= \lim_{S \to \infty} \exp\left(- i \frac{ q^{\mu} \epsilon_{q}^{\nu -} S_{1\mu\nu}}
{p_1 \cdot \epsilon^{-}_{q}} \right) \langle {\bf 2} {\bf 1} \rangle^{2S},
$$
on three-point kinematics with $q$ incoming, and
$$
m^{2 S} \epsilon_{2,\mu_1 \ldots \mu_S} \epsilon_{1}^{\mu_1 \ldots \mu_S}
= \lim_{S \to \infty} \exp\left( i \frac{ q^{\mu} \epsilon_{q}^{\nu -} S_{1\mu\nu}}
{p_1 \cdot \epsilon^{-}_{q}} \right) \langle {\bf 2} {\bf 1} \rangle^{2S},
$$
with $q$ outgoing. This properly generalizes the double angle brackets employed in the formula of the impulse for the case of particles with spin, as mentioned previously.

The massless momenta $\ell_1, \ell_3$ now have the following parametrization (for 
$\textrm{LS}^{(a)}_{\gamma}$):

\begin{equation}
\begin{split}
 \ell_{3 \dot{\alpha} \alpha }(y) &= | \ell_3 \bigr] \langle \ell_3 |
= \frac{1}{x_1 + 1} \left( | \eta \bigr] (x_1^2-1)y  + (1+ x_1 y) | \lambda \bigr] \right) 
\frac{1}{x_1 + 1} \left( \langle \eta| (x_1^2-1) - \frac{1}{y} (1+ x_1 y)  \langle \lambda | \right) 
\\&= \left( -\frac{(x_1-1) (x_1 y+1)}{x_1+1} \right) | \eta\bigr] \langle \lambda | 
+ \left( \frac{(x_1-1) (x_1 y+1)}{x_1+1} \right) | \lambda\bigr] \langle \eta | 
\\& \hspace{7.7cm} + \left( -\frac{(x_1 y+1)^2}{(x_1+1)^2 y} \right) | \lambda\bigr] \langle \lambda | 
+ y (x_1-1)^2 | \eta\bigr] \langle \eta | 
\end{split}
\end{equation}
\begin{equation}
\begin{split}
\ell_{1 \dot{\alpha} \alpha}(y)& = | \ell_1 \bigr] \langle \ell_1 |
= \frac{1}{x_1 + 1} \left( - | \eta \bigr] x_1  (x_1^2-1)y + (1- x_1^2 y)  | \lambda \bigr] \right) 
\frac{1}{x_1 + 1} \left( \langle \eta| \frac{(x_1^2-1)}{x_1} + \frac{(1-y)}{y}  \langle \lambda | \right) 
\\&
= \left( \frac{(x_1-1) x_1 (y-1)}{x_1+1} \right) | \eta\bigr] \langle \lambda | 
+ \left( -\frac{(x_1-1) \left(x_1^2 y-1\right)}{x_1 (x_1+1)}\right) | \lambda\bigr] \langle \eta | 
\\&\hspace{7cm} + \left( \frac{(y-1) \left(x_1^2 y-1\right)}{(x_1+1)^2 y} \right) | \lambda\bigr] \langle \lambda | 
- y (x_1-1)^2 | \eta\bigr] \langle \eta | .
\end{split}
\end{equation}

Virtually the same results also hold for $\textrm{LS}^{(b)}_{\gamma}$ except for an overall minus sign and the usual replacement $1 \leftrightarrow 2$ for external particles. The expressions for $(a)$ ($(b)$) are valid for all $S_1$ ($S_2$). Now let us employ the exponential form of $3$-particle amplitudes as given in Ref.~\cite{Guevara:19a}.
\bea
A_{S}({\bf 1}, {\bf 2}, 3^{+}) &=&  \sqrt{2} e 
\frac{1}{M^{2S-1}} x
\bigl[ {\bf 2} |^{2S} \exp\left( i \frac{p_{3}^{\mu} \epsilon_{+}^{\nu} J_{\mu\nu}}{p \cdot \epsilon_{+}} \right)  | {\bf 1} \bigr]^{2S}
\nn\\
A_{S}({\bf 1}, {\bf 2}, 3^{-}) &=& - \sqrt{2} e 
\frac{1}{M^{2S-1}} \frac{1}{x}
\langle {\bf 2} |^{2S} \exp\left( i \frac{p_{3}^{\mu} \epsilon_{-}^{\nu} J_{\mu\nu}}{p \cdot \epsilon_{-}} \right)  | {\bf 1} \rangle^{2S}
\eea
where $p = (p_1-p_2)/2$ and $J_{\mu\nu}$ is the angular-momentum operator whose explicit expression in terms of massive spinors can be found in Ref.~\cite{Guevara:19a}. This operator acts naturally on the product states $|{\bf 1} \rangle^{2S}$ or $|{\bf 2} \rangle^{2S}$. The Compton scattering amplitude for different helicities has also an exponential form~\cite{Guevara:19a}
\bea
A_{S_2}({\bf 2},\ell_3^{+}, \ell_1^{-}, {\bf 2}^{\prime}) &=&  
\frac{2 e^2}{M_2^{2S_2}} \frac{\langle \ell_1 | {\bf 2} | \ell_3 \bigr]^{2}}
{[ 2 p_2^{\prime} \cdot \ell_1] [ 2 p_2 \cdot \ell_1 ]}
\langle {\bf 2}^{\prime} |^{2S_2} \exp\left( i \frac{\ell_{1}^{\mu} \bar{\epsilon}_{\ell_{1}}^{\nu} J_{2 \mu\nu}}
{\bar{p}_2 \cdot \bar{\epsilon}_{\ell_{1}}} \right)  | {\bf 2} \rangle^{2S_2}
\nn\\
A_{S_1}({\bf 1},-\ell_3^{+}, -\ell_1^{-}, {\bf 1}^{\prime}) &=& 
\frac{2 e^2}{M_1^{2S_1}} \frac{\langle \ell_1 | {\bf 1} | \ell_3 \bigr]^{2}}
{[ 2 p_1^{\prime} \cdot \ell_1 ] [ 2 p_1 \cdot \ell_1 ]}
\langle {\bf 1}^{\prime} |^{2S_1} \exp\left( - i \frac{\ell_{1}^{\mu} \bar{\epsilon}_{\ell_{1}}^{\nu \, *} 
J_{1 \mu\nu}}
{\bar{p}_1 \cdot \bar{\epsilon}^{*}_{\ell_{1}}} \right)  | {\bf 1} \rangle^{2S_1}
\eea
and
\bea
A_{S_2}({\bf 2},\ell_3^{-}, \ell_1^{+}, {\bf 2}^{\prime}) &=& 
\frac{2 e^2}{M_2^{2S_2}} \frac{\langle \ell_3 | {\bf 2} | \ell_1 \bigr]^{2}}
{[ 2 p_2^{\prime} \cdot \ell_1] [ 2 p_2 \cdot \ell_1 ]}
\bigl[ {\bf 2}^{\prime} |^{2S_2} \exp\left( i \frac{\ell_{1}^{\mu} \bar{\epsilon}_{\ell_{1}}^{\nu} J_{2 \mu\nu}}
{\bar{p}_2 \cdot \bar{\epsilon}_{\ell_{1}}} \right)  | {\bf 2} \bigr]^{2S_2}
\nn\\
A_{S_1}({\bf 1},-\ell_3^{-}, -\ell_1^{+}, {\bf 1}^{\prime}) &=& 
\frac{2 e^2}{M_1^{2S_1}} \frac{\langle \ell_3 | {\bf 1} | \ell_1 \bigr]^{2}}
{[ 2 p_1^{\prime} \cdot \ell_1 ] [ 2 p_1 \cdot \ell_1 ]}
\bigl[ {\bf 1}^{\prime} |^{2S_1} \exp\left( - i \frac{\ell_{1}^{\mu} \bar{\epsilon}_{\ell_{1}}^{\nu \, *} J_{1 \mu\nu}}
{\bar{p}_1 \cdot \bar{\epsilon}^{*}_{\ell_{1}}} \right)  | {\bf 1} \bigr]^{2S_1}
\eea
where $\bar{p}_2 = (p_2 - p_2^{\prime})/2$ and $\bar{p}_1 = (p_1 - p_1^{\prime})/2$. The bar over the polarization vector for $\ell_1$ is a reminder that its reference momentum was fixed to be $\ell_3$. Now, since
\begin{equation}
 \frac{x_1}{\left(x_1^2-1\right)} = \frac{M_1}{\sqrt{-t}} \frac{1}{\sqrt{4-t/M_1^2}}
\end{equation} 
and $t = \hbar^2 \bar{q}^2$, the integrand must be independent of $\hbar$ in order to produce a classical contribution to the impulse. This implies taking the limit $x_1 \to 1$ in the above integrands. After a series of algebraic manipulations, we arrive at
\bea
\textrm{LS}^{(a+-)}_{\gamma} \biggl|_{x_1 \to 1} &=& - \frac{e^4}{2 \hbar }
\frac{M_1}{ \sqrt{-\bar{q}^2}}  
\frac{1}{ ( s - M_1^2 - M_2^2 )^2 - 4 M_1^2 M_2^2}
\frac{1}{2 \pi i} \oint_{\Gamma_a} \frac{dy}{y}
\left(\frac{ f(s, y)}
{1-y^2}\right)^2
\nn\\
&\times& 
\biggl\langle
\frac{1}{M_1^{4S_1}}
\bigl[ \ell^{I} |^{2S_1} \exp\left( i \frac{\ell_{1}^{\mu} \epsilon_{\ell_1}^{\nu } J_{1^{\prime}\mu\nu}}
{p_1^{\prime} \cdot \epsilon_{\ell_1}} \right)  | {\bf 1}^{\prime} \bigr]^{2S_1}
\langle \ell_{I} |^{2S_1} \exp\left(- i \frac{\ell_{3}^{\mu} \epsilon_{\ell_3}^{\nu } J_{1\mu\nu}}
{p_1 \cdot \epsilon_{\ell_3}} \right)  | {\bf 1} \rangle^{2S_1}
\nn\\
&\times& \frac{1}{M_2^{2S_2}} 
\langle {\bf 2}^{\prime} |^{2S_2} \exp\left( i \frac{\ell_{1}^{\mu} \bar{\epsilon}_{\ell_{1}}^{\nu} J_{2 \mu\nu}}
{\bar{p}_2 \cdot \bar{\epsilon}_{\ell_{1}}} \right)  | {\bf 2} \rangle^{2S_2}
\biggr\rangle
\eea
and
\bea
\textrm{LS}^{(b+-)}_{\gamma} \biggl|_{x_2 \to 1} &=& - \frac{e^4}{2 \hbar }
\frac{M_2}{ \sqrt{-\bar{q}^2}}  
\frac{1}{ ( s - M_1^2 - M_2^2 )^2 - 4 M_1^2 M_2^2}
 \frac{1}{2 \pi i} \oint_{\Gamma_b} \frac{dy}{y}
\left(\frac{ f(s, y)}
{1-y^2}\right)^2
\nn\\
&\times& 
\biggl\langle
\frac{1}{M_2^{4S_2}}
\bigl[ \ell^{I} |^{2S_2} \exp\left( - i \frac{\ell_{1}^{\mu} \epsilon_{\ell_1}^{\nu} J_{2^{\prime}\mu\nu}}
{p_2^{\prime} \cdot \epsilon_{\ell_1}} \right)  | {\bf 2}^{\prime} \bigr]^{2S_2}
\langle  \ell_{I} |^{2S_2} \exp\left( i \frac{\ell_{3}^{\mu} \epsilon_{\ell_3}^{\nu} J_{2\mu\nu}}
{p_2 \cdot \epsilon_{\ell_3}} \right)  | {\bf 2} \rangle^{2S_2}
\nn\\
&\times& \frac{1}{M_1^{2S_1}} 
\langle {\bf 1}^{\prime} |^{2S_1} \exp\left( - i \frac{\ell_{1}^{\mu} \bar{\epsilon}_{\ell_{1}}^{\nu} 
J_{1 \mu\nu}}
{\bar{p}_1 \cdot \bar{\epsilon}_{\ell_{1}}} \right)  | {\bf 1} \rangle^{2S_1} 
\biggr\rangle 
\eea

where we have defined  
\begin{equation}
f(s, y)\equiv  2y (s - M_1^2 - M_2^2) 
- {(1+y^2)}  \sqrt{( s - M_1^2 - M_2^2 )^2 - 4 M_1^2 M_2^2} \,  .
\end{equation}

Now, using the fact that the exponential form in the $3$-particle amplitudes can always be recast into a product of spinors~\cite{Guevara:19a}, we can obtain
\bea
\hspace{-7mm}
\textrm{LS}^{(a+-)}_{\gamma} \biggl|_{x_1 \to 1} &=& - \frac{e^4}{2 \hbar }
\frac{M_1}{ \sqrt{-\bar{q}^2}}  
\frac{1}{ ( s - M_1^2 - M_2^2 )^2 - 4 M_1^2 M_2^2}
 \frac{1}{2 \pi i} \oint_{\Gamma_a} \frac{dy}{y}
\left(\frac{ f(s, y)}
{1-y^2}\right)^2
\nn\\
&\times& 
\biggl\langle
\frac{1}{M_1^{2S_1}}
\langle {\bf 1}^{\prime} |^{2S_1} \exp \left(- i \frac{\ell_{3}^{\mu} \epsilon_{\ell_3}^{\nu} J_{1\mu\nu}}
{p_1 \cdot \epsilon_{\ell_3}} \right)  | {\bf 1} \rangle^{2S_1}
 \frac{1}{M_2^{2S_2}} 
\langle {\bf 2}^{\prime} |^{2S_2} \exp\left( i \frac{\ell_{1}^{\mu} \bar{\epsilon}_{\ell_{1}}^{\nu} J_{2 \mu\nu}}
{p_2 \cdot \bar{\epsilon}_{\ell_{1}}} \right)  | {\bf 2} \rangle^{2S_2}
\biggr\rangle
\eea
and
\bea
\hspace{-7mm}
\textrm{LS}^{(b+-)}_{\gamma} \biggl|_{x_2 \to 1} &=& - \frac{e^4}{2 \hbar }
\frac{M_2}{ \sqrt{-\bar{q}^2}}  
\frac{1}{ ( s - M_1^2 - M_2^2 )^2 - 4 M_1^2 M_2^2}
\frac{1}{2 \pi i} \oint_{\Gamma_b} \frac{dy}{y}
\left(\frac{ f(s, y)}
{1-y^2}\right)^2
\nn\\
&\times& 
\biggl\langle
\frac{1}{M_2^{2S_2}}
\langle {\bf 2}^{\prime} |^{2S_2} \exp\left(i \frac{\ell_{3}^{\mu} \epsilon_{\ell_3}^{\nu} J_{2\mu\nu}}
{p_2 \cdot \epsilon_{\ell_3}} \right)  | {\bf 2} \rangle^{2S_2}
 \frac{1}{M_1^{2S_1}} 
\langle {\bf 1}^{\prime} |^{2S_1} \exp\left( - i \frac{\ell_{1}^{\mu} \bar{\epsilon}_{\ell_{1}}^{\nu} 
J_{1 \mu\nu}}
{p_1 \cdot \bar{\epsilon}_{\ell_{1}}} \right)  | {\bf 1} \rangle^{2S_1} 
\biggr\rangle.
\eea

Essentially one now employs the same technique as carried out in Ref.~\cite{Guevara:19a}, to which we refer the reader for specific details of the calculation. These contributions to the leading singularities are finally given by 
\begin{equation}
\begin{split}
 \textrm{LS}^{(a+-)}_{\gamma} \biggl|_{x_1 \to 1} &= - \frac{e^4}{2 \hbar }
\frac{M_1}{ \sqrt{-\bar{q}^2}}  
\frac{1}{ ( s - M_1^2 - M_2^2 )^2 - 4 M_1^2 M_2^2}
\\&\times  \frac{1}{2 \pi i} \oint_{\Gamma_a}  \frac{dy}{y}
\left(\frac{f(s, y)}{1-y^2}\right)^2
\exp\left( \frac{1+y^2}{2y} \bar{q} \cdot a_1 \right)  
\exp\left[ \left( \frac{  2  M_1 M_2 e^{\phi} (1-y)^2 }{ f(s, y)} + 1 \right) \bar{q} \cdot a_2 \right]
\end{split}
\end{equation}

and
\begin{equation}
\begin{split}
 \textrm{LS}^{(b+-)}_{\gamma} \biggl|_{x_2 \to 1} &= - \frac{e^4}{2 \hbar }
\frac{M_2}{ \sqrt{-\bar{q}^2}}  
\frac{1}{ ( s - M_1^2 - M_2^2 )^2 - 4 M_1^2 M_2^2}
\\&\times  \frac{1}{2 \pi i} \oint_{\Gamma_b}  \frac{dy}{y}
\left(\frac{f(s, y)}{1-y^2}\right)^2
\exp\left( \frac{1+y^2}{2y} \bar{q} \cdot a_2 \right)  
\exp\left[ \left( \frac{  2  M_1 M_2 e^{\phi} (1-y)^2 }{ f(s, y)} + 1 \right) \bar{q} \cdot a_1 \right].
\end{split}
\end{equation}

We recall again that the expression for the $a$ ($b$) topology above is valid for all $a_1$ ($a_2$), but up to second order in $a_2$ ($a_1$). For such calculations, we have used the following expression for the classical spin and the classical rescaled spin vector
$$
S^{\mu\nu} = \epsilon^{\mu\nu\alpha\beta} p_{\alpha} \frac{a_{\beta}}{\hbar},
\,\,\,
a_{\lambda} = \frac{\hbar}{2 M^2} \epsilon_{\lambda\mu\nu\alpha} S^{\mu\nu} p^{\alpha} .
$$
Together with an evaluation of a Gram determinant for $3$-particle kinematics, this gives us
$$
a_1 \cdot \bar{q} = \pm i \frac{ q^{\mu} \epsilon_{q}^{\nu \pm} S_{1\mu\nu}}
{p_1 \cdot \epsilon^{\pm}_{q}}
\to 
\frac{ q^{\mu} \epsilon_{q}^{\nu -} J_{1\mu\nu}}
{p_1 \cdot \epsilon^{-}_{q}} = - 2 \frac{ q^{\mu} \epsilon_{q}^{\nu -} S_{1\mu\nu}}
{p_1 \cdot \epsilon^{-}_{q}}
$$
for $q$ incoming, and
$$
a_1 \cdot \bar{q} = \mp i \frac{ q^{\mu} \epsilon_{q}^{\nu \pm} S_{1\mu\nu}}
{p_1 \cdot \epsilon^{\pm}_{q}}
\to 
\frac{ q^{\mu} \epsilon_{q}^{\nu -} J_{1\mu\nu}}
{p_1 \cdot \epsilon^{-}_{q}} = - 2 \frac{ q^{\mu} \epsilon_{q}^{\nu -} S_{1\mu\nu}}
{p_1 \cdot \epsilon^{-}_{q}}
$$
for $q$ outgoing. More details of such manipulations can be found in Ref.~\cite{Guevara:19a}. 

Now we need to discuss the contours $\Gamma_{a,b}$. The triangle topology is, as well known, associated with choosing a contour around zero or infinity. To unify both descriptions, one resorts to the following change of variables
$$
z = \frac{1+y^2}{2y} .
$$
In this way, both contours are mapped to $z = \infty$. On the other hand, one needs to sum over internal helicities to obtain the full contribution to the triangle leading singularity. It is easy to see from the above expressions that, in the limit $x_1 \to 1$, the conjugate helicity configuration yields the same residue. As for the equal-helicity configurations, it is also easy to see, from the $3$-point and Compton amplitudes, that they produce the same result as the spinless case multiplied by spin factors such as 
$\langle {\bf 1} {\bf 1}^{\prime} \rangle^{2S_1} \bigl[ {\bf 2} {\bf 2}^{\prime} \bigr]^{2S_2}/(M_1^{2S_1} M_2^{2S_2})$ -- so one can employ similar considerations to show that the corresponding expressions has zero residue at both $0, \infty$. Hence we can display the full triangle leading singularity in the limit $x_1 \to 1$:
 
\begin{equation}
\begin{split}
\textrm{LS}^{(a)}_{\triangle,\gamma} \biggl|_{x_1 \to 1} &= - \frac{e^4}{\hbar }
\frac{M_1}{ \sqrt{-\bar{q}^2}}  
\frac{1}{ ( s - M_1^2 - M_2^2 )^2 - 4 M_1^2 M_2^2}
\\& \times \frac{1}{2 \pi i} \oint_{\Gamma_{\infty}} dz
\frac{ (h(s, z))^2}{(z^2-1)^{3/2}}
\exp\left( z \, \bar{q} \cdot a_1 \right)
\exp\left[ \left( \frac{  2  M_1 M_2 e^{\phi} (z-1) }{h(s, z)} + 1 \right) \bar{q} \cdot a_2 \right]
\end{split}
\end{equation}
and
\begin{equation}
\begin{split}
\textrm{LS}^{(b)}_{\triangle,\gamma} \biggl|_{x_2 \to 1} &= - \frac{e^4}{\hbar }
\frac{M_2}{ \sqrt{-\bar{q}^2}}  
\frac{1}{ ( s - M_1^2 - M_2^2 )^2 - 4 M_1^2 M_2^2}
\\& \times \frac{1}{2 \pi i} \oint_{\Gamma_{\infty}} dz
\frac{ (h(s, z))^2}{(z^2-1)^{3/2}}
\exp\left( z \, \bar{q} \cdot a_2 \right)
\exp\left[ \left( \frac{  2  M_1 M_2 e^{\phi} (z-1) }{h(s, z)} + 1 \right) \bar{q} \cdot a_1 \right]
\end{split}
\end{equation}
with
\begin{equation}
 h(s, z)\equiv (s - M_1^2 - M_2^2)
-  z \sqrt{( s - M_1^2 - M_2^2 )^2 - 4 M_1^2 M_2^2} \,  .
\end{equation}
The contour is $\Gamma_{\infty} = \{ |z| = R \}$, where $R>1/v$ is large but finite. The technical reason behind this choice can be found in Ref.~\cite{Guevara:19a}. 

As well known, such expressions allow us to display explicitly the classical contribution coming from the triangle diagram of the one-loop amplitude. In few words, the leading singularity associated with the triangle topology is the double discontinuity across the $t$-channel, which projects away quantum corrections and contact interactions. As described in Ref.~\cite{Cachazo:17}, this double discontinuity allows us to verify that the leading singularity can be taken to be self-similar, leading us to
\beq
A^{(1)}_{\triangle,\textrm{class}} =  \frac{1}{4} \left( \textrm{LS}^{(a)}_{\triangle,\gamma} 
+ \textrm{LS}^{(a)}_{\triangle,\gamma} \right) \biggl|_{x_1 \to 1} .
\eeq
One finally finds that
\bea
\hspace{-5mm}
{\cal A}^{(1)}_{\triangle,\textrm{class}} &=& - \frac{1}{4\hbar} \frac{1}{ \sqrt{-\bar{q}^2}} 
\left\{  \frac{M_1}{ ( s - M_1^2 - M_2^2 )^2 - 4 M_1^2 M_2^2}
\oint_{\Gamma_{\infty}} \frac{dz}{2 \pi i} 
\frac{(h(s, z))^2}{(z^2-1)^{3/2}}
\right.
\nn\\
&\times& \left. 
\exp\left( z \, \bar{q} \cdot a_1 \right)
\exp\left[ \left( \frac{  2  M_1 M_2 e^{\phi} (z-1) }{h(s, z)} + 1 \right) \bar{q} \cdot a_2 \right] + {\cal O}(a_2^3)
\right.
\nn\\
&+& \left. \frac{M_2}{ ( s - M_1^2 - M_2^2 )^2 - 4 M_1^2 M_2^2}
 \oint_{\Gamma_{\infty}} \frac{dz}{2 \pi i}
\frac{ (h(s, z))^2}{(z^2-1)^{3/2}}
\right.
\nn\\
&\times& \left.  
\exp\left( z \, \bar{q} \cdot a_2 \right)
\exp\left[ \left( \frac{  2  M_1 M_2 e^{\phi} (z-1) }{h(s, z)} + 1 \right) \bar{q} \cdot a_1 \right] + {\cal O}(a_1^3)
\right\} 
\eea
where factors of $e/\sqrt{\hbar}$ were extracted. Therefore, we now have the triangle contribution to the NLO impulse in the $\sqrt{\textrm{Kerr}}$ case. Observe that universal terms in the soft theorem cannot reproduce contributions to all orders in the spins of the particles.

Now let us turn to the box contribution. Unfortunately, the calculation of the associated coefficients is not as straightforward as in the spinless case, since the tensor reduction of the integrals is more involved. In any case, one can still obtain the classical contribution from the associated numerators by observing that, from the effective field theory perspective, the box contribution is an infrared-divergent iteration of tree-level exchange~\cite{Bern:19}. This implies that the box numerator should equal the product of two tree-level numerators associated with the tree-level one-photon exchange calculated earlier. We find that
\bea
\hspace{-3mm}
i A^{(1)}_{\Box} &=&
4 e^4 \int {\hat{d} ^D \ell}  \frac{B_{\gamma}(\ell, p_1,p_2) B_{\gamma}(-\ell-p_1+p_1^{\prime}, p_1,p_2)}
{ \ell^2 [(\ell + p_1)^2 - M_1^2] (\ell+p_1-p_1^{\prime})^2  [ (\ell-p_2)^2 - M_2^2]}
\nn\\
i A^{(1)}_{\boxtimes} &=& 
4 e^4 \int {\hat{d} ^D \ell}  
\frac{ B_{\gamma}(\ell, p_1,p_2) B_{\gamma}(-\ell-p_1+p_1^{\prime}, p_1,p_2)}
{\ell^2 [(\ell + p_1)^2 - M_1^2](\ell+p_1-p_1^{\prime})^2 [ (\ell+p_2^{\prime})^2 - M_2^2] }
\eea
where 
\beq
B_{\gamma}(q, p_1,p_2) =  
2 p_{1}^{\mu} \exp\left[- i \frac{(a_1+a_2)}{\hbar} * q \right]_{\mu}^{\ \nu} p_{2 \nu}. 
\eeq
We have dropped any explicit term in the numerators that will not produce a classical contribution. However, observe there are still remainder sources of $\hbar$ factors buried in the denominators. One obtains
\beq
i {\cal A}^{(1)}_{\Box}(1,2 \to 1^{\prime}, 2^{\prime}) \bigg|_{\textrm{Classical}}
+ i {\cal A}^{(1)}_{\boxtimes}(1,2 \to 1^{\prime}, 2^{\prime}) \bigg|_{\textrm{Classical}}
= i {\cal A}_{-2} + i {\cal A}_{-1} + {\cal O}(\hbar^0)
\eeq
where
\bea
i {\cal A}_{-2} &=& 
\frac{4}{\hbar^{2+2\epsilon}} 
\int {\hat{d} ^D \bar{\ell}}  
 \frac{ B_{\gamma}(\bar{\ell}, p_1,p_2) B_{\gamma}(-\bar{\ell}+\bar{q}, p_1,p_2) }
 {\bar{\ell}^2 (\bar{\ell} - \bar{q})^2 
( 2 \bar{\ell} \cdot p_1 + i\varepsilon ) 
 (- 2 \bar{\ell} \cdot p_2  + i\varepsilon )}
\nn\\
&+&  \frac{4}{\hbar^{2+2\epsilon}} 
\int {\hat{d} ^D \bar{\ell}}  
 \frac{ B_{\gamma}(\bar{\ell}, p_1,p_2) B_{\gamma}(-\bar{\ell}+\bar{q}, p_1,p_2) }
 {\bar{\ell}^2 (\bar{\ell} - \bar{q})^2 
( 2 \bar{\ell} \cdot p_1 + i\varepsilon ) 
 (2 \bar{\ell} \cdot p_2  + i\varepsilon )}
\eea
and
\bea
i {\cal A}_{-1}  &=& \frac{1}{2 \hbar^{1+2\epsilon}} 
 \int {\hat{d} ^D \bar{\ell}} \,
\frac{ B_{\gamma}(\bar{\ell}, p_1,p_2) B_{\gamma}(-\bar{\ell}+\bar{q}, p_1,p_2) }
{\bar{\ell}^2 (\bar{\ell} - \bar{q})^2 
( \bar{\ell} \cdot p_1  + i\varepsilon ) 
(\bar{\ell} \cdot p_2 - i\varepsilon )} 
\left( \frac{\bar{\ell}^2}{\bar{\ell} \cdot p_1 + i\varepsilon} \right)
+ \{ 2 \leftrightarrow -2 \}
\nn\\
&+& \frac{1}{2\hbar^{1+2\epsilon}} 
\int {\hat{d} ^D \bar{\ell}} \,
\frac{ B_{\gamma}(\bar{\ell}, p_1,p_2) B_{\gamma}(-\bar{\ell}+\bar{q}, p_1,p_2) }
{\bar{\ell}^2 (\bar{\ell} - \bar{q})^2 
( \bar{\ell} \cdot p_1 + i\varepsilon ) 
 (\bar{\ell} \cdot p_2 - i\varepsilon )}
\frac{\bar{\ell}^2}{- \bar{\ell} \cdot p_2  + i\varepsilon} 
\nn\\
&-& \frac{1}{2 \hbar^{1+2\epsilon}} 
\int {\hat{d} ^D \bar{\ell}} \, \frac{ B_{\gamma}(\bar{\ell}, p_1,p_2) B_{\gamma}(-\bar{\ell}+\bar{q}, p_1,p_2) }
{\bar{\ell}^2 (\bar{\ell} - \bar{q})^2 
( \bar{\ell} \cdot p_1 + i\varepsilon ) 
(\bar{\ell} \cdot p_2  + i\varepsilon )}
\frac{ ( \bar{\ell} - \bar{q})^2 - \bar{q}^2 }{\bar{\ell} \cdot p_2 + i\varepsilon} .
\eea
 Let us turn to ${\cal A}_{-2}$. We find that
\beq
i {\cal A}_{-2} = - \frac{i}{\hbar^{2+2\epsilon}} 
\int {\hat{d} ^D \bar{\ell}} \,
 \frac{ B_{\gamma}(\bar{\ell}, p_1,p_2) B_{\gamma}(-\bar{\ell}+\bar{q}, p_1,p_2) }
 {\bar{\ell}^2 (\bar{\ell} - \bar{q})^2 
( \bar{\ell} \cdot p_1 + i\varepsilon ) } \hat{\delta}( \bar{\ell} \cdot p_2) .
\eeq
where we used the delta-function trick 
\begin{equation}\label{deltatrick}
\frac{1}{x - i\varepsilon} - \frac{1}{x + i\varepsilon} = i \hat{\delta}(x). 
\end{equation} 
 
Now, use a trick employed in~\cite{Kosower:19}, we exploit the linear change of variables 
$\bar{\ell}^{\prime} = \bar{q} - \bar{\ell}$ and use the on-shell conditions $\bar{q} \cdot p_1 = - \hbar \bar{q}^2/2$ and $\bar{q} \cdot p_2 =  \hbar \bar{q}^2/2$. Afterwards, we Laurent expand in $\hbar$, keep terms up to ${\cal O}(1/\hbar)$ and finally average over equivalent forms:
\beq
i {\cal A}_{-2} = - \frac{1}{2\hbar^{2}} 
\int {\hat{d} ^4 \bar{\ell}} \,
 \frac{ B_{\gamma}(\bar{\ell}, p_1,p_2) B_{\gamma}(-\bar{\ell}+\bar{q}, p_1,p_2) }
 {\bar{\ell}^2 (\bar{\ell} - \bar{q})^2} 
 \hat{\delta}( \bar{\ell} \cdot p_2) \hat{\delta}( \bar{\ell} \cdot p_1) 
 + {\cal O}(1/\hbar) .
\eeq
Observe that we have already taken $D \to 4$.

\begin{figure}[H]
\begin{center}

\tikzset{every picture/.style={line width=0.75pt}} 

\begin{tikzpicture}[x=0.85pt,y=0.85pt,yscale=-1,xscale=1]

\draw [color={rgb, 255:red, 208; green, 2; blue, 27 }  ,draw opacity=1 ] [dash pattern={on 4.5pt off 4.5pt}]  (240.67,1149.17) -- (276,1149.17) ;
\draw    (258.33,1110.67) -- (291.89,1070.65) ;
\draw [shift={(275.11,1090.66)}, rotate = 129.98] [fill={rgb, 255:red, 0; green, 0; blue, 0 }  ][line width=0.08]  [draw opacity=0] (5.36,-2.57) -- (0,0) -- (5.36,2.57) -- cycle    ;
\draw    (258.33,1110.67) .. controls (256.68,1112.35) and (255.02,1112.37) .. (253.33,1110.72) .. controls (251.65,1109.07) and (249.98,1109.09) .. (248.33,1110.77) .. controls (246.68,1112.45) and (245.01,1112.46) .. (243.33,1110.81) .. controls (241.65,1109.16) and (239.98,1109.18) .. (238.33,1110.86) .. controls (236.68,1112.54) and (235.01,1112.56) .. (233.33,1110.91) .. controls (231.65,1109.26) and (229.98,1109.28) .. (228.33,1110.96) .. controls (226.68,1112.64) and (225.01,1112.66) .. (223.33,1111.01) .. controls (221.65,1109.36) and (219.99,1109.38) .. (218.34,1111.06) .. controls (216.69,1112.74) and (215.02,1112.76) .. (213.34,1111.11) .. controls (211.66,1109.46) and (209.99,1109.48) .. (208.34,1111.16) .. controls (206.69,1112.84) and (205.02,1112.86) .. (203.34,1111.21) .. controls (201.66,1109.56) and (199.99,1109.58) .. (198.34,1111.26) .. controls (196.69,1112.94) and (195.02,1112.96) .. (193.34,1111.31) -- (190.67,1111.33) -- (190.67,1111.33) ;
\draw    (204.67,1118.92) -- (225.17,1118.92) ;
\draw [shift={(228.17,1118.92)}, rotate = 180] [fill={rgb, 255:red, 0; green, 0; blue, 0 }  ][line width=0.08]  [draw opacity=0] (5.36,-2.57) -- (0,0) -- (5.36,2.57) -- cycle    ;
\draw    (290.67,1229.18) -- (258.33,1187.67) ;
\draw [shift={(274.5,1208.42)}, rotate = 52.08] [fill={rgb, 255:red, 0; green, 0; blue, 0 }  ][line width=0.08]  [draw opacity=0] (5.36,-2.57) -- (0,0) -- (5.36,2.57) -- cycle    ;
\draw    (258.33,1187.67) -- (258.33,1110.67) ;
\draw    (258.33,1187.67) .. controls (256.68,1189.35) and (255.02,1189.37) .. (253.33,1187.72) .. controls (251.65,1186.07) and (249.98,1186.09) .. (248.33,1187.77) .. controls (246.68,1189.45) and (245.01,1189.46) .. (243.33,1187.81) .. controls (241.65,1186.16) and (239.98,1186.18) .. (238.33,1187.86) .. controls (236.68,1189.54) and (235.01,1189.56) .. (233.33,1187.91) .. controls (231.65,1186.26) and (229.98,1186.28) .. (228.33,1187.96) .. controls (226.68,1189.64) and (225.01,1189.66) .. (223.33,1188.01) .. controls (221.65,1186.36) and (219.99,1186.38) .. (218.34,1188.06) .. controls (216.69,1189.74) and (215.02,1189.76) .. (213.34,1188.11) .. controls (211.66,1186.46) and (209.99,1186.48) .. (208.34,1188.16) .. controls (206.69,1189.84) and (205.02,1189.86) .. (203.34,1188.21) .. controls (201.66,1186.56) and (199.99,1186.58) .. (198.34,1188.26) .. controls (196.69,1189.94) and (195.02,1189.96) .. (193.34,1188.31) -- (190.67,1188.33) -- (190.67,1188.33) ;
\draw    (209.67,1196.42) -- (230.17,1196.42) ;
\draw [shift={(233.17,1196.42)}, rotate = 180] [fill={rgb, 255:red, 0; green, 0; blue, 0 }  ][line width=0.08]  [draw opacity=0] (5.36,-2.57) -- (0,0) -- (5.36,2.57) -- cycle    ;
\draw    (190.67,1188.33) -- (190.67,1111.33) ;
\draw [color={rgb, 255:red, 208; green, 2; blue, 27 }  ,draw opacity=1 ] [dash pattern={on 4.5pt off 4.5pt}]  (208.33,1149.83) -- (173,1149.83) ;
\draw    (190.67,1111.33) -- (157.11,1071.31) ;
\draw [shift={(173.89,1091.32)}, rotate = 50.02] [fill={rgb, 255:red, 0; green, 0; blue, 0 }  ][line width=0.08]  [draw opacity=0] (5.36,-2.57) -- (0,0) -- (5.36,2.57) -- cycle    ;
\draw    (158.33,1229.84) -- (190.67,1188.33) ;
\draw [shift={(174.5,1209.09)}, rotate = 127.92] [fill={rgb, 255:red, 0; green, 0; blue, 0 }  ][line width=0.08]  [draw opacity=0] (5.36,-2.57) -- (0,0) -- (5.36,2.57) -- cycle    ;

\draw (213,1200) node [anchor=north west][inner sep=0.75pt]    {$\ell$};
\draw (199,1125.5) node [anchor=north west][inner sep=0.75pt]    {$q-\ell$};
\draw (282.5,1192) node [anchor=north west][inner sep=0.75pt]    {$p_{1}$};
\draw (282,1084) node [anchor=north west][inner sep=0.75pt]    {$p'_{1}$};
\draw (148.5,1194) node [anchor=north west][inner sep=0.75pt]    {$p_{2}$};
\draw (148.5,1090) node [anchor=north west][inner sep=0.75pt]    {$p'_{2}$};

\end{tikzpicture}
\caption{The two-particle cut box.}
\label{cutbox}
\end{center}
\end{figure}
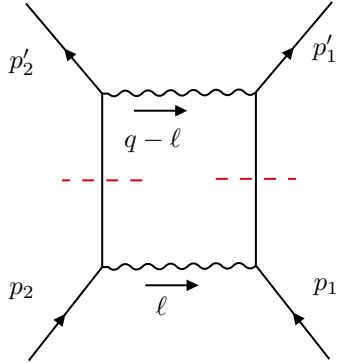

Now, to cancel the ${\cal O}(1/\hbar^2)$ terms, we need to consider the two-particle cut of the one-loop box, as depicted in Fig.~\ref{cutbox}, to which we now turn. Here the two-particle cut of the one-loop box is just the $s$-channel cut of the one-loop box weighted by the loop momentum $w$.  Recall that in this calculation we take $p_1'$ and $p_2'$ as outgoing; then we extract factors of $e/\sqrt{\hbar}$ and use similar considerations as above. There are two contributions that we should consider. The cut of the matter propagators produce the same contributions coming from each one of them, {and we find that the analytic expression associated with Fig.~\ref{cutbox} reads}
\beq
\frac{4}{\hbar^{4}}
\frac{ B_{\gamma}(\bar{w}, p_1,p_2) B_{\gamma}(-\bar{w}+\bar{q}, p_1^{\prime},p_2^{\prime}) }
{\bar{w}^2 (\bar{w} - \bar{q})^2} .
\eeq
We now relabel $w \to \ell$ for ease of comparison; one has that
\bea
{\cal I}^{\mu (1)}_{\gamma,\textrm{Cut Box}} &=& 
- i \frac{4}{\hbar^{2}} 
\int {\hat{d} ^4 \bar{\ell}} \,\hat{\delta}( 2 \bar{\ell} \cdot p_1 + \hbar \bar{\ell}^2) 
\hat{\delta}( 2 \bar{\ell} \cdot p_2 - \hbar \bar{\ell}^2) \bar{\ell}^{\mu}
\frac{ B_{\gamma}(\bar{\ell}, p_1,p_2) B_{\gamma}(-\bar{\ell}+\bar{q},p_1^{\prime},p_2^{\prime}) }
{\bar{\ell}^2 (\bar{\ell} - \bar{q})^2} .
\eea
Observe that there is still an additional factor of $\hbar$ to be multiplied into the final expression for the weighted cut box. As above, expand in $\hbar$, and truncate after order $1/\hbar$, so that
\beq
{\cal I}^{\mu (1)}_{\gamma,\textrm{Cut Box}} = {\cal C}_{-2} ^{\mu} + {\cal C}_{-1} ^{\mu}
\eeq
where
\bea
{\cal C}_{-2}^{\mu} &=& - \frac{i}{\hbar^{2}}
\int {\hat{d} ^4 \bar{\ell}} \,\hat{\delta}( \bar{\ell} \cdot p_1 ) 
\hat{\delta}( \bar{\ell} \cdot p_2 ) 
\frac{\bar{\ell}^{\mu}}{\bar{\ell}^2 (\bar{\ell} - \bar{q})^2}
B_{\gamma}(\bar{\ell}, p_1,p_2) B_{\gamma}(-\bar{\ell}+\bar{q}, p_1,p_2)
\nn\\
&=& - \frac{i}{2 \hbar^{2}} 
\int {\hat{d} ^4 \bar{\ell}} \,\frac{q^{\mu}}{\bar{\ell}^2 (\bar{\ell} - \bar{q})^2}
\hat{\delta}( \bar{\ell} \cdot p_1 ) 
\hat{\delta}( \bar{\ell} \cdot p_2 ) 
B_{\gamma}(\bar{\ell}, p_1,p_2) B_{\gamma}(-\bar{\ell}+\bar{q}, p_1,p_2)
+ {\cal O}(1/\hbar)
\eea
and
\beq
{\cal C}_{-1}^{\mu} =  \widetilde{{\cal C}}^{(1) \mu} +  \widetilde{{\cal C}}^{(2) \mu}
\eeq
with
\bea
\widetilde{{\cal C}}^{(1) \mu} &=& i \frac{2}{\hbar} 
\int {\hat{d} ^4 \bar{\ell}} \,\hat{\delta}( \bar{\ell} \cdot p_1) 
\hat{\delta}( \bar{\ell} \cdot p_2) \bar{\ell}^{\mu}
\frac{ B_{\gamma}(\bar{\ell}, p_1,p_2) p_{1}^{\rho} 
\exp\Bigl[- i (a_1+a_2) * (-\bar{\ell}+\bar{q}) \Bigr]_{\rho}^{\ \sigma} \bar{q}_{\sigma} }
{\bar{\ell}^2 (\bar{\ell} - \bar{q})^2} 
\nn\\
&-& i \frac{2}{\hbar} 
\int {\hat{d} ^4 \bar{\ell}} \,\hat{\delta}( \bar{\ell} \cdot p_1) 
\hat{\delta}( \bar{\ell} \cdot p_2) \bar{\ell}^{\mu}
\frac{ B_{\gamma}(\bar{\ell}, p_1,p_2)  \bar{q}^{\rho} 
\exp\Bigl[- i (a_1+a_2) * ( -\bar{\ell} + \bar{q} ) \Bigr]_{\rho}^{\ \sigma} p_{2 \sigma} }
{\bar{\ell}^2 (\bar{\ell} - \bar{q})^2} 
\eea
and
\beq
\widetilde{{\cal C}}^{(2) \mu} =
 - \frac{i}{2\hbar}
\int {\hat{d} ^4 \bar{\ell}} \,
\bar{\ell}^2 \frac{\bar{\ell}^{\mu}}{\bar{\ell}^2 (\bar{\ell} - \bar{q})^2}
B_{\gamma}(\bar{\ell}, p_1,p_2) B_{\gamma}(-\bar{\ell}+\bar{q}, p_1,p_2)
\Bigl( \hat{\delta}^{\prime}( \bar{\ell} \cdot p_1) 
\hat{\delta}( \bar{\ell} \cdot p_2 ) 
- \hat{\delta}( \bar{\ell} \cdot p_1) 
\hat{\delta}^{\prime}( \bar{\ell} \cdot p_2 ) \Bigr) .
\eeq
In ${\cal C}_{-2}$ we have performed the linear change of variables $\bar{\ell}^{\prime} = \bar{q} - \bar{\ell}$, used the on-shell conditions $\bar{q} \cdot p_1 = - \hbar \bar{q}^2/2$ and $\bar{q} \cdot p_2 =  \hbar \bar{q}^2/2$, Laurent expanded in $\hbar$, keeping terms up to ${\cal O}(1/\hbar)$, and finally averaged over equivalent forms. It is easy to see that both singular terms $ q^{\mu} {\cal A}_{-2}$ and 
${\cal C}_{-2}^\mu $ cancel out in the expression for the classical impulse, leaving only well-defined contributions.

Following Ref.~\cite{Kosower:19}, one must now collect all ${\cal O}(1/\hbar)$ terms to obtain cleaner expressions for the box, crossed box and cut-box contributions. We find that
\bea
\hspace{-8mm}
i {\cal A}_{-1}  + [i {\cal A}_{-2}]_{{\cal O}(1/\hbar)} &=&
\frac{i}{2\hbar^{1+2\epsilon}} 
\int {\hat{d} ^D \bar{\ell}} \,
\frac{ \bar{\ell} \cdot (\bar{\ell} - \bar{q}) \hat{\delta}(\bar{\ell} \cdot p_2) }{\bar{\ell}^2 (\bar{\ell} - \bar{q})^2 
( \bar{\ell} \cdot p_1  + i\varepsilon )^2 } 
B_{\gamma}(\bar{\ell}, p_1,p_2) B_{\gamma}(-\bar{\ell}+\bar{q}, p_1,p_2)
\nn\\
&+&
\frac{i}{ 2 \hbar^{1+2\epsilon}} 
\int {\hat{d} ^D \bar{\ell}} \,
\frac{ \bar{\ell} \cdot (\bar{\ell} - \bar{q}) \hat{\delta}(\bar{\ell} \cdot p_1) }
{\bar{\ell}^2 (\bar{\ell} - \bar{q})^2 ( \bar{\ell} \cdot p_2  - i\varepsilon )^2 } 
B_{\gamma}(\bar{\ell}, p_1,p_2) B_{\gamma}(-\bar{\ell}+\bar{q}, p_1,p_2)
\nn\\
&-& \frac{1}{4 \hbar^{1+2\epsilon}} 
\int {\hat{d} ^D \bar{\ell}} \,
( 2 \bar{\ell} \cdot \bar{q} - \bar{\ell}^2 ) \frac{ B_{\gamma}(\bar{\ell}, p_1,p_2) 
B_{\gamma}(-\bar{\ell}+\bar{q}, p_1,p_2) }{\bar{\ell}^2 (\bar{\ell} - \bar{q})^2}
\nn\\
&\times& \Bigl( \hat{\delta}^{\prime}( \bar{\ell} \cdot p_1) 
\hat{\delta}( \bar{\ell} \cdot p_2 ) 
- \hat{\delta}( \bar{\ell} \cdot p_1) 
\hat{\delta}^{\prime}( \bar{\ell} \cdot p_2 ) \Bigr) 
\eea
and
\bea
\hspace{-8mm}
\widetilde{{\cal C}}^{(2)  \mu} + [{\cal C}_{-2}^{\mu}]_{{\cal O}(1/\hbar)} &=& 
- \frac{i}{2 \hbar} 
\int {\hat{d} ^4 \bar{\ell}} \,
\bar{\ell} \cdot (\bar{\ell} - \bar{q}) \frac{\bar{\ell}^{\mu}}{\bar{\ell}^2 (\bar{\ell} - \bar{q})^2}
B_{\gamma}(\bar{\ell}, p_1,p_2) B_{\gamma}(-\bar{\ell}+\bar{q}, p_1,p_2)
\nn\\
&\times& \Bigl( \hat{\delta}^{\prime}( \bar{\ell} \cdot p_1) 
\hat{\delta}( \bar{\ell} \cdot p_2 ) 
- \hat{\delta}( \bar{\ell} \cdot p_1) 
\hat{\delta}^{\prime}( \bar{\ell} \cdot p_2 ) \Bigr)
\nn\\
&-& \frac{i}{4 \hbar} 
\int {\hat{d} ^4 \bar{\ell}} \,
( 2 \bar{\ell} \cdot \bar{q} - \bar{\ell}^2 ) \frac{\bar{q}^{\mu}}{\bar{\ell}^2 (\bar{\ell} - \bar{q})^2}
B_{\gamma}(\bar{\ell}, p_1,p_2) B_{\gamma}(-\bar{\ell}+\bar{q}, p_1,p_2)
\nn\\
&\times& \Bigl( \hat{\delta}^{\prime}( \bar{\ell} \cdot p_1) 
\hat{\delta}( \bar{\ell} \cdot p_2 ) 
- \hat{\delta}( \bar{\ell} \cdot p_1) 
\hat{\delta}^{\prime}( \bar{\ell} \cdot p_2 ) \Bigr).
\eea
The expression given above for the remaining contribution $\widetilde{{\cal C}}^{(1)}$ is already in good shape so it does not require any further polishing.

{There is one important point that ought to be addressed in this one-loop calculation.  When calculating scattering amplitudes, infrared-divergent integrals usually appear. In such cases one may resort to the standard dimensional regularization, and we take space-time dimensions to be $d = 4-2\epsilon$. Now infrared singularities are identified as poles in $\epsilon$, which can interfere with ${\cal O}(\epsilon)$ terms, introducing in this way finite contributions. It is well known that in the full quantum theory such terms arise. One thus has to confirm that ${\cal O}(\epsilon)$ terms in the integrand do not generate extra contributions in our case. Basically, one can pinpoint two sources of contamination: Gram determinants with five independent vectors and factors of dimension arising from the trace of the metric. The former cannot arise, as the one-loop amplitude encloses only four independent momenta. On the other hand, the latter can occur, but only in bubble diagrams, which do not contribute to the classical limit. In other words, what distinguishes four-dimensional methods from cuts in arbitrary dimensions is the emergence of rational terms possessing improper correct scaling associated with small transferred momentum which is required by the classical limit. This subtle point is extensively discussed in Ref.~\cite{Bern:19}, to which we refer the reader for more details (even though the analysis put forward in such a reference was carried out for scalar particles, it is easy to see that the same holds for spinning particles). See also Ref.~\cite{Bern:2020buy}.}

\subsection{Electromagnetic impulse and the scattering angle}\label{sectionem}

We can now gather all of our terms to obtain our final result for the linear impulse, up to NLO, for a non-aligned spin configuration. Inserting the above results in Eq.~(\ref{impulse}), we find that 
\bea\label{ii1}
\hspace{-5mm}
\Delta p^\mu_{\text{LO}, 1} &=& i e^2
 \int {\hat{d} ^4 \bar{q}} \,\hat{\delta}( \bar{q} \cdot u_1 ) \hat{\delta}( \bar{q} \cdot u_2 )
e^{- i b \cdot {\bar q}}  
\frac{\bar{q}^{\mu}}{ \bar{q}^2 } \,
u_{1}^{\rho} \exp(- i (a_1+a_2)* \bar{q})_{\rho}^{\ \sigma} u_{2 \sigma}
\nn\\
&=& \frac{e^2}{2\pi \sinh\phi} \, \textrm{Re} 
\left[\frac{\cosh\phi \, b_{\perp}^{\mu} - i \epsilon^{\mu\nu\alpha\beta} b_{\perp \nu} u_{1 \alpha} u_{2 \beta}}
{b_{\perp}^{2}}
\right]
\eea
and
\bea\label{ii2}
\Delta p^\mu_{\text{NLO}, 1} &=& - i \frac{e^4}{2} 
 \int {\hat{d} ^4 \bar{q}} \,\hat{\delta}( \bar{q} \cdot u_1 ) \hat{\delta}( \bar{q} \cdot u_2 )
e^{- i b \cdot {\bar q}} \bar{q}^{\mu}
\frac{1}{8 \sqrt{-\bar{q}^2}}
\nn\\
&\times& 
\left\{  \frac{1}{M_2 \beta^2}
\oint_{\Gamma_{\infty}} \frac{dz}{2 \pi i} 
\frac{ ( 1 - \beta z )^2}{(z^2-1)^{3/2}}
\exp\left( z \, \bar{q} \cdot a_1 \right)
\exp\left[ \left( \frac{  e^{\phi} (z-1) }{\gamma ( 1 -  \beta z )} + 1 \right) \bar{q} \cdot a_2 \right] 
+ {\cal O}(a_2^3)
\right.
\nn\\
&+& \left.  \frac{1}{M_1 \beta^2}
\oint_{\Gamma_{\infty}} \frac{dz}{2 \pi i} 
\frac{ ( 1 - \beta z )^2}{(z^2-1)^{3/2}}
\exp\left(  z \, \bar{q} \cdot a_2 \right)
\exp\left[ \left( \frac{  e^{\phi} (z-1) }{\gamma ( 1 -  \beta z )} + 1 \right) \bar{q} \cdot a_1 \right] + {\cal O}(a_1^3)
\right\} 
\nn\\
&+& \frac{i e^4}{2} \int {\hat{d} ^4 \bar{q}} \,\int {\hat{d} ^4 \bar{\ell}} \,
 \frac{ 1 }{\bar{\ell}^2 (\bar{\ell} - \bar{q})^2} 
 \hat{\delta}( \bar{q} \cdot u_1 ) \hat{\delta}( \bar{q} \cdot u_2 )
e^{- i b \cdot {\bar q}} \,
u_{1}^{\alpha} [e^{- i (a_1+a_2)* \bar{\ell}}]_{\alpha}^{\ \nu} u_{2 \nu}  
\nn\\
&\times& \biggl[  \bar{q}^{\mu} 
\left( \frac{\hat{\delta}( \bar{\ell} \cdot u_2 )}{M_1}  \frac{\bar{\ell} \cdot (\bar{\ell} - \bar{q})}
{(\bar{\ell} \cdot u_1 + i \varepsilon)^2} 
+ \frac{\hat{\delta}( \bar{\ell} \cdot u_1 )}{M_2}  \frac{\bar{\ell} \cdot (\bar{\ell} - \bar{q})}
{(\bar{\ell} \cdot u_2 - i \varepsilon)^2} \right) 
u_{1}^{\beta} [e^{- i (a_1+a_2)* (\bar{q} - \bar{\ell})}]_{\beta}^{\ \rho} u_{2 \rho}
\nn\\
&-& i  \bar{\ell}^{\mu} \bar{\ell} \cdot (\bar{\ell} - \bar{q})
\left( \frac{\hat{\delta}^{\prime}( \bar{\ell} \cdot u_1) 
\hat{\delta}( \bar{\ell} \cdot u_2 )}{M_1} 
- \frac{\hat{\delta}( \bar{\ell} \cdot u_1) 
\hat{\delta}^{\prime}( \bar{\ell} \cdot u_2 )}{M_2} \right)
u_{1}^{\beta} [e^{- i (a_1+a_2)* (\bar{q} - \bar{\ell})}]_{\beta}^{\ \rho} u_{2 \rho}
\nn\\
&+&  2 i \bar{\ell}^{\mu}
\left( \frac{ u_{1}^{\beta} [e^{- i (a_1+a_2) * (-\bar{\ell}+\bar{q})}]_{\beta}^{\ \rho} \bar{q}_{\rho}  }{M_2}
- \frac{ \bar{q}^{\beta} [e^{- i (a_1+a_2) * ( -\bar{\ell} + \bar{q} )}]_{\beta}^{\ \rho} u_{2 \rho}  }{M_1} \right)
\hat{\delta}( \bar{\ell} \cdot u_1) \hat{\delta}( \bar{\ell} \cdot u_2)
\biggr]
\nn\\
\eea
where we used the fact that the integrals over the particle wave functions effectively replace the momenta $p_j$ by their classical values $M_j u_j$. $b_\perp$ is defined by
$b_{\perp} = \Pi[ b+i(a_1+a_2)]$, $\Pi$ being the projector onto the plane orthogonal to $u_1$ and $u_2$, given by~\cite{Maybee:19,Vines:18}
\beq\label{proj1}
\Pi^{\mu}_{\ \nu} = \eta^{\mu}_{\ \nu} + \frac{1}{\gamma^2-1}
\Bigl( u_1^{\mu}(u_{1 \nu} - \gamma u_{2\nu}) + u_2^{\mu}(u_{2 \nu} - \gamma u_{1\nu}) \Bigr)  =  \frac{{\epsilon^{\mu\alpha}} (u_1, u_2){\epsilon_{\nu\alpha}} (u_1, u_2)}{\gamma^2-1}
\eeq
where we have defined, for a generic tensor $T$
\begin{equation}
T(a, b, c, d)\equiv T_{\mu\nu\alpha\beta} a^\mu b^\nu c^\alpha d^\beta,\,\,\,T_\mu (a, b, c)\equiv T_{\mu\nu\alpha\beta} a^\nu b^\alpha c^\beta, \,\,\, T_{\mu\nu}(a, b)= T_{\mu\nu\alpha\beta}a^\alpha b^\beta .
\end{equation}
Finally, observe that the last term in $\Delta p^\mu_{\text{NLO}, 1}$ is absent for scalar particles.

As a simple corroboration of the above results, one must check whether the final momentum of the outgoing particle after the classical scattering process is on shell. In the present context, this means that the expression
\beq
(\Delta p_1)^2 + 2 p_1 \cdot \Delta p_1 = 0
\eeq
must hold to all orders in perturbation theory. At order $e^4$, one must find that
\beq
(\Delta p _{\text{LO}, 1})^2 + 2 p_1 \cdot\Delta p _{\text{NLO}, 1} = 0.
\eeq
It is easy to see that this is true. First, from the tree-level calculation, we find that
\bea
(\Delta p _{\text{LO}, 1})^2 &=& - e^4
 \int {\hat{d} ^4 \bar{q}} \,\int {\hat{d} ^4 \bar{q}'} \,
 \hat{\delta}( \bar{q} \cdot u_1 ) \hat{\delta}( \bar{q} \cdot u_2 )
 \hat{\delta}( \bar{q}^{\prime} \cdot u_1 ) 
\hat{\delta}( \bar{q}^{\prime} \cdot u_2 )
e^{- i b \cdot ({\bar q} + {\bar q}^{\prime})}  
\nn\\
&\times& \frac{\bar{q} \cdot \bar{q}^{\prime}}{ \bar{q}^2 \bar{q}^{\prime 2} }
\, u_{1}^{\mu} \exp(- i (a_1+a_2)* \bar{q})_{\mu}^{\ \nu} u_{2 \nu}
\, u_{1}^{\alpha} \exp(- i (a_1+a_2)* \bar{q}^{\prime})_{\alpha}^{\ \beta} u_{2 \beta} .
\eea
On the other hand, from the one-loop calculation, we get
\bea
2 p_1 \cdot \Delta p_{\text{NLO}, 1} &=& 
e^4 \int {\hat{d} ^4 \bar{q}} \,
 \hat{\delta}( \bar{q} \cdot u_1 ) \hat{\delta}( \bar{q} \cdot u_2 )
e^{- i b \cdot {\bar q}}
\int {\hat{d} ^4 \bar{\ell}} \,
 \frac{ \bar{\ell} \cdot (\bar{\ell} - \bar{q}) }{\bar{\ell}^2 (\bar{\ell} - \bar{q})^2} 
\bar{\ell} \cdot u_1 
\hat{\delta}^{\prime}( \bar{\ell} \cdot u_1) 
\hat{\delta}( \bar{\ell} \cdot u_2 )
\nn\\
&& \qquad\qquad\times
u_{1}^{\mu} [e^{- i (a_1+a_2)* \bar{\ell}}]_{\mu}^{\ \nu} u_{2 \nu}
u_{1}^{\alpha} [e^{- i (a_1+a_2)* (\bar{q} - \bar{\ell})}]_{\alpha}^{\ \beta} u_{2 \beta}
\nn\\
&=& - e^4 \int {\hat{d} ^4 \bar{q}} \,
\int {\hat{d} ^4 \bar{\ell}} \,
 \hat{\delta}( \bar{q} \cdot u_1 ) \hat{\delta}( \bar{q} \cdot u_2 )
\hat{\delta}( \bar{\ell} \cdot u_1) \hat{\delta}( \bar{\ell} \cdot u_2 )
e^{- i b \cdot {\bar q}}
\frac{ \bar{\ell} \cdot (\bar{\ell} - \bar{q}) }{\bar{\ell}^2 (\bar{\ell} - \bar{q})^2} 
\nn\\
&& \qquad\qquad\times
u_{1}^{\mu} [e^{- i (a_1+a_2)* \bar{\ell}}]_{\mu}^{\ \nu} u_{2 \nu}
u_{1}^{\alpha} [e^{- i (a_1+a_2)* (\bar{q} - \bar{\ell})}]_{\alpha}^{\ \beta} u_{2 \beta}
\nn\\
&=& e^4 \int {\hat{d} ^4 \bar{q}'} \, 
\int {\hat{d} ^4 \bar{\ell}} \,
 \hat{\delta}( \bar{q}^{\prime} \cdot u_1 ) \hat{\delta}( \bar{q}^{\prime} \cdot u_2 )
\hat{\delta}( \bar{\ell} \cdot u_1) \hat{\delta}( \bar{\ell} \cdot u_2 )
e^{- i b \cdot (\bar{q}^{\prime} + \bar{\ell})}
\nn\\
&& \qquad\qquad\times
\frac{ \bar{\ell} \cdot \bar{q}^{\prime} }{\bar{\ell}^2 \bar{q}^{\prime 2}} 
u_{1}^{\mu} [e^{- i (a_1+a_2)* \bar{\ell}}]_{\mu}^{\ \nu} u_{2 \nu}
\, u_{1}^{\alpha}[e^{- i (a_1+a_2)* \bar{q}^{\prime}}]_{\alpha}^{\ \beta} u_{2 \beta}
\eea
where in the last line we set $\bar{q}^{\prime} = \bar{q} - \bar{\ell}$. Summing up both expressions, we arrive at the desired result.

Let us now examine the transverse contribution to our final deflection formula. We consider this to be the one coming from the triangle leading singularity calculation. One can give further grounds to this argument by thinking about the holomorphic classical limit perspective where double discontinuities in the $t$-channel contain all the classical scattering information. Since box topologies do not have double discontinuities in the $t$-channel, they do not contribute to the classical pieces. Therefore, we do not expect box integrals nor cut-box terms to contribute to the transverse part of the impulse.

 We can rewrite the triangle terms as
\begin{equation}
\begin{split}
\Delta p^\mu_{\text{NLO}, 1, \triangle}  &= - i \frac{e^4}{16 M_2 \beta^2} 
\oint_{\Gamma_{\infty}} \frac{dz}{2 \pi i} 
\frac{ ( 1 - \beta z )^2}{(z^2-1)^{3/2}}
 \int \hat{d}^4 \bq\,   \hat{\delta} ( \bar{q} \cdot u_1 )  \hat{\delta}( \bar{q} \cdot u_2 )
e^{- i ( b + i {a}_{12} (z)) \cdot {\bar q}} 
\frac{\bar{q}^{\mu}}{\sqrt{-\bar{q}^2}}
\\&\hspace{9cm}
+\mathcal{O}(a_2^3)+(1\leftrightarrow 2)
\end{split}
\end{equation}
defining 
\begin{equation}
{a}_{ij}^\mu (z) \equiv z a_i ^\mu + \left( \frac{  e^{\phi} (z-1) }{\gamma ( 1 -  \beta z )} + 1 \right)  a_j ^\mu .
\end{equation}
Note that this can be recasted in a form which makes parity invariance more explicit (odd spin terms always come with a Levi-Civita tensor). In fact,  on-shell of the $\bar{q}$ integral we  have   $ \sinh \phi\,
\bar{q}^\mu = {i}\epsilon^\mu (\bar{q}, u_1, u_2)$ 
at the exponent \cite{Arkani-Hamed:20}, so that the triangle deflection can be put  in the following form
\begin{equation}
\begin{split}
\Delta p^\mu_{\text{NLO}, 1, \triangle}  &= - i \frac{e^4}{16 M_2 \beta^2} 
\oint_{\Gamma_{\infty}} \frac{dz}{2 \pi i} 
\frac{ ( 1 - \beta z )^2}{(z^2-1)^{3/2}}
 \int \hat{d}^4 \bq\,   \hat{\delta} ( \bar{q} \cdot u_1 )  \hat{\delta}( \bar{q} \cdot u_2 )
e^{- i ( b +\mathfrak{a}_{12} (z)) \cdot {\bar q}} 
\frac{\bar{q}^{\mu}}{\sqrt{-\bar{q}^2}}
\\&\hspace{9cm}
+\mathcal{O}(a_2^3)+(1\leftrightarrow 2)
\\&=
\frac{e^4}{32 \pi M_2 \beta^2 \sqrt{\gamma^2 - 1}} 
\oint_{\Gamma_{\infty}} \frac{dz}{2 \pi i} 
\frac{ ( 1 - \beta z )^2}{(z^2-1)^{3/2}}
\frac{( b^{\mu} + {\Pi^\mu}_\nu \mathfrak{a}_{12} ^\nu (z))}{|b + \Pi \mathfrak{a}_{12} (z)|^{3}} 
+\mathcal{O}(a_2^3)+(1\leftrightarrow 2)
\end{split}
\end{equation}
where we wrote ${\sinh\phi}\,\mathfrak{a}_{ij}^\mu (z)\equiv {\epsilon^\mu (a_{ij}(z), u_1, u_2)}.$

In particular, in the scalar probe limit, $M_1/M_2 \to 0$ and $a_1 \to 0$,   we find the triangle LS deflection to be
\bea
\Delta p^\mu_{\text{NLO}, 1, \triangle}  =  \frac{e^4}{32 \pi M_1 \beta^2 \sqrt{\gamma^2 - 1}} 
\oint_{\Gamma_{\infty}} \frac{dz}{2 \pi i} 
\frac{ ( 1 - \beta z )^2}{(z^2-1)^{3/2}}
\frac{( b^{\mu} +  \Pi^{\mu}_{\ \nu} \mathfrak{a}^\nu _2 (z) )}{|b +  \Pi\mathfrak{a} _2 (z)|^{3}} ,  \,\,\,\,\, \,\,\,\mathfrak{a}_{2}^\mu (z)\equiv z \frac{\epsilon^\mu (a_{2} , u_1, u_2)}{\sinh\phi}.
\eea
There is no branch cut singularity in the integral now, hence the contour $\Gamma_{\infty}$ is simply a contour enclosing the pole at infinity. Note how the NJ shift is still manifesting itself in our formulae, yet in a more elaborated form. We have not been able to recast the leading singularity contour integral in terms of elementary functions, but it is easy to expand this in powers of the spin and to take the residue at infinity.

For completeness, let us now explicitly evaluate the aligned scattering angle up to NLO.
It is not particularly difficult to see that box and cut-box terms will not contribute to the  angle defined by 
\beq
\theta = \frac{b \cdot \Delta p_1}{b |{\bf p}_1|} .
\eeq

Since $b$ is such that $b\cdot u_1=b\cdot u_2=0$, any term in the impulse proportional to the velocities will not contribute to $\theta$. Moreover, in the aligned-spin configuration, the rescaled spin vectors will also be orthogonal to $b$~\cite{Vines:18}. We work in the rest frame of particle 2 and also orientate the spatial coordinates so that particle 1 is moving along the $z$ axis. This allows us to calculate the associated integrals in much the same way as in Ref.~\cite{Guevara:19a}:
\bea\label{rootkerrdp}
&& - i \int {\hat{d} ^4 \bar{q}} \,\hat{\delta}( \bar{q} \cdot u_1 ) \hat{\delta}( \bar{q} \cdot u_2 )
e^{- i b \cdot {\bar q}} 
\frac{b \cdot \bar{q}}{b \sqrt{-\bar{q}^2}}
\nn\\
&& \qquad\qquad\times \oint_{\Gamma_{\infty}} \frac{dz}{2 \pi i} 
\frac{ ( 1 - \beta z )^2}{(z^2-1)^{3/2}}
\exp\left( z \, \bar{q} \cdot a_1 \right)
\exp\left[ \left( \frac{  e^{\phi} (z-1) }{\gamma ( 1 -  \beta z )} + 1 \right) \bar{q} \cdot a_2 \right] 
\nn\\
&& = \frac{1}{2 \pi \gamma \beta} \frac{\partial}{\partial b}
\oint_{\Gamma_{\infty}} \frac{dz}{2 \pi i} 
\frac{ ( 1 - \beta z )^2}{(z^2-1)^{3/2}}
\int \frac{d^2 {\bf \bar{q}}}{2\pi |{\bf \bar{q}}|} 
\exp\left[ i {\bf \bar{q}} \cdot \left( {\bf b} - z \hat{\bf p} \times {\bf a}_1 
- \frac{z-\beta}{1-\beta z} \hat{\bf p} \times {\bf a}_2 \right) \right] 
\nn\\
&& = \frac{1}{2 \pi \gamma \beta} \frac{\partial}{\partial b}
\oint_{\Gamma_{\infty}} \frac{dz}{2 \pi i} 
\frac{ ( 1 - \beta z )^2}{(z^2-1)^{3/2}}
\bigg| b - z a_1 - \frac{z-\beta}{1-\beta z} a_2 \bigg|^{-1}
\eea
where $\hat{\bf p}$ is the unit vector in the direction of the relative momentum. The denominator has two roots, $z_{\pm}$ such that $z_+ \to \infty$ and $z_- \to 1/v$ in the spinless limit. Therefore the contour integral is given by the residues at $z_+$ and $\infty$. Since the residue at infinity vanishes, one has to calculate only the residue at $z_+$. The final result, including also the LO term, reads:
\bea
\theta^{(0)} &=& \frac{e^2}{2 \pi b }
\frac{\sqrt{M_1^2 + M_2^2 + 2 M_1 M_2 \gamma}}{M_1 M_2 (\sinh \phi)^2} \, 
\textrm{Re} \left[\frac{\cosh\phi \, b \cdot b_{\perp} 
- i \epsilon( b, b_{\perp }, u_{1 }, u_{2})}
{b_{\perp}^{2}}
\right]
\nn\\
\theta^{(1)} &=& - \frac{e^4}{\pi \gamma^2 M_1^2 M_2^2} \sqrt{M_1^2 + M_2^2 + 2 M_1 M_2 \gamma} \,
\frac{\partial}{\partial b} \Bigl[ M_1 \bigl( {\cal E}(a_1, a_2,3) + {\cal O}(a_2^{3}) \bigr)
+ M_2 \bigl( {\cal E}(a_2, a_1,3) + {\cal O}(a_1^{3}) \bigr) \Bigr]
\nn\\
\eea
where
\bea
{\cal E}(a_1, a_2,n) &=&
\frac{( \iota + \kappa - 2 a_1 )^n}
{4 \beta  \kappa [ (\iota + \kappa)^2 - (2  a_1 \beta)^2 ]^{3/2}}
\nn\\
\iota &=& \beta b +  a_1 +  a_2
\nn\\
\kappa &=& \sqrt{ \iota^2 - 4  a_1 \beta ( a_2 \beta + b )}
\eea
and we used that the center-of-mass three-momentum is given by
$$
|{\bf p}_1| = \frac{M_1 M_2 \sqrt{\gamma^2 - 1}}{\sqrt{M_1^2 + M_2^2 + 2 M_1 M_2 \gamma}} 
\to M_1 \sqrt{\gamma^2 - 1} .
$$
The second result is valid in the probe limit. In this limit, the expressions for the scattering angle up to NLO reduce to
\bea\label{angolettino}
\theta^{(0)} &\approx& \frac{e^2}{2 \pi b }
\frac{1}{M_1 \gamma^2 \beta^2} \, 
\textrm{Re} \left[\frac{\cosh\phi \, b \cdot b_{\perp} 
- i \epsilon( b_{}, b_{\perp }, u_{1 }, u_{2 })}
{b_{\perp}^{2}}
\right]
\nn\\
\theta^{(1)} &\approx& - \frac{e^4}{\pi M_1^2 \gamma^2 } \,
\frac{\partial}{\partial b}  {\cal E}(a_2, 0,3)  
\eea
where we have taken $a_1 = 0$ and therefore now $b_{\perp} = \Pi[ b+i a_2]$. Notice that in such limits we obtain expressions which are valid to all orders in the spin of the heavy $\sqrt{\textrm{Kerr}}$ source (the spurious poles dropped), but under the aligned-spin constraint $b\cdot a=0$.

\section{Kerr black hole impulse from quantum scattering amplitudes}\label{GRRRR}

In this section we repeat the same steps as in the EM chapters, but for the case of gravity. It will be clear to the reader that the gravity side of the story will not be too dissimilar from its gauge theory counterpart.

Gravitationally interacting spinning black holes with masses $M_1, M_2$ can be described by a system of two spin-$S_k$ fields $\Phi_k$, $k=1,2$ minimally coupled to gravity:
\beq
S = \int d^{4}x \sqrt{-g} \left( \frac{1}{16 \pi G_N} R + {\cal L}_{\textrm{int}}[\Phi_{k}, g_{\mu\nu}] \right)
\eeq
where the first term is the Einstein-Hilbert contribution. Here we consider the point-particle approximation and as in the previous case we exclude local interactions between matter fields. We are interested in calculating the total change in the classical momentum of a Kerr black hole during the collision with another Kerr black hole. We will follow the same recipe as above, namely, we will first calculate tree-level and one-loop quantum scattering amplitudes, focusing on the pieces that are relevant to the computation of our classical observable. So now let us turn to the calculation of such objects.

\subsection{Calculation of the classical contribution from scattering amplitudes}

  The tree-level gravity amplitudes can be obtained by considering the KLT relations. One finds~\cite{Arkani-Hamed:17,Johansson:19}
\bea
M_{S}({\bf 1}, {\bf 2}, 3^{++}) &=& \frac{i \kappa}{2} 
\frac{\langle {\bf 1} {\bf 2} \rangle^{2S}}{M^{2S-2}} x^2
= \frac{i \kappa}{2} 
\frac{1}{M^{2S-2}} x^2
\bigl[ {\bf 2} |^{2S} \exp\left( i \frac{p_{3}^{\mu} \epsilon_{+}^{\nu} J_{\mu\nu}}{p \cdot \epsilon_{+}} \right)  | {\bf 1} \bigr]^{2S}
\nn\\
M_{S}({\bf 1}, {\bf 2}, 3^{--}) &=&  \frac{i \kappa}{2}
\frac{\bigl[ {\bf 1} {\bf 2} \bigr]^{2S}}{M^{2S-2}} \frac{1}{x^2}
= \frac{i \kappa}{2}
\frac{1}{M^{2S-2}} \frac{1}{x^2}
\langle {\bf 2} |^{2S} \exp\left( i \frac{p_{3}^{\mu} \epsilon_{-}^{\nu} J_{\mu\nu}}{p \cdot \epsilon_{-}} \right)  | {\bf 1} \rangle^{2S}
\eea
where $\kappa^2 = 32 \pi G_N$. The four-point amplitude corresponding to the scattering of a spin-$S_1$ particle with mass $M_1$ with a spin-$S_2$ particle with mass $M_2$ with exchange of gravitons is given by (see Fig.~\ref{treelevel})
\beq
M^{(0)}(1,2 \to 1^{\prime}, 2^{\prime}) = -  M_1^2 M_2^2 \frac{\kappa^2}{4 q^2} 
\left( \frac{\langle {\bf 2} {\bf 2}^{\prime} \rangle^{2S_2}}{M_2^{2S_2}} 
\frac{\bigl[ {\bf 1} {\bf 1}^{\prime} \bigr]^{2S_1}}{M_1^{2S_1}} e^{- 2 \phi}
+ \frac{\bigl[ {\bf 2} {\bf 2}^{\prime} \bigr]^{2S_2}}{M_2^{2S_2}} 
 \frac{\langle {\bf 1} {\bf 1}^{\prime} \rangle^{2S_1}}{M_1^{2S_1}}  e^{2 \phi} \right) .
\eeq
The classical contribution can be extracted as in the electromagnetic case:
\bea
{\cal M}^{(0)}(1,2 \to 1^{\prime}, 2^{\prime}) &=&  
-  M_1^2 M_2^2 \frac{1}{4 \hbar^2 \bar{q}^2} 
\left[ e^{ \bar{q} \cdot a_1} \, e^{ \bar{q} \cdot a_2}  e^{- 2 \phi} 
 + e^{ -\bar{q} \cdot a_1} \, e^{ -\bar{q} \cdot a_2}  e^{ 2 \phi} \right] 
\nn\\
&=&  -  M_1^2 M_2^2 \frac{1}{2 \hbar^2 \bar{q}^2} 
\cosh[\bar{q} \cdot (a_1+a_2) - 2 \phi]
\nn\\
&=&  -  \frac{1}{4 \hbar^2 \bar{q}^2} 
\biggl[ \Bigl( ( s - M_1^2 - M_2^2 )^2 - 2 M_1^2 M_2^2 \Bigr) \cosh[\bar{q} \cdot (a_1+a_2)]  
\nn\\
&-& ( s - M_1^2 - M_2^2 ) \sqrt{( s - M_1^2 - M_2^2 )^2 - 4 M_1^2 M_2^2} \sinh[\bar{q} \cdot (a_1+a_2)]
\biggr] .
\eea
As in the previous case, we can recast this amplitude in a more convenient form using the Gram determinant constraint as given above and on-shell conditions; we find that
\bea
&& \cosh[ \bar{q} \cdot (a_1+a_2)]  \cosh 2\phi - \sinh[ \bar{q} \cdot (a_1+a_2)]  \sinh 2\phi 
\nn\\
&& = P_{\mu\nu\alpha\beta} u_{1}^{\mu} u_{1}^{\nu} \left[ 2 u_{2}^{\alpha} u_{2}^{\beta} 
-  i \epsilon^{(\alpha}_{\ \lambda\rho\sigma} u_{2}^{\beta)} u_{2}^{\lambda} 
(a_1 + a_2)^{\rho} \bar{q}^{\sigma}
+ \frac{1}{2!} \Bigl( \bar{q} \cdot (a_1+a_2) \Bigr)^2 2 u_{2}^{\alpha} u_{2}^{\beta}
+ \cdots \right]
\nn\\
&& =  P_{\mu\nu\alpha\beta} \, u_{1}^{\mu} u_{1}^{\nu} 
\exp(- i (a_1+a_2)* \bar{q})^{(\alpha}_{\ \lambda} u_{2}^{\beta)} u_{2}^{\lambda}
\eea
where $P_{\mu\nu\alpha\beta}$ is the usual de Donder trace reverser
\begin{equation}
P_{\mu\nu\rho\sigma}=\frac{1}{2}\left(
\eta_{\mu\rho}\eta_{\nu\sigma}+ \eta_{\mu\sigma}\eta_{\nu\rho}-\eta_{\mu\nu}\eta_{\rho\sigma}
\right).
\end{equation}
Therefore
\beq
{\cal M}^{(0)}(1,2 \to 1^{\prime}, 2^{\prime}) = - M_1^2 M_2^2 \frac{1}{4 \hbar^2 \bar{q}^2} \,
2 P_{\mu\nu\alpha\beta} \, u_{1}^{\mu} u_{1}^{\nu} 
\exp(- i (a_1+a_2)* \bar{q})^{(\alpha}_{\ \lambda} u_{2}^{\beta)} u_{2}^{\lambda} .
\eeq
Tree-level Compton scattering amplitudes can be easily calculated by using the double-copy approach. The associated YM amplitudes will have a similar form as those of the photon amplitudes calculated previously. In the YM case, the Compton amplitudes will also have a massless pole in the $t$-channel, besides the massive pole in the $u$-channel. One finds~\cite{Arkani-Hamed:17,Johansson:19,Bjerrum-Bohr:14}:
{\bea
M_{S}({\bf 1},2^{++}, 3^{--}, {\bf 4}) &=& -i \frac{\kappa^2}{4} 
 \frac{\langle 3| {\bf 1} |2 \bigl]^{4-2S}}{[(p_1 + p_2)^2 - \mu^2] [(p_1 + p_3)^2 - \mu^2] (p_2 + p_3)^2}
\Bigl( \langle {\bf 4} 3 \rangle \bigl[ {\bf 1} 2 \bigr] + \langle {\bf 1} 3 \rangle \bigl[ {\bf 4} 2 \bigr] \Bigr)^{2S}
\,\,\,\,\,\,
\nn\\
&=& -i \frac{\kappa^2}{4 \mu^{2S}} 
\frac{\langle 3| {\bf 1} |2 \bigl]^{4}}{[(p_1 + p_2)^2 - \mu^2][(p_1 + p_3)^2 - \mu^2] (p_2 + p_3)^2}
\langle {\bf 4} |^{2S} \exp\left( i \frac{p_{3}^{\mu} \bar{\epsilon}_{3}^{\nu} J_{1 \mu\nu}}
{\bar{p} \cdot \bar{\epsilon}_{3}} \right)  | {\bf 1} \rangle^{2S}
\nn\\
M_{S}({\bf 1},2^{++}, 3^{++}, {\bf 4}) &=& - i \mu^{4-2S} \frac{\kappa^2}{4 }  
\frac{\bigl[ 23 \bigr]^4 }{(p_2 + p_3)^2}
\frac{\langle {\bf 1} {\bf 4} \rangle^{2S}}{[(p_1 + p_2)^2 - \mu^2] [(p_1 + p_3)^2 - \mu^2]} 
\eea}which holds for $ S \leq 2$ and $\bar{p} = (p_1 - p_4)/2$. The bar over the polarization vector for $p_3$ is a reminder that its reference momentum was fixed to be $p_2$. The above Compton amplitudes can also be derived from the double-copy formula
\beq{
M^{\textrm{tree}}(1_{S},2,3,4_{S}) = i \left( \frac{\kappa}{2} \right)^2 s_{23} 
(-1)^{ \left \lfloor{S}\right \rfloor - \left \lfloor{S_1}\right \rfloor - \left \lfloor{S_2}\right \rfloor + 1}
A^{\textrm{tree}}[1_{S_1},2,3,4_{S_1}] 
A^{\textrm{tree}}[1_{S_2},3,2,4_{S_2}] , \,\,\,
S = S_1 + S_2 }
\label{DC}
\eeq
as proposed in Ref.~\cite{Johansson:19}. There is also an apparent double-copy tension already highlighted in Ref.~\cite{Johansson:19} regarding the first amplitude, which ceases to be local for $S>2$. Again this is connected to the fact that for $S \leq 2$ we obtain universal terms in the soft expansion and 
$S > 2$ are related to non-universal contributions. For a discussion of the gravitational Compton amplitudes for arbitrary spin, see Ref.~\cite{Chung:19}.

Now let us discuss the one-loop amplitude. As in the electromagnetic case, we will resort to the calculation of leading singularities in order to obtain the triangle contribution. We will use the same parametrization as before. We wish to compute
\beq
\textrm{LS}^{(a+-)}_{h} = \frac{x_1}{4 M_1^2 \left(x_1^2-1\right)} 
\frac{1}{2 \pi i} \oint_{\Gamma_a} \frac{dy}{y}
\Bigl\langle
M_{S_1}(-{\bf 1}^{\prime}, \boldsymbol\ell, - \ell_1^{+}) 
M_{S_1}({\bf 1}, -\boldsymbol\ell, -\ell_3^{-}) 
M_{S_2}({\bf 2},\ell_3^{+}, \ell_1^{-}, -{\bf 2}^{\prime})
\Bigr\rangle
\eeq
and
\beq
\textrm{LS}^{(b+-)}_{h} =  \frac{x_2}{4 M_2^2 \left(x_2^2-1\right)} 
\frac{1}{2 \pi i} \oint_{\Gamma_b} \frac{dy}{y}
\Bigl\langle
M_{S_2}(-{\bf 2}^{\prime},\boldsymbol\ell, \ell_1^{+})
M_{S_2}({\bf 2}, -\boldsymbol\ell, \ell_3^{-}) 
M_{S_1}({\bf 1},-\ell_3^{+}, -\ell_1^{-}, -{\bf 1}^{\prime})
\Bigr\rangle .
\eeq
We find that
\bea
\hspace{-7mm}
\textrm{LS}^{(a+-)}_{h}  \biggl|_{x_1 \to 1} &=&
- i \frac{\kappa^4}{2048 \hbar} 
\frac{M_1}{\sqrt{-\bar{q}^2}} 
\frac{1}{( s - M_1^2 - M_2^2 )^2 - 4 M_1^2 M_2^2}
 \frac{1}{2 \pi i} \oint_{\Gamma_a} \frac{dy}{y^3}
\frac{ (f(s, y))^4}
{(1-y^2)^2}
\nn\\
&\times& 
\biggl\langle
\frac{1}{M_1^{2S_1}}
\langle {\bf 1}^{\prime} |^{2S_1} \exp\left(- i \frac{\ell_{3}^{\mu} \epsilon_{\ell_3}^{\nu} J_{1\mu\nu}}
{p_1 \cdot \epsilon_{\ell_3}} \right)  | {\bf 1} \rangle^{2S_1}
 \frac{1}{M_2^{2S_2}} 
\langle {\bf 2}^{\prime} |^{2S_2} \exp\left( i \frac{\ell_{1}^{\mu} \bar{\epsilon}_{\ell_{1}}^{\nu} J_{2 \mu\nu}}
{p_2 \cdot \bar{\epsilon}_{\ell_{1}}} \right)  | {\bf 2} \rangle^{2S_2}
\biggr\rangle
\eea
and
\bea
\hspace{-7mm}
\textrm{LS}^{(b+-)}_{h}  \biggl|_{x_2 \to 1} &=&
- i \frac{\kappa^4}{2048 \hbar} 
\frac{M_2}{\sqrt{-\bar{q}^2}} 
\frac{1}{( s - M_1^2 - M_2^2 )^2 - 4 M_1^2 M_2^2}
 \frac{1}{2 \pi i} \oint_{\Gamma_b} \frac{dy}{y^3}
\frac{ (f(s, y))^4}
{(1-y^2)^2}
\nn\\
&\times& 
\biggl\langle
\frac{1}{M_2^{2S_2}}
\langle {\bf 2}^{\prime} |^{2S_2} \exp\left( i \frac{\ell_{3}^{\mu} \epsilon_{\ell_3}^{\nu} J_{2\mu\nu}}
{p_2 \cdot \epsilon_{\ell_3}} \right)  | {\bf 2} \rangle^{2S_2}
 \frac{1}{M_1^{2S_1}} 
\langle {\bf 1}^{\prime} |^{2S_1} \exp\left( - i \frac{\ell_{1}^{\mu} \bar{\epsilon}_{\ell_{1}}^{\nu} J_{1 \mu\nu}}
{p_1 \cdot \bar{\epsilon}_{\ell_{1}}} \right)  | {\bf 1} \rangle^{2S_1}
\biggr\rangle 
\eea
note that we will be using the same definitions for $f(s, y)$ and  $h(s, z)$ that we used in the electromagnetic case. Note also how the same function  $f(s, y)$ appears here to the power 4 rather than 2, which is an instance of the double copy: $ f(s, y)^{2 |h|}$, $|h|=1, 2$.

Then, proceeding analogously to the EM case we are able to verify that
 
\begin{equation}
\begin{split}
 \textrm{LS}^{(a+-)}_{h} \biggl|_{x_1 \to 1} &= - i \frac{\kappa^4}{2048 \hbar} 
\frac{M_1}{ \sqrt{-\bar{q}^2}}  
\frac{1}{ ( s - M_1^2 - M_2^2 )^2 - 4 M_1^2 M_2^2}
\\&\times \frac{1}{2 \pi i} \oint_{\Gamma_a} \frac{dy}{y^3}
\frac{(f(s, y))^4}{(1-y^2)^2}
\exp\left( \frac{1+y^2}{2y} \bar{q} \cdot a_1 \right)
\exp\left[ \left( \frac{  2  M_1 M_2 e^{\phi} (1-y)^2 }{f(s, y)} + 1 \right) \bar{q} \cdot a_2 \right]
\end{split}
\end{equation}
and

\begin{equation}
\begin{split}
 \textrm{LS}^{(a+-)}_{h} \biggl|_{x_2 \to 1} &= - i \frac{\kappa^4}{2048 \hbar} 
\frac{M_2}{ \sqrt{-\bar{q}^2}}  
\frac{1}{ ( s - M_1^2 - M_2^2 )^2 - 4 M_1^2 M_2^2}
\\&\times \frac{1}{2 \pi i} \oint_{\Gamma_b} \frac{dy}{y^3}
\frac{(f(s, y))^4}{(1-y^2)^2}
\exp\left( \frac{1+y^2}{2y} \bar{q} \cdot a_2 \right)
\exp\left[ \left( \frac{  2  M_1 M_2 e^{\phi} (1-y)^2 }{f(s, y)} + 1 \right) \bar{q} \cdot a_1 \right].
\end{split}
\end{equation}
We remark that the expression for the $a$ ($b$) topology above is valid for all $a_1$ ($a_2$), but up to fourth order in $a_2$ ($a_1$). On the other hand, again equal-helicity configurations will produce a vanishing result -- one can use the same technique as before. Choosing the contour associated with the triangle topology and employing the same change of variables as above, we obtain the following classical contribution coming from the full triangle leading singularity: 
 
\bea
\textrm{LS}^{(a)}_{\triangle,\gamma} \biggl|_{x_1 \to 1} &=& - i \frac{\kappa^4}{256 \hbar}
\frac{M_1}{ \sqrt{-\bar{q}^2}}  
\frac{1}{ ( s - M_1^2 - M_2^2 )^2 - 4 M_1^2 M_2^2}
\nn\\
&\times&\frac{1}{2 \pi i} \oint_{\Gamma_{\infty}} dz
\frac{ ( h(s, z))^4}{(z^2-1)^{3/2}}
\exp\left( z \, \bar{q} \cdot a_1 \right)
\exp\left[ \left( \frac{  2  M_1 M_2 e^{\phi} (z-1) }{ h(s, z)} + 1 \right) \bar{q} \cdot a_2 \right]
\eea
and
\bea
\hspace{-5mm}
\textrm{LS}^{(b)}_{\triangle,\gamma} \biggl|_{x_2 \to 1} &=& - i \frac{\kappa^4}{256 \hbar}
\frac{M_2}{ \sqrt{-\bar{q}^2}}  
\frac{1}{ ( s - M_1^2 - M_2^2 )^2 - 4 M_1^2 M_2^2}
\nn\\
&\times&\times \frac{1}{2 \pi i} \oint_{\Gamma_{\infty}} dz
\frac{ (h(s, z))^4}{(z^2-1)^{3/2}}
\exp\left( z \, \bar{q} \cdot a_2 \right) 
\exp\left[ \left( \frac{  2  M_1 M_2 e^{\phi} (z-1) }{h(s, z)} + 1 \right) \bar{q} \cdot a_1 \right].
\eea
Such expressions allow us to display explicitly the classical contribution coming from the triangle diagram of the one-loop amplitude. One finally finds that
\bea
\hspace{-5mm}
{\cal M}^{(1)}_{\triangle,\textrm{class}} &=& - i \frac{1}{1024 \hbar}
 \frac{1}{ \sqrt{-\bar{q}^2}} 
\left\{  \frac{M_1}{ ( s - M_1^2 - M_2^2 )^2 - 4 M_1^2 M_2^2}
\oint_{\Gamma_{\infty}} \frac{dz}{2 \pi i} 
\frac{ (h(s, z))^4}{(z^2-1)^{3/2}}
\right.
\nn\\
&&\qquad\qquad\times \left. 
\exp\left( z \, \bar{q} \cdot a_1 \right)
\exp\left[ \left( \frac{  2  M_1 M_2 e^{\phi} (z-1) }{h(s, z)} + 1 \right) \bar{q} \cdot a_2 \right] + {\cal O}(a_2^5)
\right.
\nn\\
&+& \left. \frac{M_2}{ ( s - M_1^2 - M_2^2 )^2 - 4 M_1^2 M_2^2}
 \oint_{\Gamma_{\infty}} \frac{dz}{2 \pi i}
\frac{(h(s, z))^4}{(z^2-1)^{3/2}}
\right.
\nn\\
&&\qquad\qquad\times \left.  
\exp\left( z \, \bar{q} \cdot a_2 \right)
\exp\left[ \left( \frac{  2  M_1 M_2 e^{\phi} (z-1) }{h(s, z)} + 1 \right) \bar{q} \cdot a_1 \right] + {\cal O}(a_1^5)
\right\} .
\eea
 Therefore, we now have the triangle contribution to the NLO impulse in the Kerr case. As in the previous case, universal terms in the soft theorem cannot reproduce contributions to all orders in the spins of the particles.

Now let us consider the box and crossed box contribution. Essentially the calculation follows the same considerations as in the photon case. One finds that
\bea
\hspace{-3mm}
i M^{(1)}_{\Box} &=&
4 \kappa^4 \int {\hat{d} ^D \ell}  \frac{B_{h}(\ell, p_1,p_2) B_{h}(-\ell-p_1+p_1^{\prime}, p_1,p_2)}
{ \ell^2 [(\ell + p_1)^2 - M_1^2] (\ell+p_1-p_1^{\prime})^2  [ (\ell-p_2)^2 - M_2^2]}
\nn\\
i M^{(1)}_{\boxtimes} &=& 
4 \kappa^4 \int {\hat{d} ^D \ell}  
\frac{ B_{h}(\ell, p_1,p_2) B_{h}(-\ell-p_1+p_1^{\prime}, p_1,p_2) }
{\ell^2 [(\ell + p_1)^2 - M_1^2](\ell+p_1-p_1^{\prime})^2 [ (\ell+p_2^{\prime})^2 - M_2^2] }
\eea
where 
\beq
B_h(q, p_1,p_2) = \frac{1}{4} P_{\mu\nu\alpha\beta} \, p_{1}^{\mu} p_{1}^{\nu} 
\exp\Bigl(- i \frac{(a_1+a_2)}{\hbar}* q \Bigr)^{(\alpha}_{\ \lambda} p_{2}^{\beta)} p_{2}^{\lambda}
\eeq
and again we have dropped any explicit term in the numerators that will not produce a classical contribution. Now it is easy to see that, by following similar steps as those outlined above, we obtain similar contributions at one-loop as those found in the electromagnetic case. For completeness, let us quote the final results:
\beq
i {\cal M}^{(1)}_{\Box}(1,2 \to 1^{\prime}, 2^{\prime}) \bigg|_{\textrm{Classical}}
+ i {\cal M}^{(1)}_{\boxtimes}(1,2 \to 1^{\prime}, 2^{\prime}) \bigg|_{\textrm{Classical}}
= i {\cal M}_{-2} + i {\cal M}_{-1} + {\cal O}(\hbar^0)
\eeq
where
\bea
i {\cal M}_{-2} &=& 
- \frac{1}{2\hbar^{2}} 
\int {\hat{d} ^4 \bar{\ell}} \,
 \frac{ B_{h}(\bar{\ell}, p_1,p_2) B_{h}(-\bar{\ell}+\bar{q}, p_1,p_2) }{\bar{\ell}^2 (\bar{\ell} - \bar{q})^2} 
 \hat{\delta}( \bar{\ell} \cdot p_2) \hat{\delta}( \bar{\ell} \cdot p_1) 
 + {\cal O}(1/\hbar)
\eea
and
\bea
\hspace{-8mm}
i {\cal M}_{-1}  &=&
\frac{1}{2 \hbar^{1+2\epsilon}} 
 \int {\hat{d} ^D \bar{\ell}} \,
\frac{ B_h(\bar{\ell}, p_1,p_2) B_h(-\bar{\ell}+\bar{q}, p_1,p_2) }{\bar{\ell}^2 (\bar{\ell} - \bar{q})^2 
( \bar{\ell} \cdot p_1  + i\varepsilon ) 
(\bar{\ell} \cdot p_2 - i\varepsilon )} 
\left( \frac{\bar{\ell}^2}{\bar{\ell} \cdot p_1 + i\varepsilon} \right)
+ \{ 2 \leftrightarrow -2 \}
\nn\\
&+& \frac{1}{2\hbar^{1+2\epsilon}} 
\int {\hat{d} ^D \bar{\ell}} \,
\frac{ B_h(\bar{\ell}, p_1,p_2) B_h(-\bar{\ell}+\bar{q}, p_1,p_2) }{\bar{\ell}^2 (\bar{\ell} - \bar{q})^2 
( \bar{\ell} \cdot p_1 + i\varepsilon ) 
 (\bar{\ell} \cdot p_2 - i\varepsilon )}
\frac{\bar{\ell}^2}{- \bar{\ell} \cdot p_2  + i\varepsilon} 
\nn\\
&-& \frac{1}{2 \hbar^{1+2\epsilon}} 
\int {\hat{d} ^D \bar{\ell}} \, \frac{ B_h(\bar{\ell}, p_1,p_2) B_h(-\bar{\ell}+\bar{q}, p_1,p_2) }
{\bar{\ell}^2 (\bar{\ell} - \bar{q})^2 
( \bar{\ell} \cdot p_1 + i\varepsilon ) 
(\bar{\ell} \cdot p_2  + i\varepsilon )}
\frac{ ( \bar{\ell} - \bar{q})^2 - \bar{q}^2 }{\bar{\ell} \cdot p_2 + i\varepsilon} .
\eea
For the cut-box, we find that
\beq
{\cal I}^{\mu (1)}_{h,\textrm{Cut Box}} = {\cal D}_{-2}^{\mu} + {\cal D}_{-1} ^{\mu}
\eeq
where
\beq
{\cal D}_{-2}^{\mu}  = - \frac{i}{2 \hbar^{2}} 
\int {\hat{d} ^4 \bar{\ell}} \,\frac{\bar{q}^{\mu}}{\bar{\ell}^2 (\bar{\ell} - \bar{q})^2}
\hat{\delta}( \bar{\ell} \cdot p_1 ) 
\hat{\delta}( \bar{\ell} \cdot p_2 ) 
B_{h}(\bar{\ell}, p_1,p_2) B_{h}(-\bar{\ell}+\bar{q}, p_1,p_2)
+ {\cal O}(1/\hbar)
\eeq
and
\beq
{\cal D}_{-1}^{\mu} =  \widetilde{{\cal D}}^{(1)\mu} +  \widetilde{{\cal D}}^{(2)\mu}
\eeq
with
\bea
\widetilde{{\cal D}}^{(1)\mu} &=& i \frac{1}{4\hbar} 
\int {\hat{d} ^4 \bar{\ell}} \,\hat{\delta}( \bar{\ell} \cdot p_1) 
\hat{\delta}( \bar{\ell} \cdot p_2) \bar{\ell}^{\mu}
\frac{ B_{h}(\bar{\ell}, p_1,p_2) }
{\bar{\ell}^2 (\bar{\ell} - \bar{q})^2} 
\Bigl[ P_{\mu\nu\alpha\beta} \, p_{1}^{\mu} p_{1}^{\nu} 
\exp\bigl(- i (a_1+a_2)* ( -\bar{\ell} + \bar{q} ) \bigr)^{(\alpha}_{\ \lambda} \bar{q}^{\beta)} p_{2}^{\lambda}
\nn\\
&+& P_{\mu\nu\alpha\beta} \, p_{1}^{\mu} p_{1}^{\nu} 
\exp\bigl(- i (a_1+a_2)* ( -\bar{\ell} + \bar{q} ) \bigr)^{(\alpha}_{\ \lambda} p_{2}^{\beta)} \bar{q}^{\lambda}
\Bigr]
\nn\\
&-& i \frac{1}{4\hbar} 
\int {\hat{d} ^4 \bar{\ell}} \,\hat{\delta}( \bar{\ell} \cdot p_1) 
\hat{\delta}( \bar{\ell} \cdot p_2) \bar{\ell}^{\mu}
\frac{ B_{h}(\bar{\ell}, p_1,p_2) }
{\bar{\ell}^2 (\bar{\ell} - \bar{q})^2}  
\Bigl[ P_{\mu\nu\alpha\beta} \, \bar{q}^{\mu} p_{1}^{\nu} 
\exp\bigl(- i (a_1+a_2)* ( -\bar{\ell} + \bar{q} ) \bigr)^{(\alpha}_{\ \lambda} p_{2}^{\beta)} p_{2}^{\lambda}
\nn\\
&+& P_{\mu\nu\alpha\beta} \, p_{1}^{\mu} \bar{q}^{\nu} 
\exp\bigl(- i (a_1+a_2)* ( -\bar{\ell} + \bar{q} ) \bigr)^{(\alpha}_{\ \lambda} p_{2}^{\beta)} p_{2}^{\lambda}
\Bigr]
\eea
and
\beq
\widetilde{{\cal D}}^{(2)\mu} =
 - \frac{i}{2\hbar}
\int {\hat{d} ^4 \bar{\ell}} \,
\bar{\ell}^2 \frac{\bar{\ell}^{\mu}}{\bar{\ell}^2 (\bar{\ell} - \bar{q})^2}
B_{h}(\bar{\ell}, p_1,p_2) B_{h}(-\bar{\ell}+\bar{q}, p_1,p_2)
\Bigl( \hat{\delta}^{\prime}( \bar{\ell} \cdot p_1) 
\hat{\delta}( \bar{\ell} \cdot p_2 ) 
- \hat{\delta}( \bar{\ell} \cdot p_1) 
\hat{\delta}^{\prime}( \bar{\ell} \cdot p_2 ) \Bigr) .
\eeq
All ${\cal O}(1/\hbar)$ terms can be grouped in a similar fashion as in the electromagnetic case~\cite{Kosower:19}:
\bea
\hspace{-8mm}
i {\cal M}_{-1}  + [i {\cal M}_{-2}]_{{\cal O}(1/\hbar)} &=&
\frac{i}{2\hbar^{1+2\epsilon}} 
\int {\hat{d} ^D \bar{\ell}} \,
\frac{ \bar{\ell} \cdot (\bar{\ell} - \bar{q}) \hat{\delta}(\bar{\ell} \cdot p_2) }{\bar{\ell}^2 (\bar{\ell} - \bar{q})^2 
( \bar{\ell} \cdot p_1  + i\varepsilon )^2 } 
B_{h}(\bar{\ell}, s) B_{h}(-\bar{\ell}+\bar{q}, s)
\nn\\
&+&
\frac{i}{ 2 \hbar^{1+2\epsilon}} 
\int {\hat{d} ^D \bar{\ell}} \,
\frac{ \bar{\ell} \cdot (\bar{\ell} - \bar{q}) \hat{\delta}(\bar{\ell} \cdot p_1) }
{\bar{\ell}^2 (\bar{\ell} - \bar{q})^2 ( \bar{\ell} \cdot p_2  - i\varepsilon )^2 } 
B_{h}(\bar{\ell}, s) B_{h}(-\bar{\ell}+\bar{q}, s)
\nn\\
&-& \frac{1}{4 \hbar^{1+2\epsilon}} 
\int {\hat{d} ^D \bar{\ell}} \,
( 2 \bar{\ell} \cdot \bar{q} - \bar{\ell}^2 ) \frac{ B_{h}(\bar{\ell}, s) B_{h}(-\bar{\ell}+\bar{q}, s) }
{\bar{\ell}^2 (\bar{\ell} - \bar{q})^2}
\nn\\
&\times& \Bigl( \hat{\delta}^{\prime}( \bar{\ell} \cdot p_1) 
\hat{\delta}( \bar{\ell} \cdot p_2 ) 
- \hat{\delta}( \bar{\ell} \cdot p_1) 
\hat{\delta}^{\prime}( \bar{\ell} \cdot p_2 ) \Bigr) 
\eea
and
\bea
\hspace{-8mm}
\widetilde{{\cal D}}^{(2)\mu} + [{\cal D}^{\mu}_{-2}]_{{\cal O}(1/\hbar)} &=& 
- \frac{i}{2 \hbar} 
\int {\hat{d} ^4 \bar{\ell}} \,
\bar{\ell} \cdot (\bar{\ell} - \bar{q}) \frac{\bar{\ell}^{\mu}}{\bar{\ell}^2 (\bar{\ell} - \bar{q})^2}
B_{h}(\bar{\ell}, s) B_{h}(-\bar{\ell}+\bar{q}, s)
\nn\\
&\times& \Bigl( \hat{\delta}^{\prime}( \bar{\ell} \cdot p_1) 
\hat{\delta}( \bar{\ell} \cdot p_2 ) 
- \hat{\delta}( \bar{\ell} \cdot p_1) 
\hat{\delta}^{\prime}( \bar{\ell} \cdot p_2 ) \Bigr)
\nn\\
&-& \frac{i}{4 \hbar} 
\int {\hat{d} ^4 \bar{\ell}} \,
( 2 \bar{\ell} \cdot \bar{q} - \bar{\ell}^2 ) \frac{\bar{q}^{\mu}}{\bar{\ell}^2 (\bar{\ell} - \bar{q})^2}
B_{h}(\bar{\ell}, s) B_{h}(-\bar{\ell}+\bar{q}, s)
\nn\\
&\times& \Bigl( \hat{\delta}^{\prime}( \bar{\ell} \cdot p_1) 
\hat{\delta}( \bar{\ell} \cdot p_2 ) 
- \hat{\delta}( \bar{\ell} \cdot p_1) 
\hat{\delta}^{\prime}( \bar{\ell} \cdot p_2 ) \Bigr).
\eea
In possession of such expressions, we can now proceed to calculate the impulse, up to NLO.

\subsection{Gravity impulse}

Collecting the results above, one can easily deduce the following contributions at LO and NLO to the impulse, in the non-aligned spin configuration -- we simply substitute the above results in Eq.~(\ref{impulse}). 
We find that  
{\bea
\Delta p^\mu_{\text{LO}, 1}  &=& - \frac{i \kappa^2}{16} M_1 M_2
 \int {\hat{d} ^4 \bar{q}} \,\hat{\delta}( \bar{q} \cdot u_1 ) \hat{\delta}( \bar{q} \cdot u_2 )
e^{- i b \cdot {\bar q}}  
\frac{\bar{q}^{\mu}}{ \bar{q}^2 }
\left[ e^{ \bar{q} \cdot a_1} \, e^{ \bar{q} \cdot a_2}  e^{- 2 \phi} 
 + e^{ -\bar{q} \cdot a_1} \, e^{ -\bar{q} \cdot a_2}  e^{ 2 \phi} \right]
\nn\\
&=& - \frac{i \kappa^2}{8} M_1 M_2
 \int {\hat{d} ^4 \bar{q}} \,\hat{\delta}( \bar{q} \cdot u_1 ) \hat{\delta}( \bar{q} \cdot u_2 )
e^{- i b \cdot {\bar q}}  
\frac{\bar{q}^{\mu}}{ \bar{q}^2 }
P[u_1,u_1, U\cdot u_2,u_2](\bq)
\nn\\
&=&-\frac{2 M_1 M_2 G_N}{\sinh\phi}
\textrm{Re} 
\left[\frac{\cosh 2\phi \, b_{\perp}^{\mu} +2i \cosh \phi \,\epsilon^{\mu}( b_{\perp }, u_{1 }, u_{2 })}
{b_{\perp}^{2}}
\right]
\eea
and
\bea\label{loopdp}
\Delta p^\mu_{\text{NLO}, 1}  &=& \frac{i\kappa^4 (M_1 M_2)^2}{128} 
 \int {\hat{d} ^4 \bar{q}} \,\hat{\delta}( \bar{q} \cdot u_1 ) \hat{\delta}( \bar{q} \cdot u_2 )
e^{- i b \cdot {\bar q}}
\frac{ \bar{q}^{\mu}}{8 \sqrt{-\bar{q}^2}}
\nn\\
&\times& 
\left\{  \frac{\gamma^2}{M_2 \beta^2}
\oint_{\Gamma_{\infty}} \frac{dz}{2 \pi i} 
\frac{ ( 1 -  \beta z )^4}{(z^2-1)^{3/2}}
\exp\left( z \, \bar{q} \cdot a_1 \right)
\exp\left[ \left( \frac{  e^{\phi} (z-1) }{\gamma ( 1-\beta  z)} + 1 \right) \bar{q} \cdot a_2 \right] + {\cal O}(a_2^5)
\right.
\nn\\
&+& \left. \frac{\gamma^2}{M_1 \beta^2}
\oint_{\Gamma_{\infty}} \frac{dz}{2 \pi i} 
\frac{ ( 1 -  \beta z )^4}{(z^2-1)^{3/2}}
\exp\left( z \, \bar{q} \cdot a_2 \right)
\exp\left[ \left( \frac{  e^{\phi} (z-1) }{\gamma ( 1-\beta  z)} + 1 \right) \bar{q} \cdot a_1 \right] + {\cal O}(a_1^5)
\right\} 
\nn\\
&+& \frac{i \kappa^4 (M_1 M_2)^2}{128} \int {\hat{d} ^4 \bar{q}} \,\int {\hat{d} ^4 \bar{\ell}} \,
 \frac{ 1 }{\bar{\ell}^2 (\bar{\ell} - \bar{q})^2} 
 \hat{\delta}( \bar{q} \cdot u_1 ) \hat{\delta}( \bar{q} \cdot u_2 )
e^{- i b \cdot {\bar q}}
\,
P[u_1,u_1, U\cdot u_2,u_2](\bl)\nn\\
&\times& \Biggl\{  \bar{q}^{\mu} 
\left( \frac{\hat{\delta}( \bar{\ell} \cdot u_2 )}{M_1}  \frac{\bar{\ell} \cdot (\bar{\ell} - \bar{q})}
{(\bar{\ell} \cdot u_1 + i \varepsilon)^2} 
+ \frac{\hat{\delta}( \bar{\ell} \cdot u_1 )}{M_2}  \frac{\bar{\ell} \cdot (\bar{\ell} - \bar{q})}
{(\bar{\ell} \cdot u_2 - i \varepsilon)^2} \right)
P[u_1,u_1, U\cdot u_2,u_2](\bq-\bl)\nn\\
&-& i  \bar{\ell}^{\mu} \bar{\ell} \cdot (\bar{\ell} - \bar{q})
\left( \frac{\hat{\delta}^{\prime}( \bar{\ell} \cdot u_1) 
\hat{\delta}( \bar{\ell} \cdot u_2 )}{M_1} 
- \frac{\hat{\delta}( \bar{\ell} \cdot u_1) 
\hat{\delta}^{\prime}( \bar{\ell} \cdot u_2 )}{M_2} \right)
P[u_1,u_1, U\cdot u_2,u_2](\bq-\bl)\nn\\
&+&   2 i \bar{\ell}^{\mu} 
\biggl[ \frac{1}{M_2}
\Bigl( P[u_1,u_1, U\cdot u_2,\bq](\bq-\bl)+ P[u_1,u_1, U\cdot \bq,u_2](\bq-\bl)
\Bigr)
\nn\\
&-& \frac{ 1 }{M_1} \Bigl(  P[u_1,\bq, U\cdot u_2,u_2](\bq-\bl)+ P[\bq ,u_1, U\cdot  u_2,u_2](\bq-\bl)\Bigr)
\biggr]
\hat{\delta}( \bar{\ell} \cdot u_1) \hat{\delta}( \bar{\ell} \cdot u_2)
\Biggr\} .
\eea
Here, we have employed  the pieces of notation\footnote{$U_{\alpha\beta}(\bq)$ is similar to what we will use later in \eqref{not}.}
\begin{equation}
{U}_{\alpha\beta}(\bq)\equiv\left[ \exp(- i (a_1+a_2)* \bar{q})\right]_{\alpha\beta} ,\,\,\,\, P_{\rho\sigma\alpha\beta} \, a^{\rho}b^{\sigma} 
 {U}^{(\alpha}_{\ \lambda}(\bq) c^{\beta)}d^{\lambda} \equiv P[a,b,U\cdot d,c] (\bq)
\end{equation}
which help tidying up our expressions a bit.}

 This result is valid to all orders in the spin of the heavy Kerr source (particle 2), extending some of the finite spin results already present in the literature \cite{mogullspin, Liu:2021zxr}. Likewise, it is also valid in the non-aligned spin configuration. We have verified that the above formulae agree exactly with the results quoted in Ref.~\cite{Liu:2021zxr} at linear order in spin.

As in  electrodynamics, one important check is to verify that the final momentum  is on shell. In other words we must confirm the relation
\beq
(\Delta p_{\text{LO}, 1} )^2 + 2 p_1 \cdot\Delta p_{\text{NLO}, 1} = 0.
\eeq
  Indeed
\bea
(\Delta p_{\text{LO}, 1})^2 &=& - \frac{\kappa^4 M_1^2 M_2^2}{64}
 \int {\hat{d} ^4 \bar{q}} \,\int {\hat{d} ^4 \bar{q}'} \,
 \hat{\delta}( \bar{q} \cdot u_1 ) \hat{\delta}( \bar{q} \cdot u_2 )
 \hat{\delta}( \bar{q}^{\prime} \cdot u_1 ) 
\hat{\delta}( \bar{q}^{\prime} \cdot u_2 )
e^{- i b \cdot ({\bar q} + {\bar q}^{\prime})}  
\nn\\
&&\qquad\qquad \qquad \times\,\,\,\, \frac{\bar{q} \cdot \bar{q}^{\prime}}{ \bar{q}^2 \bar{q}^{\prime 2} }
P[u_1,u_1, U\cdot u_2,u_2](\bq)P[u_1,u_1, U\cdot u_2,u_2](\bq')
\eea
and
\bea
2 p_1 \cdot \Delta p_{\text{NLO}, 1} &=& 
\frac{\kappa^4 M_1^2 M_2^2}{64} \int {\hat{d} ^4 \bar{q}} \,
 \hat{\delta}( \bar{q} \cdot u_1 ) \hat{\delta}( \bar{q} \cdot u_2 )
e^{- i b \cdot {\bar q}}
\int {\hat{d} ^4 \bar{\ell}} \,
 \frac{ \bar{\ell} \cdot (\bar{\ell} - \bar{q}) }{\bar{\ell}^2 (\bar{\ell} - \bar{q})^2} 
\bar{\ell} \cdot u_1 
\hat{\delta}^{\prime}( \bar{\ell} \cdot u_1) 
\hat{\delta}( \bar{\ell} \cdot u_2 )
\nn\\
&&\qquad\qquad \qquad \times\,\,\,\, 
P[u_1,u_1, U\cdot u_2,u_2](\bl)P[u_1,u_1, U\cdot u_2,u_2](\bq-\bl)\nn\\
&=& - \frac{\kappa^4 M_1^2 M_2^2}{64} \int {\hat{d} ^4 \bar{q}} \,
\int {\hat{d} ^4 \bar{\ell}} \,
 \hat{\delta}( \bar{q} \cdot u_1 ) \hat{\delta}( \bar{q} \cdot u_2 )
\hat{\delta}( \bar{\ell} \cdot u_1) \hat{\delta}( \bar{\ell} \cdot u_2 )
e^{- i b \cdot {\bar q}}
\frac{ \bar{\ell} \cdot (\bar{\ell} - \bar{q}) }{\bar{\ell}^2 (\bar{\ell} - \bar{q})^2} 
\nn\\
&&\qquad\qquad \qquad \times\,\,\,\,  
P[u_1,u_1, U\cdot u_2,u_2](\bl)P[u_1,u_1, U\cdot u_2,u_2](\bq-\bl)\nn\\
&=& \frac{\kappa^4 M_1^2 M_2^2}{64} \int {\hat{d} ^4 \bar{q}'} \, 
\int {\hat{d} ^4 \bar{\ell}} \,
 \hat{\delta}( \bar{q}^{\prime} \cdot u_1 ) \hat{\delta}( \bar{q}^{\prime} \cdot u_2 )
\hat{\delta}( \bar{\ell} \cdot u_1) \hat{\delta}( \bar{\ell} \cdot u_2 )
e^{- i b \cdot (\bar{q}^{\prime} + \bar{\ell})}
\nn\\
&&\qquad\qquad \qquad \times\,\,\,\,
\frac{ \bar{\ell} \cdot \bar{q}^{\prime} }{\bar{\ell}^2 \bar{q}^{\prime 2}} 
P[u_1,u_1, U\cdot u_2,u_2](\bl)P[u_1,u_1, U\cdot u_2,u_2](\bq')\eea
where in the last line we set $\bar{q}^{\prime} = \bar{q} - \bar{\ell}$. So we see that the final momentum is on shell as it must be.

A discussion analogous to the electromagnetic case allows us to argue that again the transverse part of the impulse is connected with the triangle contribution. In particular, the transverse NLO impulse in the scalar probe limit reads
\begin{equation}
\Delta p^\mu_{\text{NLO}, 1, \triangle} = - \frac{ \kappa^4 (M_1 M_2)^2}{128} \frac{\gamma^2}
{16\pi M_1 \beta^2 \sinh\phi} 
\oint_{\Gamma_{\infty}} \frac{dz}{2 \pi i} 
\frac{ ( 1 - \beta z )^4}{(z^2-1)^{3/2}}
\frac{( b^{\mu} +  \Pi^{\mu}_{\ \nu} \mathfrak{a}^\nu _2 (z) )}{|b +  \Pi\mathfrak{a} _2 (z)|^{3}} 
\end{equation}
where
\begin{equation}
\mathfrak{a}_{2}^\mu (z)\equiv  z \frac{\epsilon^\mu (a_2 , u_1, u_2)}{\sinh\phi}  {}, \,\,\,\, \,\,\,\,\,  
b_{\perp}^\mu  = {\Pi^\mu}_\nu (b+i a_2)^\nu
\end{equation}

Again, we have not been able to recast the integral in terms of elementary functions, yet it is immediate to expand it to the desired spin order  and extract the residue at infinity. 

{\subsection{Direct evaluation of the gravity spin-orbit term}}

{Let us now gain some intuition on \eqref{loopdp} by solving the one loop integrals in a specific scenario -- we work in the scalar probe limit taking $a_1\to 0,\,\,\, M_2\to \infty$ and we work up to linear order in the source spin $a_2$. Even though our considerations here apply to the gravity case, one can easily carry out similar calculations for the electromagnetic case worked out previously.}

{We start by computing the transverse part of the momentum change. The contour integral becomes
\begin{equation}
\begin{split}
\oint_{\Gamma_{\infty}} \frac{dz}{2 \pi i} 
\frac{ ( 1 - \beta z )^4}{(z^2-1)^{3/2}}
\frac{( b^{\mu} +  \Pi^{\mu}_{\ \nu} \mathfrak{a}^\nu _2 (z) )}{|b +  \Pi\mathfrak{a} _2 (z)|^{3}} &\approx 
\oint_{\Gamma_{\infty}} \frac{dz}{2 \pi i} 
\frac{ ( 1 - \beta z )^4}{(z^2-1)^{3/2}}
\left(
\frac{b^\mu}{b^3}+z
\frac{ \epsilon^\mu (a_2 , u_1, u_2)}{b^3 \sinh\phi} -3z
\frac{b^\mu  \epsilon  (b, a_2 , u_1, u_2)}{b^5 \sinh\phi} 
\right) .
\end{split}
\end{equation}
There are two types of residues here, namely
\begin{equation}
\oint_{\Gamma_{\infty}} \frac{dz}{2 \pi i} 
\frac{ ( 1 - \beta z )^4}{(z^2-1)^{3/2}}=-\frac{3}{2}\frac{\beta^2}{\gamma^2} (5\gamma^2-1),\,\,\,\,\, \frac{1}{\sinh\phi}
\oint_{\Gamma_{\infty}} \frac{dz}{2 \pi i} 
\frac{ z( 1 - \beta z )^4}{(z^2-1)^{3/2}}= \frac{2}{\gamma^3} (5\gamma^2-3)
\end{equation} 
giving
\begin{equation}
\Delta p^\mu_{\text{NLO}, 1, \triangle} = G^2\frac{M_1 M_2^2}{b^3}\left( \frac{3\pi}{4} \frac{5\gamma^2-1}{\sqrt{\gamma^2-1}} b^\mu+ \pi\frac{\gamma(5\gamma^2-3)}{({\gamma^2-1})^{3/2}}\left(3  
\frac{b^\mu  \epsilon  (b, a_2 , u_1, u_2)}{b^2} -{
\epsilon^\mu (a_2 , u_1, u_2)}{}
\right)
\right).
\end{equation}
What about the longitudinal terms? We have that
 \bea\label{longexact}
\Delta p^\mu_{\text{NLO} \parallel}  &=& {8i\pi^2 G^2  M_1 M_2^2} \int {\hat{d} ^4 \bar{q}} \,\int {\hat{d} ^4 \bar{\ell}} \,
 \frac{ 1 }{\bar{\ell}^2 (\bar{\ell} - \bar{q})^2} 
 \hat{\delta}( \bar{q} \cdot u_1 ) \hat{\delta}( \bar{q} \cdot u_2 )
e^{- i b \cdot {\bar q}}
\,
P[u_1,u_1, U\cdot u_2,u_2](\bl)\nn\\
&\times& \Biggl\{ \left( \bar{q}^{\mu} 
 {\hat{\delta}( \bar{\ell} \cdot u_2 )}  \frac{\bar{\ell} \cdot (\bar{\ell} - \bar{q})}
{(\bar{\ell} \cdot u_1 + i \varepsilon)^2} 
- i  \bar{\ell}^{\mu} \bar{\ell} \cdot (\bar{\ell} - \bar{q})
 {\hat{\delta}^{\prime}( \bar{\ell} \cdot u_1)  
\hat{\delta}( \bar{\ell} \cdot u_2 )}  \right)
P[u_1,u_1, U\cdot u_2,u_2](\bq-\bl)\nn\\
&-&   2 i \bar{\ell}^{\mu} 
\biggl(  P[u_1,\bq, U\cdot u_2,u_2](\bq-\bl)+ P[\bq ,u_1, U\cdot  u_2,u_2](\bq-\bl)\Bigr)
\hat{\delta}( \bar{\ell} \cdot u_1) \hat{\delta}( \bar{\ell} \cdot u_2)
\Biggr\} .
\nn\\
\eea
We start with the zero-order spin term. This reads, on-shell\footnote{We find that the Box integral $\sim\displaystyle{\int}\frac{\hat{d}^4   \bl}{ \bl^2 (\bq-\bl)^2}  \frac{\bar{\ell} \cdot (\bar{\ell} - \bar{q})}
{(\bar{\ell} \cdot u_1 + i \varepsilon)^2} {\hat{\delta}( \bar{\ell} \cdot u_2 )}   $ evaluates to zero in 4 spacetime dimensions,  for both the $\mathcal{O }(a_2^0)$ and $\mathcal{O }(a_2^1)$ terms, so we do not consider it here.} of the integrals
\bea\label{longa0}
\Delta p^\mu_{\text{NLO} \parallel}  &\approx& {8\pi^2 G^2  M_1 M_2^2}  \int  
 \frac{{\hat{d} ^4 \bar{q}}  \, {\hat{d} ^4 \bar{\ell}}}{\bar{\ell}^2 (\bar{\ell} - \bar{q})^2} 
 \hat{\delta}( \bar{q} \cdot u_1 ) \hat{\delta}( \bar{q} \cdot u_2 )
e^{- i b \cdot {\bar q}} \bar{\ell}^{\mu} \bar{\ell} \cdot (\bar{\ell} - \bar{q})
 {\hat{\delta}^{\prime}( \bar{\ell} \cdot u_1)  
\hat{\delta}( \bar{\ell} \cdot u_2 )}   (2\gamma^2-1)^2 
\nn\\
\eea
because in this limit the numerator contractions evaluate to 
\begin{equation}
 P[\bq ,u_1, U\cdot  u_2,u_2] \to 0,\,\,\,\,\,\,
 P[u_1 ,u_1, U\cdot  u_2,u_2] \to 2\gamma^2-1.
\end{equation}
Let us compute \eqref{longa0} in full detail, since the tricks we need will be useful for the spin-orbit computation too. First note that the following identity  holds:
\begin{equation}\label{loop vel}
 \int  
 \frac{ {\hat{d} ^4 \bar{\ell}}}{\bar{\ell}^2 (\bar{\ell} - \bar{q})^2} 
  \bar{\ell}^{\mu} \bar{\ell} \cdot (\bar{\ell} - \bar{q})
 {\hat{\delta}^{\prime}( \bar{\ell} \cdot u_1)  
\hat{\delta}( \bar{\ell} \cdot u_2 )} =\tilde{u}^\mu \int  
 \frac{ {\hat{d} ^4 \bar{\ell}}}{\bar{\ell}^2 (\bar{\ell} - \bar{q})^2}   \bar{\ell} \cdot (\bar{\ell} - \bar{q})
 {\hat{\delta} ( \bar{\ell} \cdot u_1)  
\hat{\delta}( \bar{\ell} \cdot u_2 )} 
\end{equation}
where $\tilde{u}^\mu = \frac{u_1^\mu -\gamma u_2^\mu}{\gamma^2-1}.$
This can be shown, for instance, by expanding the loop integral in a basis of vectors and contracting to find the coefficients. Such identity allows us to rewrite  \eqref{longa0} in a more symmetric way as
\begin{equation}
\begin{split}
\Delta p^\mu_{\text{NLO} \parallel} &= {8\pi^2 G^2  M_1 M_2^2}  (2\gamma^2-1)^2  	\tilde{u}^\mu \int  
 \frac{{\hat{d} ^4 \bar{q}}  \, {\hat{d} ^4 \bar{\ell}}}{\bar{\ell}^2 (\bar{\ell} - \bar{q})^2} 
 \hat{\delta}( \bar{q} \cdot u_1 ) \hat{\delta}( \bar{q} \cdot u_2 )
e^{- i b \cdot {\bar q}}  \bar{\ell} \cdot (\bar{\ell} - \bar{q})
 {\hat{\delta} ( \bar{\ell} \cdot u_1)  
\hat{\delta}( \bar{\ell} \cdot u_2 )} \\&=
 {8\pi^2 G^2  M_1 M_2^2}   (2\gamma^2-1)^2  	\tilde{u}^\mu \,\mathcal{I}\cdot \mathcal{I}
\end{split}
\end{equation}
where we could recast the two integrations as the square of a vector after changing $\bq-\bl\to \bq$. We can then readily evaluate the Fourier transform 
\begin{equation}
\mathcal{I}^\mu= \int  
 {{\hat{d} ^4 \bar{q}}   } \,
 \hat{\delta}( \bar{q} \cdot u_1 ) \hat{\delta}( \bar{q} \cdot u_2 )
e^{- i b \cdot {\bar q}} \frac{\bq^\mu}{\bq^2}=\frac{i}{2\pi \sinh\phi} \frac{b^\mu}{b^2}
\end{equation}
using (see for instance \cite{Herrmann:2021tct})
\beq\label{master}
i \int {\hat{d}^D \bar{q}}\,
\hat{\delta}( \bar{q} \cdot u_1 )\hat{\delta}( \bar{q} \cdot u_2 ) e^{- i b \cdot {\bar q}}
\frac{q^{\mu}}{\bar{q}^{2\alpha}}
= - \frac{1}{\sinh \phi} \frac{\Gamma(D/2-\alpha)}{2^{2\alpha-1} \pi^{(D-2)/2} \Gamma(\alpha)}
\frac{b^{\mu}}{b^{D-2\alpha}}.
\eeq
We finally get to 
\begin{equation}
\Delta p^\mu_{\text{NLO} \parallel} =\frac{2 G^2  M_1 M_2^2}{b^2}  \frac{(2\gamma^2-1)^2}{(\gamma^2-1)^2}  	(u_1^\mu -\gamma u_2^\mu)
\end{equation}
in agreement with the literature  \cite{Herrmann:2021tct}.}

{The spin-orbit longitudinal term in \eqref{longexact} requires a bit more care. This spin order is here encoded by the linearised  matrix exponential
\begin{equation}
U_{\alpha\beta}(\bq)\approx \eta_{\alpha \beta}-i \epsilon_{\alpha\beta} (a_2, \bq).
\end{equation}
Using this we find that, after some algebra
 \bea 
\Delta p^\mu_{\text{NLO} \parallel}  &\approx& {16i\pi^2 G^2  M_1 M_2^2}\,\gamma (2\gamma^2-1) \int {\hat{d} ^4 \bar{q}} \,\int {\hat{d} ^4 \bar{\ell}} \,
 \frac{ 1 }{\bar{\ell}^2 (\bar{\ell} - \bar{q})^2} 
 \hat{\delta}( \bar{q} \cdot u_1 ) \hat{\delta}( \bar{q} \cdot u_2 )
e^{- i b \cdot {\bar q}}
\nn\\
&\times& \Biggl\{   \bar{\ell}^{\mu} \bar{\ell} \cdot (\bar{\ell} - \bar{q})
 {\hat{\delta}^{\prime}( \bar{\ell} \cdot u_1)  
\hat{\delta}( \bar{\ell} \cdot u_2 )}   
\epsilon  (  \bq, a_2,  u_1, u_2)+   2  \bar{\ell}^{\mu} 
 \epsilon(a_2, u_2, \bq, \bl)
\hat{\delta}( \bar{\ell} \cdot u_1) \hat{\delta}( \bar{\ell} \cdot u_2)
\Biggr\} \nn\\
&=& 
{16i\pi^2 G^2  M_1 M_2^2}\,\gamma (2\gamma^2-1) \int {\hat{d} ^4 \bar{q}} \,\int {\hat{d} ^4 \bar{\ell}} \,
 \frac{ 1 }{\bar{\ell}^2 (\bar{\ell} - \bar{q})^2} 
 \hat{\delta}( \bar{q} \cdot u_1 ) \hat{\delta}( \bar{q} \cdot u_2 )
e^{- i b \cdot {\bar q}}
\nn\\
&\times& \Biggl\{    \bar{\ell} \cdot (\bar{\ell} - \bar{q})
\tilde{u}^\mu 
\epsilon  (  \bq, a_2,  u_1, u_2)+   2  \bar{\ell}^{\mu} 
 \epsilon(a_2, u_2, \bq, \bl)
\Biggr\}\hat{\delta}( \bar{\ell} \cdot u_1) \hat{\delta}( \bar{\ell} \cdot u_2)
\eea
having used  \eqref{loop vel} again. Let us now follow the same line of thinking of the scalar term. Changing $\bq-\bl \to \bq$ the two integrations separate,  after a short calculation we find that our expression can be recast into
 \bea 
\Delta p^\mu_{\text{NLO} \parallel} 
&=& 
{16i\pi^2 G^2  M_1 M_2^2}\,\gamma (2\gamma^2-1)  \left( \tilde{ u}^\mu 
\epsilon_\alpha (a_2, u_1, u_2) I^\alpha+ 2 \epsilon_{\alpha\beta} (a_2, u_2)I^{\mu \alpha\beta}
\right)
\nn\\
\eea
where we've hidden inside the $I$ tensors the remaining integrations, now factorized. 
The first one reads
\begin{equation}
\begin{split}
I^\alpha &=
2 \int {\hat{d} ^4 \bar{q}} \, \hat{\delta}( \bar{q} \cdot u_1 ) \hat{\delta}( \bar{q} \cdot u_2 )\frac{ e^{-ib\cdot \bar{q}}}{\bq^2} \bq^\alpha \bq^\beta\int 
 { {\hat{d} ^4 \bar{\ell}}   } \,  \hat{\delta }(\bar{\ell}\cdot u_1)\hat{\delta} (\bar{\ell}\cdot u_2)  \frac{ e^{-ib\cdot \bl}}{\bl^2}  \bl_\beta.
\end{split}
\end{equation}
We note that the two Fourier transforms can be obtained by direct differentiation of 
the master integral \eqref{master}
\begin{equation} 
\int {\hat{d} ^4 \bar{q}}   \,
    \hat{\delta }(\bar{q}\cdot u_1)\hat{\delta} (\bar{q}\cdot u_2)  \frac{ e^{-ib\cdot \bq}}{\bq^{2k}}   \bq^{\mu_1}\cdots\bq^{\mu_n}  =  i^n\partial^{\mu_1}_\perp \cdots \partial^{\mu_n}_\perp \int {\hat{d} ^4 \bar{q}}   \,
    \hat{\delta }(\bar{q}\cdot u_1)\hat{\delta} (\bar{q}\cdot u_2)  \frac{ e^{-ib\cdot \bq}}{\bq^{2k}}, \,\,\,\, \partial_\perp^\mu\equiv{\Pi^\mu }_\nu \partial^\nu_b
\end{equation}
and hence 
\begin{equation}
\begin{split}
 \epsilon_\alpha (a_2, u_1, u_2) I^\alpha&=
\frac{2i}{(2\pi)^2} \frac{u_1^\mu -\gamma\, u^\mu _2}{(\sinh \phi)^4}  \epsilon_\alpha( u_1, u_2, a_2)     \frac{1}{b^4}\left(
    \Pi^{\alpha\beta}-2 \frac{b^\alpha b^\beta}{b^2}
    \right)b_\beta\\&
    =
-\frac{2i}{(2\pi)^2} \frac{u_1^\mu -\gamma\, u^\mu _2}{(\sinh \phi)^4}  \frac{\epsilon(b, u_1, u_2, a_2)    }{b^4}. 
\end{split}
\end{equation}
The remaining piece is instead
\begin{equation}
\begin{split}
  I^{\mu \alpha\beta}&=  \int {\hat{d} ^4 \bar{q}} \, \hat{\delta}( \bar{q} \cdot u_1 ) \hat{\delta}( \bar{q} \cdot u_2 )\frac{ e^{-ib\cdot \bar{q}}}{\bq^2} \,\bq^{\alpha}
\int {\hat{d} ^4 \bar{\ell}}   \,
    \hat{\delta }(\bar{\ell}\cdot u_1)\hat{\delta} (\bar{\ell}\cdot u_2)  \frac{ e^{-ib\cdot \bl}}{\bl^2} \bl^\mu   \bl^{\beta}\\&=\frac{i}{(2\pi\sinh\phi )^2} \frac{b^\alpha}{b^4} \left(
    \Pi^{\mu\beta}-2 \frac{b^\mu b^\beta}{b^2} 
    \right)
    \end{split}
\end{equation}
yielding
\begin{equation}
\begin{split}
I^{\mu\alpha\beta}  \epsilon_{\alpha\beta}(a_2, u_2)&=
  \frac{i}{(2\pi\sinh\phi )^2\, b^4} {\Pi^{\mu \beta}}\epsilon_\beta(b, a_2, u_2)\\&=   \frac{i}{(2\pi\sinh\phi )^2\, b^4}  \left(\epsilon^\mu (b, a_2, u_2)+
\frac{u_1^\mu-\gamma u_2^\mu}{(\sinh \phi)^2}\epsilon (u_1, a_2, b, u_2)
\right).
\end{split}
\end{equation}
Finally we find that
\begin{equation}
\Delta p^\mu_{\text{NLO} \parallel}\approx 
 \frac{16  G^2  M_1 M_2^2}{     b^4} \frac{ {\gamma}(2\gamma^2-1)}{\gamma^2-1}
\left(
 \frac{u_1^\mu -\gamma\, u^\mu _2}{\gamma^2-1}  {\epsilon(b, u_1, u_2, a_2)    }  +  \frac{1}{2}   \epsilon^\mu (b, a_2, u_2) 
  \right)
\end{equation}
and putting everything together one obtains
\begin{equation}
\begin{split}
\Delta p^\mu_{\text{NLO}} = G^2\frac{M_1 M_2^2}{b^3}&\left[ \frac{3\pi}{4} \frac{5\gamma^2-1}{\sqrt{\gamma^2-1}} b^\mu+ 2b  \frac{(2\gamma^2-1)^2}{(\gamma^2-1)^2}  	(u_1^\mu -\gamma u_2^\mu)
\right.\\&+\left.  \pi\frac{\gamma(5\gamma^2-3)}{({\gamma^2-1})^{3/2}}\left(3  
\frac{b^\mu  \epsilon  (b, a_2 , u_1, u_2)}{b^2} -{
\epsilon^\mu (a_2 , u_1, u_2)}{}
\right)
\right.
\\&+
\frac{16 {\gamma}(2\gamma^2-1)}{\gamma^2-1}\frac{1}{b}
\left(\left.
 \frac{u_1^\mu -\gamma\, u^\mu _2}{\gamma^2-1}  {\epsilon(b, u_1, u_2, a_2)    }  +  \frac{1}{2}   \epsilon^\mu (b, a_2, u_2) 
  \right)\right]+\mathcal{O}(a^2_2)
\end{split}
\end{equation}
in agreement with \cite{Liu:2021zxr}.}

{\subsection{Scattering angle for aligned spins}}
 
We can also calculate the scattering angle for aligned spins from our formulae, recovering the results given in \cite{Guevara:19a}. {In such case the motion is confined on a two dimensional plane and the dynamics specified entirely by the scattering angle. We only quote the results here, the residue evaluation is analogous to the electromagnetic one outlined in section  \eqref{sectionem}. The only difference is in the integrand numerator $(1-\beta z)^2\to (1-\beta z)^2\times(1-\beta z)^2$ for gravity, which again is an instance of the double copy.} We have
\bea
\theta^{(0)} &=& \frac{2 G_N M_1 M_2}{b}
\frac{\sqrt{M_1^2 + M_2^2 + 2 M_1 M_2 \gamma}}{M_1 M_2 \gamma^2 \beta^2} \,
\textrm{Re}\left[
\Bigl( (2\gamma^2-1) b_{\nu} - 2 i \gamma \epsilon_{\mu\nu\alpha\beta} b_{\mu} u_1^{\alpha} u_{2}^{\beta}
\Bigr)
\frac{b^{\nu}_{\perp}}{b^{2}_{\perp}}
 \right]
\nn\\
\theta^{(1)} &=& - \pi G_N^2 
\sqrt{M_1^2 + M_2^2 + 2 M_1 M_2 \gamma} \,
\frac{\partial}{\partial b} 
\Bigl[ M_1 \bigl( {\cal G}(a_1, a_2,5) + {\cal O}(a_2^{5}) \bigr)
+ M_2 \bigl( {\cal G}(a_2, a_1,5) + {\cal O}(a_1^{5}) \bigr) \Bigr]
\nn\\
{\cal G}( a_1,  a_2,n) &=& \frac{1}{2  a_1^2}
\Bigl[ - b + {\cal E}( a_1,  a_2,n) \Bigr].
\eea
{In the scalar probe limit} and $a_1 \to 0,\,\,\, M_1/M_2 \to 0$, one finds that  %
\bea\label{gravprob}
\theta^{(0)} &\approx& \frac{2 G_N M_1 M_2}{b}
\frac{1}{M_1\gamma^2 \beta^2} \,
\textrm{Re}\left[
\Bigl( (2\gamma^2-1) b_{  \nu} - 2 i \gamma \epsilon_{\mu\nu\alpha\beta} b_{\mu} u_1^{\alpha} u_{2}^{\beta}
\Bigr)
\frac{b^{\nu}_{\perp}}{b^{2}_{\perp}}
 \right]
\nn\\
\theta^{(1)} &\approx& - \pi (G_NM_1 M_2)^2 \frac{1}{M_1^2} \,
\frac{\partial}{\partial b} {\cal G}(a_2, 0,5) .
\eea
{We note again that such limits yields expressions valid to all orders in the spin of the heavy Kerr source.  This is because the two sets of spurious Compton poles that appear in \eqref{loopdp} drop: $\frac{O(a_2^5)}{M_2}\to 0$ because of the source heavy mass $M_2\to\infty$, and  $\frac{O(a_1^5)}{M_1}= 0$ since we are taking $a_1=0$.}

\section{Some considerations on the probe limit and   the Kerr-Schild gauge}\label{probesss}

In this final sections we propose an alternative method that can be used to derive probe limit deflections. Namely, we consider a light spin-less body scattering off a heave spinning source. To do so, we will make use of Kerr-Schild (KS) coordinates. Even if, as it will turn out, KS expressions are computationally difficult to deal with we believe this is interesting in itself since the KS gauge is not often employed in QFT computations. Besides, we will be able to establish a beautiful connection with the classical Double copy of \cite{Monteiro:2014cda}, but from a momentum space point of view. 

Let us start with a lightning review of Kerr-Schild coordinates, which are most often introduced in the context of gravity. A KS spacetime metric $g_{\mu\nu}$ which solves the \textit{full} Einstein equation can be written as 
\begin{equation}
g_{\mu\nu}=\eta_{\mu\nu}+\phi k_\mu k_\nu
\end{equation}
where $\phi$ and $k_\mu$ are a (space-time dependent\footnote{We will actually specify to the case where both $\phi$ and $k_\mu$ also don't depend on time: $\phi=\phi(\vec{x})$, $k_\mu=k_\mu (\vec{x})$.}) scalar and vector with the following properties
\begin{equation}\label{prop}
k^\mu k^\nu \eta_{\mu\nu}=0= k^\mu k^\nu g_{\mu\nu},\,\,\,\, \, \eta_{\alpha\beta}k^\alpha \partial^\beta k^\mu=0.
\end{equation}
So $k_\mu$ is null and geodesic with respect to the flat space metric (which is what we will use to contract up and down indices in what follows). These simple but powerful properties imply that, for instance
\begin{equation}
g^{\mu\nu}=\eta^{\mu\nu}-\phi k^\mu k^\nu, \,\,\,\,\, \sqrt{-g}=1
\end{equation}
hold as exact statements. One can also see that \eqref{prop} have the effect of linearising Einstein's field equations. One striking consequence of this is that if one defines a gauge potential as 
\begin{equation}
A^\mu=\phi k^\mu
\end{equation}  
then the field equations for the metric reduce to the Maxwell equations for $A^\mu$. This is the essence of the classical double copy \cite{Monteiro:2014cda} between gauge theory and Einstein gravity. Note indeed that the ({exact}) graviton field is explicitly symmetric and traceless. 
However, this is just a gist of the classical double copy tale, for more details and recent developments see, for instance \cite{Luna:2016due, Goldberger:2016iau, Luna:2017dtq, Luna:2015paa, Cardoso:2016ngt, Luna:2016hge, Adamo:2017nia, Anastasiou:2018rdx, Lee:2018gxc, Plefka:2018dpa, Luna:2018dpt, Cho:2019ype, Godazgar:2019ikr, Bah:2019sda, Alawadhi:2019urr, Goldberger:2019xef, Luna:2020adi, Cristofoli:2020hnk, Adamo:2022rmp, Borsten:2020xbt, Chacon:2020fmr, Godazgar:2020zbv, White:2020sfn, Berman:2020xvs, Lescano:2020nve, Monteiro:2020plf, Monteiro:2021ztt}.

\subsection{Gauge theory probe limit}

Inspired by the discussion above, we start by considering a Lagrangian of a charged scalar  which is minimally coupled to a KS static background. Here we will use $m$ to indicate the mass of the probe
\begin{equation}
\mathcal{L}= (D^\mu \Phi)^\dagger D_\mu \Phi- m^2 \Phi^\dagger \Phi, \,\,\, D_\mu=\partial_\mu -ieA_\mu,\,\,\,\, A_\mu=\phi k_\mu.
\end{equation}
This Lagrangian will model (when we will take the classical limit of the amplitudes it generates) the geodesic motion\footnote{We will neglect self interactions.} of the light scalar on the heavy background entailed by the gauge potential. We will specify the specific expression of $A_\mu$ only later when needed. For now we just observe that if $A^2=0$ then $\mathcal{L}=\mathcal{L}_0+\mathcal{L}_{int}$, where $\mathcal{L}_0$ is a free, charged Klein Gordon Lagrangian\footnote{Note that we omit the kinematic part of the Lagrangian since we don't want dynamical photons/gravitons here, the background is static.}, and the interaction is strictly \textit{linear} in $A^\mu$
\begin{equation}
\mathcal{L}_{int}=ie A_{\mu} \left(\Phi^\dagger \partial^\mu \Phi-\Phi \,\partial^\mu \Phi^\dagger     \right)
\end{equation} 
with Feynman rule 
\begin{figure}[H]
\begin{center}

 
\tikzset{
pattern size/.store in=\mcSize, 
pattern size = 5pt,
pattern thickness/.store in=\mcThickness, 
pattern thickness = 0.3pt,
pattern radius/.store in=\mcRadius, 
pattern radius = 1pt}
\makeatletter
\pgfutil@ifundefined{pgf@pattern@name@_26d2ucawd}{
\pgfdeclarepatternformonly[\mcThickness,\mcSize]{_26d2ucawd}
{\pgfqpoint{0pt}{0pt}}
{\pgfpoint{\mcSize+\mcThickness}{\mcSize+\mcThickness}}
{\pgfpoint{\mcSize}{\mcSize}}
{
\pgfsetcolor{\tikz@pattern@color}
\pgfsetlinewidth{\mcThickness}
\pgfpathmoveto{\pgfqpoint{0pt}{0pt}}
\pgfpathlineto{\pgfpoint{\mcSize+\mcThickness}{\mcSize+\mcThickness}}
\pgfusepath{stroke}
}}
\makeatother
\tikzset{every picture/.style={line width=0.75pt}} 

\begin{tikzpicture}[x=0.75pt,y=0.75pt,yscale=-1,xscale=1]

\draw    (143.67,835.76) -- (99.33,794.32) ;
\draw [shift={(121.5,815.04)}, rotate = 43.07] [fill={rgb, 255:red, 0; green, 0; blue, 0 }  ][line width=0.08]  [draw opacity=0] (5.36,-2.57) -- (0,0) -- (5.36,2.57) -- cycle    ;
\draw    (99.33,794.32) -- (148.67,754.99) ;
\draw [shift={(124,774.66)}, rotate = 141.43] [fill={rgb, 255:red, 0; green, 0; blue, 0 }  ][line width=0.08]  [draw opacity=0] (5.36,-2.57) -- (0,0) -- (5.36,2.57) -- cycle    ;
\draw    (99.33,794.32) .. controls (97.7,796.02) and (96.03,796.05) .. (94.33,794.41) .. controls (92.64,792.78) and (90.97,792.81) .. (89.33,794.5) .. controls (87.7,796.19) and (86.03,796.22) .. (84.34,794.59) .. controls (82.65,792.96) and (80.98,792.99) .. (79.34,794.68) .. controls (77.7,796.37) and (76.03,796.4) .. (74.34,794.77) .. controls (72.65,793.14) and (70.98,793.17) .. (69.34,794.86) .. controls (67.7,796.55) and (66.03,796.58) .. (64.34,794.95) .. controls (62.65,793.32) and (60.98,793.35) .. (59.34,795.04) .. controls (57.7,796.73) and (56.03,796.76) .. (54.34,795.13) .. controls (52.65,793.49) and (50.98,793.52) .. (49.34,795.21) .. controls (47.7,796.9) and (46.03,796.93) .. (44.34,795.3) -- (42.08,795.34) -- (42.08,795.34) ;
\draw    (57.44,803.13) -- (74,803.52) ;
\draw [shift={(77,803.59)}, rotate = 181.37] [fill={rgb, 255:red, 0; green, 0; blue, 0 }  ][line width=0.08]  [draw opacity=0] (5.36,-2.57) -- (0,0) -- (5.36,2.57) -- cycle    ;
\draw  [pattern=_26d2ucawd,pattern size=3.2249999999999996pt,pattern thickness=0.75pt,pattern radius=0pt, pattern color={rgb, 255:red, 0; green, 0; blue, 0}] (22.25,795.34) .. controls (22.25,789.87) and (26.69,785.43) .. (32.17,785.43) .. controls (37.64,785.43) and (42.08,789.87) .. (42.08,795.34) .. controls (42.08,800.82) and (37.64,805.26) .. (32.17,805.26) .. controls (26.69,805.26) and (22.25,800.82) .. (22.25,795.34) -- cycle ;
\draw [line width=1.5]    (32.17,805.26) -- (32.11,834.79) ;
\draw [line width=1.5]    (32.11,754.79) -- (32.17,785.43) ;

\draw (141,759.66) node [anchor=north west][inner sep=0.75pt]    {$p+q$};
\draw (136,805.66) node [anchor=north west][inner sep=0.75pt]    {$p$};
\draw (60.17,809.49) node [anchor=north west][inner sep=0.75pt]    {$q$};
\draw (194.5,794.51) node [anchor=north west][inner sep=0.75pt]    {$=ie \tilde{A}(q)\cdot (2p+q)$};

\end{tikzpicture}

\end{center}
\end{figure}
where $\tilde{A}_\mu (q) $ represents the Fourier transform of the gauge potential.

Before we start gathering terms we also report, for the sake of completeness, the deflection expressions that one finds in the probe limit. These are found following the same steps first outlined in \cite{Kosower:19}, the only difference is that the incoming state is now a one particle state
\begin{equation}
|\psi_{in}\rangle\equiv |\psi\rangle= \int d \Phi(p) e^{i\frac{b\cdot p}{\hbar}} \phi(p)|p\rangle
\end{equation} 
then for the  probe deflection 
we find 
\begin{equation} 
\begin{split}
&
\Delta p^\mu_{\text{LO}} =\KMOCav{\int \hat{d}^4 q\,\hat{\delta} (2p\cdot q + q^2)  \hat{\delta}(V\cdot q) e^{-\frac{i}{\hbar}{b\cdot q}} q^\mu i\mathcal{A}_\text{LO}(q)}
\end{split}
\end{equation}
and
\begin{equation}\label{i1i2}
\begin{split}
&
\Delta p^\mu_{\text{NLO}} =\KMOCav{\int \hat{d}^4 q\,\hat{\delta} (2p\cdot q + q^2)  \hat{\delta}(V\cdot q) e^{-\frac{i}{\hbar}{b\cdot q}} q^\mu i\mathcal{A}_\text{NLO}(q)}\\&
+\KMOCav{\int\hat{d}^4 q\,\hat{\delta} (2p\cdot q + q^2) \hat{\delta}(V\cdot q )e^{-\frac{i}{\hbar}{b\cdot q}}  \int \hat{d}^4 \ell\,\hat{\delta} (2p\cdot \ell + \ell^2) \hat{\delta}(V\cdot \ell)\ell^\mu \mathcal{A}_\text{LO}(q-\ell)\mathcal{A}_\text{LO}(\ell)}
\end{split}
\end{equation}
where $V^\mu$ is the source particle's 4-velocity, which in the frame of the heavy particle can be taken to be $V^\mu=(1,\vec{0})$. Note the slight change of notation with respect to the previous sections where both particle's velocities where indicated with $u^\mu _i$.

One difference with respect to \eqref{dp} is that the source delta functions $\hat{\delta}(V\cdot q )$ are now exact, so they will not need to be expanded in the classical limit as it is usually done for $\hat{\delta}(2p\cdot q + q^2)$. These delta functions can also be understood as appearing from computing $\tilde{A}_\mu (q)$: since $A_\mu(x)=A_\mu(\vec{x})$ the transform of $A_\mu (x)$ will include a factor of $\hat{\delta}(q_0)=\hat{\delta}(q\cdot V)$, we have peeled this off the amplitudes in  eq \eqref{i1i2} but we may equivalently think of it as part of the $\tilde{A}_\mu(q)$ hoping no confusion arises. We also factor out the charge of the source, which we write as $Q$: $A_\mu\to Q A_\mu$.

Next thing we need is to figure out the $\hbar$ scaling of $\tilde{A}_\mu (q)$, which is needed to take the classical limit. This is easy once we know that $A_\mu (x)$ is a classical background. Stripping off the on-shell delta function and the charge $Q$ we are left with
\begin{equation}
\tilde{A}_\mu (q)\to \hbar^{-2} \tilde{A}_\mu (\bar{q}).
\end{equation}

The tree-level deflection is then obtained readily from the expression of the vertex and  \eqref{i1i2}. We have, rescaling momenta and couplings as usual\footnote{Note that here we use calligraphic letters in a slightly different way as in the previous sections,  as we want to use ``$A$" just  to indicate the vector potential expression. }
\begin{equation}
i\mathcal{A}(\bar{q})=\frac{2i eQ}{\hbar^3} \tilde{A}(\bar{q})\cdot p
\end{equation}
which may look odd at first sight, since we are missing the $1/q^2$ pole, however we will see that this is hiding inside $\tilde{A}$ as the Fourier transform of $\sim 1/r$. We thus get the classical leading order deflection by the KMOC average over wave packets, setting $p^\mu\approx m u^\mu$
\begin{equation}\label{okok}
{\Delta} p^\mu_\text{LO}=-ieQ \int \hat{d}^4 \bar{q}\, \hat{\delta}(u\cdot \bar{q})\hat{\delta} (V\cdot \bar{q})e^{-ib\cdot\bar{q}}{\bar{q}^\mu}\, u\cdot \tilde{A}(\bar{q}).
\end{equation} 

\subsubsection{1-loop momentum deflection}

Now let us turn our attention to the NLO deflection. We start by gathering all relevant diagrams first. At this order both $I_1^\mu$ and $I_2^\mu$ terms will be important. Conveniently, in KS gauge only 1 diagram is needed: the box diagram (the cross box one is the same as the box in the probe limit). Below we report for clarity the box and cut-box graphs with their explicit momentum routing:

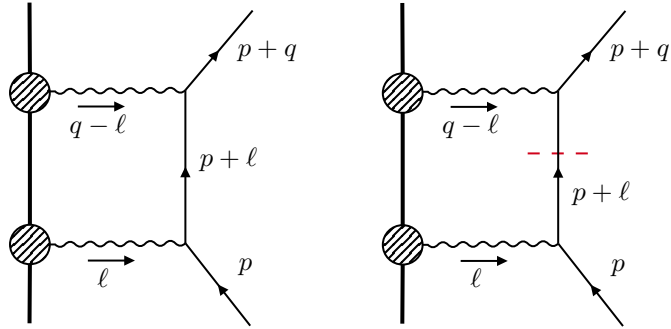
\begin{figure}[H]
\begin{center}

 
\tikzset{
pattern size/.store in=\mcSize, 
pattern size = 5pt,
pattern thickness/.store in=\mcThickness, 
pattern thickness = 0.3pt,
pattern radius/.store in=\mcRadius, 
pattern radius = 1pt}
\makeatletter
\pgfutil@ifundefined{pgf@pattern@name@_oy50tdxpz}{
\pgfdeclarepatternformonly[\mcThickness,\mcSize]{_oy50tdxpz}
{\pgfqpoint{0pt}{0pt}}
{\pgfpoint{\mcSize+\mcThickness}{\mcSize+\mcThickness}}
{\pgfpoint{\mcSize}{\mcSize}}
{
\pgfsetcolor{\tikz@pattern@color}
\pgfsetlinewidth{\mcThickness}
\pgfpathmoveto{\pgfqpoint{0pt}{0pt}}
\pgfpathlineto{\pgfpoint{\mcSize+\mcThickness}{\mcSize+\mcThickness}}
\pgfusepath{stroke}
}}
\makeatother

 
\tikzset{
pattern size/.store in=\mcSize, 
pattern size = 5pt,
pattern thickness/.store in=\mcThickness, 
pattern thickness = 0.3pt,
pattern radius/.store in=\mcRadius, 
pattern radius = 1pt}
\makeatletter
\pgfutil@ifundefined{pgf@pattern@name@_hzqwiqu7h}{
\pgfdeclarepatternformonly[\mcThickness,\mcSize]{_hzqwiqu7h}
{\pgfqpoint{0pt}{0pt}}
{\pgfpoint{\mcSize+\mcThickness}{\mcSize+\mcThickness}}
{\pgfpoint{\mcSize}{\mcSize}}
{
\pgfsetcolor{\tikz@pattern@color}
\pgfsetlinewidth{\mcThickness}
\pgfpathmoveto{\pgfqpoint{0pt}{0pt}}
\pgfpathlineto{\pgfpoint{\mcSize+\mcThickness}{\mcSize+\mcThickness}}
\pgfusepath{stroke}
}}
\makeatother

 
\tikzset{
pattern size/.store in=\mcSize, 
pattern size = 5pt,
pattern thickness/.store in=\mcThickness, 
pattern thickness = 0.3pt,
pattern radius/.store in=\mcRadius, 
pattern radius = 1pt}
\makeatletter
\pgfutil@ifundefined{pgf@pattern@name@_zr0kwm7gl}{
\pgfdeclarepatternformonly[\mcThickness,\mcSize]{_zr0kwm7gl}
{\pgfqpoint{0pt}{0pt}}
{\pgfpoint{\mcSize+\mcThickness}{\mcSize+\mcThickness}}
{\pgfpoint{\mcSize}{\mcSize}}
{
\pgfsetcolor{\tikz@pattern@color}
\pgfsetlinewidth{\mcThickness}
\pgfpathmoveto{\pgfqpoint{0pt}{0pt}}
\pgfpathlineto{\pgfpoint{\mcSize+\mcThickness}{\mcSize+\mcThickness}}
\pgfusepath{stroke}
}}
\makeatother

 
\tikzset{
pattern size/.store in=\mcSize, 
pattern size = 5pt,
pattern thickness/.store in=\mcThickness, 
pattern thickness = 0.3pt,
pattern radius/.store in=\mcRadius, 
pattern radius = 1pt}
\makeatletter
\pgfutil@ifundefined{pgf@pattern@name@_etqi5qa6b}{
\pgfdeclarepatternformonly[\mcThickness,\mcSize]{_etqi5qa6b}
{\pgfqpoint{0pt}{0pt}}
{\pgfpoint{\mcSize+\mcThickness}{\mcSize+\mcThickness}}
{\pgfpoint{\mcSize}{\mcSize}}
{
\pgfsetcolor{\tikz@pattern@color}
\pgfsetlinewidth{\mcThickness}
\pgfpathmoveto{\pgfqpoint{0pt}{0pt}}
\pgfpathlineto{\pgfpoint{\mcSize+\mcThickness}{\mcSize+\mcThickness}}
\pgfusepath{stroke}
}}
\makeatother
\tikzset{every picture/.style={line width=0.75pt}} 

\begin{tikzpicture}[x=0.75pt,y=0.75pt,yscale=-1,xscale=1]

\draw    (199.33,919.48) -- (232.89,879.46) ;
\draw [shift={(216.11,899.47)}, rotate = 129.98] [fill={rgb, 255:red, 0; green, 0; blue, 0 }  ][line width=0.08]  [draw opacity=0] (5.36,-2.57) -- (0,0) -- (5.36,2.57) -- cycle    ;
\draw    (199.33,919.48) .. controls (197.68,921.16) and (196.02,921.18) .. (194.33,919.53) .. controls (192.65,917.88) and (190.98,917.9) .. (189.33,919.58) .. controls (187.68,921.26) and (186.01,921.28) .. (184.33,919.63) .. controls (182.65,917.98) and (180.98,918) .. (179.33,919.68) .. controls (177.68,921.36) and (176.01,921.38) .. (174.33,919.73) .. controls (172.65,918.08) and (170.98,918.09) .. (169.33,919.77) .. controls (167.68,921.45) and (166.01,921.47) .. (164.33,919.82) .. controls (162.65,918.17) and (160.99,918.19) .. (159.34,919.87) .. controls (157.69,921.55) and (156.02,921.57) .. (154.34,919.92) .. controls (152.66,918.27) and (150.99,918.29) .. (149.34,919.97) .. controls (147.69,921.65) and (146.02,921.67) .. (144.34,920.02) .. controls (142.66,918.37) and (140.99,918.39) .. (139.34,920.07) .. controls (137.69,921.75) and (136.02,921.77) .. (134.34,920.12) -- (131.67,920.15) -- (131.67,920.15) ;
\draw    (145.67,927.73) -- (166.17,927.73) ;
\draw [shift={(169.17,927.73)}, rotate = 180] [fill={rgb, 255:red, 0; green, 0; blue, 0 }  ][line width=0.08]  [draw opacity=0] (5.36,-2.57) -- (0,0) -- (5.36,2.57) -- cycle    ;
\draw  [pattern=_oy50tdxpz,pattern size=3.2249999999999996pt,pattern thickness=0.75pt,pattern radius=0pt, pattern color={rgb, 255:red, 0; green, 0; blue, 0}] (111.83,920.8) .. controls (111.83,915.32) and (116.27,910.88) .. (121.75,910.88) .. controls (127.23,910.88) and (131.67,915.32) .. (131.67,920.8) .. controls (131.67,926.27) and (127.23,930.71) .. (121.75,930.71) .. controls (116.27,930.71) and (111.83,926.27) .. (111.83,920.8) -- cycle ;
\draw    (231.67,1037.99) -- (199.33,996.48) ;
\draw [shift={(215.5,1017.23)}, rotate = 52.08] [fill={rgb, 255:red, 0; green, 0; blue, 0 }  ][line width=0.08]  [draw opacity=0] (5.36,-2.57) -- (0,0) -- (5.36,2.57) -- cycle    ;
\draw    (199.33,996.48) -- (199.33,919.48) ;
\draw [shift={(199.33,957.98)}, rotate = 90] [fill={rgb, 255:red, 0; green, 0; blue, 0 }  ][line width=0.08]  [draw opacity=0] (5.36,-2.57) -- (0,0) -- (5.36,2.57) -- cycle    ;
\draw    (199.33,996.48) .. controls (197.68,998.16) and (196.02,998.18) .. (194.33,996.53) .. controls (192.65,994.88) and (190.98,994.9) .. (189.33,996.58) .. controls (187.68,998.26) and (186.01,998.28) .. (184.33,996.63) .. controls (182.65,994.98) and (180.98,995) .. (179.33,996.68) .. controls (177.68,998.36) and (176.01,998.38) .. (174.33,996.73) .. controls (172.65,995.08) and (170.98,995.09) .. (169.33,996.77) .. controls (167.68,998.45) and (166.01,998.47) .. (164.33,996.82) .. controls (162.65,995.17) and (160.99,995.19) .. (159.34,996.87) .. controls (157.69,998.55) and (156.02,998.57) .. (154.34,996.92) .. controls (152.66,995.27) and (150.99,995.29) .. (149.34,996.97) .. controls (147.69,998.65) and (146.02,998.67) .. (144.34,997.02) .. controls (142.66,995.37) and (140.99,995.39) .. (139.34,997.07) .. controls (137.69,998.75) and (136.02,998.77) .. (134.34,997.12) -- (131.67,997.15) -- (131.67,997.15) ;
\draw    (150.67,1005.23) -- (171.17,1005.23) ;
\draw [shift={(174.17,1005.23)}, rotate = 180] [fill={rgb, 255:red, 0; green, 0; blue, 0 }  ][line width=0.08]  [draw opacity=0] (5.36,-2.57) -- (0,0) -- (5.36,2.57) -- cycle    ;
\draw  [pattern=_hzqwiqu7h,pattern size=3.2249999999999996pt,pattern thickness=0.75pt,pattern radius=0pt, pattern color={rgb, 255:red, 0; green, 0; blue, 0}] (111.83,997.15) .. controls (111.83,991.67) and (116.27,987.23) .. (121.75,987.23) .. controls (127.23,987.23) and (131.67,991.67) .. (131.67,997.15) .. controls (131.67,1002.62) and (127.23,1007.06) .. (121.75,1007.06) .. controls (116.27,1007.06) and (111.83,1002.62) .. (111.83,997.15) -- cycle ;
\draw [line width=1.5]    (121.56,875.46) -- (121.75,910.88) ;
\draw [line width=1.5]    (121.75,930.71) -- (121.75,987.23) ;
\draw [line width=1.5]    (121.75,1007.06) -- (121.56,1036.79) ;
\draw [color={rgb, 255:red, 208; green, 2; blue, 27 }  ,draw opacity=1 ] [dash pattern={on 4.5pt off 4.5pt}]  (370.11,951.33) -- (405.44,951.33) ;
\draw    (385.33,919.67) -- (418.89,879.65) ;
\draw [shift={(402.11,899.66)}, rotate = 129.98] [fill={rgb, 255:red, 0; green, 0; blue, 0 }  ][line width=0.08]  [draw opacity=0] (5.36,-2.57) -- (0,0) -- (5.36,2.57) -- cycle    ;
\draw    (385.33,919.67) .. controls (383.68,921.35) and (382.02,921.37) .. (380.33,919.72) .. controls (378.65,918.07) and (376.98,918.09) .. (375.33,919.77) .. controls (373.68,921.45) and (372.01,921.46) .. (370.33,919.81) .. controls (368.65,918.16) and (366.98,918.18) .. (365.33,919.86) .. controls (363.68,921.54) and (362.01,921.56) .. (360.33,919.91) .. controls (358.65,918.26) and (356.98,918.28) .. (355.33,919.96) .. controls (353.68,921.64) and (352.01,921.66) .. (350.33,920.01) .. controls (348.65,918.36) and (346.99,918.38) .. (345.34,920.06) .. controls (343.69,921.74) and (342.02,921.76) .. (340.34,920.11) .. controls (338.66,918.46) and (336.99,918.48) .. (335.34,920.16) .. controls (333.69,921.84) and (332.02,921.86) .. (330.34,920.21) .. controls (328.66,918.56) and (326.99,918.58) .. (325.34,920.26) .. controls (323.69,921.94) and (322.02,921.96) .. (320.34,920.31) -- (317.67,920.33) -- (317.67,920.33) ;
\draw    (331.67,927.92) -- (352.17,927.92) ;
\draw [shift={(355.17,927.92)}, rotate = 180] [fill={rgb, 255:red, 0; green, 0; blue, 0 }  ][line width=0.08]  [draw opacity=0] (5.36,-2.57) -- (0,0) -- (5.36,2.57) -- cycle    ;
\draw  [pattern=_zr0kwm7gl,pattern size=3.2249999999999996pt,pattern thickness=0.75pt,pattern radius=0pt, pattern color={rgb, 255:red, 0; green, 0; blue, 0}] (297.83,920.98) .. controls (297.83,915.51) and (302.27,911.07) .. (307.75,911.07) .. controls (313.23,911.07) and (317.67,915.51) .. (317.67,920.98) .. controls (317.67,926.46) and (313.23,930.9) .. (307.75,930.9) .. controls (302.27,930.9) and (297.83,926.46) .. (297.83,920.98) -- cycle ;
\draw    (417.67,1038.18) -- (385.33,996.67) ;
\draw [shift={(401.5,1017.42)}, rotate = 52.08] [fill={rgb, 255:red, 0; green, 0; blue, 0 }  ][line width=0.08]  [draw opacity=0] (5.36,-2.57) -- (0,0) -- (5.36,2.57) -- cycle    ;
\draw    (385.33,996.67) -- (385.33,919.67) ;
\draw [shift={(385.33,958.17)}, rotate = 90] [fill={rgb, 255:red, 0; green, 0; blue, 0 }  ][line width=0.08]  [draw opacity=0] (5.36,-2.57) -- (0,0) -- (5.36,2.57) -- cycle    ;
\draw    (385.33,996.67) .. controls (383.68,998.35) and (382.02,998.37) .. (380.33,996.72) .. controls (378.65,995.07) and (376.98,995.09) .. (375.33,996.77) .. controls (373.68,998.45) and (372.01,998.46) .. (370.33,996.81) .. controls (368.65,995.16) and (366.98,995.18) .. (365.33,996.86) .. controls (363.68,998.54) and (362.01,998.56) .. (360.33,996.91) .. controls (358.65,995.26) and (356.98,995.28) .. (355.33,996.96) .. controls (353.68,998.64) and (352.01,998.66) .. (350.33,997.01) .. controls (348.65,995.36) and (346.99,995.38) .. (345.34,997.06) .. controls (343.69,998.74) and (342.02,998.76) .. (340.34,997.11) .. controls (338.66,995.46) and (336.99,995.48) .. (335.34,997.16) .. controls (333.69,998.84) and (332.02,998.86) .. (330.34,997.21) .. controls (328.66,995.56) and (326.99,995.58) .. (325.34,997.26) .. controls (323.69,998.94) and (322.02,998.96) .. (320.34,997.31) -- (317.67,997.33) -- (317.67,997.33) ;
\draw    (336.67,1005.42) -- (357.17,1005.42) ;
\draw [shift={(360.17,1005.42)}, rotate = 180] [fill={rgb, 255:red, 0; green, 0; blue, 0 }  ][line width=0.08]  [draw opacity=0] (5.36,-2.57) -- (0,0) -- (5.36,2.57) -- cycle    ;
\draw  [pattern=_etqi5qa6b,pattern size=3.2249999999999996pt,pattern thickness=0.75pt,pattern radius=0pt, pattern color={rgb, 255:red, 0; green, 0; blue, 0}] (297.83,997.33) .. controls (297.83,991.86) and (302.27,987.42) .. (307.75,987.42) .. controls (313.23,987.42) and (317.67,991.86) .. (317.67,997.33) .. controls (317.67,1002.81) and (313.23,1007.25) .. (307.75,1007.25) .. controls (302.27,1007.25) and (297.83,1002.81) .. (297.83,997.33) -- cycle ;
\draw [line width=1.5]    (307.56,875.65) -- (307.75,911.07) ;
\draw [line width=1.5]    (307.75,930.9) -- (307.75,987.42) ;
\draw [line width=1.5]    (307.75,1007.25) -- (307.56,1036.98) ;

\draw (154,1007.81) node [anchor=north west][inner sep=0.75pt]    {$\ell$};
\draw (140,930.31) node [anchor=north west][inner sep=0.75pt]    {$q-\ell$};
\draw (224.5,1002.81) node [anchor=north west][inner sep=0.75pt]    {$p$};
\draw (205,946.81) node [anchor=north west][inner sep=0.75pt]    {$p+\ell$};
\draw (224,892.81) node [anchor=north west][inner sep=0.75pt]    {$p+q$};
\draw (340,1008) node [anchor=north west][inner sep=0.75pt]    {$\ell$};
\draw (326,930.5) node [anchor=north west][inner sep=0.75pt]    {$q-\ell$};
\draw (410.5,1003) node [anchor=north west][inner sep=0.75pt]    {$p$};
\draw (391,964) node [anchor=north west][inner sep=0.75pt]    {$p+\ell$};
\draw (410,893) node [anchor=north west][inner sep=0.75pt]    {$p+q$};

\end{tikzpicture}

\end{center}
\end{figure}

The box contribution reads   
\begin{equation}
iq^\mu \mathcal{A}_\square(q)=-i e^2Q^2 q^\mu \int \hat{d^4}{\ell} \,\hat{\delta}({\ell}\cdot V)\frac{\tilde{A}(l)\cdot (2p+{\ell}) \tilde{A}(q-\ell)\cdot (2p+q+\ell)}{2p\cdot \ell +\ell^2+ i \varepsilon}
\end{equation}
which, introducing wave vectors, becomes
\begin{equation}\label{boxgen}
i\hbar^3\, \bq ^\mu \mathcal{A}_\square (\bar{q})=\frac{-i e^2Q^2 \bq^\mu }{2\hbar} \int \hat{d^4}\bl \,\hat{\delta}( \bar{\ell}\cdot V)\frac{\tilde{A}(\bl)\cdot (2p+\hbar\bl) \tilde{A}(\bar{q}-\bar{\ell})\cdot \left(2p+\hbar(\bq+\bl)\right)}{p\cdot \bl +\hbar\,{\bl^2}/{2}+ i \varepsilon}.
\end{equation}
As expected at this order, the RHS above has a superclassical part which behaves as $\sim\frac{1}{\hbar}$ which  will need to cancel if we are to have a well defined physical observable. Let us write  this divergence
\begin{equation}
\begin{split}
i\hbar^3\, \bq ^\mu \mathcal{A}_\square (\bar{q})&=\frac{-2i e^2Q^2 \bq^\mu }{\hbar} \int \hat{d^4}\bl \,\hat{\delta}( \bar{\ell}\cdot V)\frac{\tilde{A}(\bl)\cdot p\, \tilde{A}(\bar{q}-\bar{\ell})\cdot p}{p\cdot \bl  +i \varepsilon}+\mathcal{O}(\hbar^0)\\&=
\frac{-e^2Q^2 \bq^\mu }{\hbar} \int \hat{d^4}\bl \,\hat{\delta}( \bar{\ell}\cdot V) \hat{\delta}( \bar{l}\cdot p) {\tilde{A}(\bl)\cdot p\, \tilde{A}(\bar{q}-\bar{\ell})\cdot p} +\mathcal{O}(\hbar^0)
\end{split}
\end{equation}
where we have averaged over equivalent expressions to use the delta-function trick \eqref{deltatrick}. To show that this singularity is harmless we  will need to add it to the cut box one and show that they cancel. Thus, let's focus onto the cut box amplitude, this is\footnote{Mind the slightly different use of letters, with respect to previous chapters, used to identify various superclassical/classical contributions.}
\begin{equation}
\mathcal{C} ^\mu (q)= e^2Q^2  \int \hat{d^4}\ell \,\hat{\delta}(\ell\cdot V)\hat{\delta}(2p\cdot \ell+ \ell^2)\ell^\mu  {\tilde{A}(\ell)\cdot (2p+\ell) \tilde{A}(q-\ell)\cdot (2p+q+\ell)}
\end{equation}
or, written with $\hbar$s and wave vectors
\begin{equation}\label{cboxgen}
\begin{split}
\hbar^3\mathcal{C} ^\mu (\bq)&= \frac{e^2Q^2}{\hbar}  \int \hat{d^4}\bl \,\hat{\delta}(\bl\cdot V)\hat{\delta}(2p\cdot \bl+ \bl^2)\bl^\mu  {\tilde{A}(\bl)\cdot (2p+\hbar \bl) \tilde{A}(\bq-\bl)\cdot \left(2p+\hbar(\bq+\bl))\right)}\\&= 
\frac{2e^2Q^2}{\hbar}  \int \hat{d^4}\bl \,\hat{\delta}(\bl\cdot V)\hat{\delta}(p\cdot \bl)\bl^\mu  {\tilde{A}(\bl)\cdot p  \,\tilde{A}(\bq-\bl)\cdot p}+\mathcal{O}(\hbar^0)
\end{split}
\end{equation}
but   $\bl^\mu =\frac{\bq^\mu}{2}$ inside the superclassical loop integral, so the two singularities cancel each other as they equal with opposite signs. 

The next step is to now gather all classically relevant terms from \eqref{boxgen} and \eqref{cboxgen}. Let's start from the box amplitude. The classical contributions can be obtained in 3 different ways (which are the different
possibilities of picking an additional power of $\hbar$ at the numerator): by taking a power of null momenta from the
numerator, by Laurent expanding the propagator and by doing the $	\bl \to \bq-\bl$ shift to take care of the condition $p\cdot \bq =-\hbar \bq^2/2$.  These terms are, respectively for \eqref{boxgen}
\begin{equation}\label{wee}
\begin{split}
i\hbar^3\, \bq ^\mu \mathcal{A}_\square (\bar{q})=&\,{-i e^2Q^2 \bq^\mu } \int \hat{d^4}\bl \,\hat{\delta}( \bar{\ell}\cdot V)\frac{\tilde{A}(\bl)\cdot p \tilde{A}(\bar{q}-\bar{\ell})\cdot \left(\bq+\bl\right)+ \tilde{A}(\bl)\cdot \bl \tilde{A}(\bar{q}-\bar{\ell})\cdot p}{p\cdot \bl + i \varepsilon}\\&+
 {i e^2Q^2 \bq^\mu } \int \hat{d^4}\bl \,   \hat{\delta}( \bar{\ell}\cdot V)\frac{\tilde{A}(\bl)\cdot p\, \tilde{A}(\bar{q}-\bar{\ell})\cdot p}{(p\cdot \bl  +i \varepsilon)^2} {\bl^2} 
\\&-
\frac{i e^2Q^2 \bq^\mu }{2} \int \hat{d^4}\bl \,   \hat{\delta}( \bar{\ell}\cdot V)\frac{\tilde{A}(\bl)\cdot p\, \tilde{A}(\bar{q}-\bar{\ell})\cdot p}{(p\cdot \bl-i \varepsilon)^2} {\bq^2}+\mathcal{O}(\hbar)
\\&\equiv \mathcal{A}^\mu_{  a} (\bar{q})+\mathcal{A}^\mu_{  b} (\bar{q})+\mathcal{A}^\mu_{  c} (\bar{q})+\mathcal{O}(\hbar)
\end{split}
\end{equation}
having defined each line as the 3 different contributions,  in the last equality.
On the other hand, for \eqref{cboxgen} we get 
\begin{equation}\label{nee}
\begin{split}
\hbar^3\mathcal{C} ^\mu (\bq)=&\,{ e^2Q^2   } \int \hat{d^4}\bl \,\bl^\mu \hat{\delta}( \bar{\ell}\cdot V) \hat{\delta}( \bar{\ell}\cdot p) \left( \tilde{A}(\bl)\cdot p \tilde{A}(\bar{q}-\bar{\ell})\cdot \left(\bq+\bl\right)+ \tilde{A}(\bl)\cdot \bl \tilde{A}(\bar{q}-\bar{\ell})\cdot p\right)\\&+ { e^2Q^2   } \int \hat{d^4}\bl \, {\bl^2}  \bl^\mu \hat{\delta}( \bar{\ell}\cdot V)   \hat{\delta}'( \bar{\ell}\cdot p) {\tilde{A}(\bl)\cdot p\, \tilde{A}(\bar{q}-\bar{\ell})\cdot p}\, 
\\&+
\frac{ e^2Q^2  }{2} \int \hat{d^4}\bl \,{\bq^2}(\bq-\bl)^\mu   \hat{\delta}( \bar{\ell}\cdot V)   \hat{\delta}'( \bar{\ell}\cdot p) {\tilde{A}(\bl)\cdot p\, \tilde{A}(\bar{q}-\bar{\ell})\cdot p}+\mathcal{O}(\hbar)
\\&\equiv \mathcal{C}^\mu_{  a} (\bar{q})+\mathcal{C}^\mu_{  b} (\bar{q})+\mathcal{C}^\mu_{  c} (\bar{q})+\mathcal{O}(\hbar).
\end{split}
\end{equation}
The first lines of \eqref{wee} and \eqref{nee} look similar so let's add them, we find that these two bits can be brought into the following form
\begin{equation}\label{ez1}
\mathcal{A}^\mu_{  a}+\mathcal{C}^\mu_{  a}= \,{-i2 e^2Q^2  } \int \hat{d^4}\bl \,\hat{\delta}( \bar{\ell}\cdot V){ \tilde{A}(\bl)\cdot p \,\tilde{A}(\bar{q}-\bar{\ell})\cdot \bl }\left(\frac{\bq^\mu}{p\cdot \bl + i\varepsilon}+i\bl^\mu \hat{\delta}(p\cdot \bl)
\right).
\end{equation}

Let's now move onto the remaining 4 terms. We find that $\mathcal{C}^\mu_{  b}+\mathcal{C}^\mu_{  c}$, when added together after averaging over equivalent forms, becomes
\begin{equation}\label{ez2}
\begin{split}
\mathcal{C}^\mu_{  b}+\mathcal{C}^\mu_{  c}=&\,{e^2Q^2   } \int \hat{d^4}\bl \,   \bl^\mu \hat{\delta}( \bar{\ell}\cdot V)   \hat{\delta}'( \bar{\ell}\cdot p) {\tilde{A}(\bl)\cdot p\, \tilde{A}(\bar{q}-\bar{\ell})\cdot p}\, \bl\cdot (\bl-\bq)\, 
\\&-
\frac{ e^2Q^2 \bq^\mu  }{2} \int \hat{d^4}\bl \,  \hat{\delta}( \bar{\ell}\cdot V)   \hat{\delta}'( \bar{\ell}\cdot p) {\tilde{A}(\bl)\cdot p\, \tilde{A}(\bar{q}-\bar{\ell})\cdot p} \,\bl \cdot (\bl-2 \bq).
\end{split}
\end{equation}
If we define the second integral on the RHS above to be $Z^\mu $, that is 
\begin{equation}
Z^\mu= -
\frac{ e^2Q^2 \bq^\mu  }{2} \int \hat{d^4}\bl \,  \hat{\delta}( \bar{\ell}\cdot V)   \hat{\delta}'( \bar{\ell}\cdot p) {\tilde{A}(\bl)\cdot p\, \tilde{A}(\bar{q}-\bar{\ell})\cdot p} \,\bl \cdot (\bl-2 \bq)
\end{equation}
a small calculation reveals that adding $Z^\mu $ to the remaining terms yields
\begin{equation}\label{ez3}
Z^\mu + \mathcal{A}^\mu_{  b}+\mathcal{A}^\mu_{  c} =
 {i e^2Q^2 \bq^\mu } \int \hat{d^4}\bl \,   \hat{\delta}( \bar{\ell}\cdot V)\frac{\tilde{A}(\bl)\cdot p\, \tilde{A}(\bar{q}-\bar{\ell})\cdot p}{(p\cdot \bl  +i \varepsilon)^2} \,\bl\cdot (\bl-\bq).
\end{equation}

We now have all the ingredients needed for the 1-loop deflection. Inserting  \eqref{ez1}, \eqref{ez2} and \eqref{ez3} into \eqref{i1i2} and taking the momenta $p$ to be close to their classical value we finally obtain   
\begin{equation}\label{deltaaaa}
\begin{split}
\Delta p^\mu_{\text{NLO}}=\frac{ie^2Q^2}{2m}\int \hat{{d}}^4 \bq\,& \hat{\delta}(u\cdot \bq)\hat{\delta}(V\cdot \bq)e^{-ib\cdot \bq}\int \hat{d}^4\bl\, \hat{\delta}(V\cdot \bl)\\&\times \left[
\bq^\mu \left(
{ -2}\frac{\tilde{A}(\bl)\cdot u \,\tilde{A}(\bar{q}-\bar{\ell})\cdot \bl }{u\cdot \bl + i\varepsilon}+\frac{\tilde{A}(\bl)\cdot u\, \tilde{A}(\bar{q}-\bar{\ell})\cdot u}{(u\cdot \bl  +i \varepsilon)^2} \,\bl\cdot (\bl-\bq)
\right)\right.
\\&-\left.
i\bl^\mu \left(2 \hat{\delta} (u\cdot l) {\tilde{A}(\bl)\cdot u \,\tilde{A}(\bar{q}-\bar{\ell})\cdot \bl }+
     \hat{\delta}'( \bar{\ell}\cdot u)\,  \bl\cdot (\bl-\bq) {\tilde{A}(\bl)\cdot u\, \tilde{A}(\bar{q}-\bar{\ell})\cdot u}
\right)\frac{}{}\right].
\end{split}
\end{equation}
Also note that the on-shellness of the probe is respected, as 
\begin{equation}
2p\cdot \Delta p_{\text{NLO}}= {e^2Q^2} \int \hat{{d}}^4 \bq\, \hat{\delta}(u\cdot \bq)\hat{\delta}(V\cdot \bq)e^{-ib\cdot \bq}\int \hat{d}^4\bl\, \hat{\delta}(V\cdot \bl) 
 \bl\cdot u    \hat{\delta}'( \bar{\ell}\cdot u)\,  \bl\cdot (\bl-\bq) {\tilde{A}(\bl)\cdot u\, \tilde{A}(\bar{q}-\bar{\ell})\cdot u}
\end{equation}
 becomes, using the property of distributions $x\hat{\delta}'(x)=-\hat{\delta}(x)$ 
\begin{equation}\label{oskk}
\begin{split}
2p\cdot \Delta p_{\text{NLO}}&=- {e^2Q^2} \int \hat{{d}}^4 \bq\, \hat{\delta}(u\cdot \bq)\hat{\delta}(V\cdot \bq)e^{-ib\cdot \bq}\int \hat{d}^4\bl\, \hat{\delta}(V\cdot \bl) 
  \hat{\delta}( \bar{\ell}\cdot u)\,  \bl\cdot (\bl-\bq) {\tilde{A}(\bl)\cdot u\, \tilde{A}(\bar{q}-\bar{\ell})\cdot u}\\&=
 {e^2Q^2} \int \hat{{d}}^4\bq'\, \hat{\delta}(u\cdot \bq')\hat{\delta}(V\cdot \bq')e^{-ib\cdot \bq'}\bq' _\rho\tilde{A}(\bq')\cdot u \int \hat{d}^4\bl\, \hat{\delta}(V\cdot \bl) 
\hat{\delta}( \bar{\ell}\cdot u)\,  e^{-ib\cdot \bl}  \bl^\rho {\tilde{A}(\bl)\cdot u\, }  \\&=-\Delta p_{\text{LO}}\cdot \Delta p_{\text{LO}}
\end{split}
\end{equation}
where the last expression can be achieved by shifting the variable $\bq$ to $\bq'=\bq+\bl$.

Before carrying out some numerical checks we move to the gravitational computation, which turns out to be extremely similar to the electromagnetic one.

\subsection{The probe limit for gravity}

The action of a scalar body of mass $m$, minimally coupled to a gravitational  heavy background of mass $M$ is 

\begin{equation}
{S}= \frac{1}{2}\int d^4x\, \sqrt{-g}\left(
g^{\mu\nu}(\partial_{\mu} \Phi) (\partial_{\nu} \Phi)-m^2 \Phi^2
\right)
\end{equation}
again, we start by employing the KS gauge. In this case the inverse metric reads, in its exact form 
\begin{equation}
g^{\mu\nu}=\eta^{\mu\nu}-\kappa h^{\mu\nu}
\end{equation}
and given in these coordinates $\sqrt{-g}=1$ the Lagrangian simplifies drastically to
\begin{equation}
\mathcal{L}=\mathcal{L}_0-\frac{\kappa}{2} h^{\mu\nu} (\partial_{\mu} \Phi) (\partial_{\nu} \Phi)
\end{equation}
i.e. the interaction is strictly linear, with no further corrections.  The corresponding vertex rule  is 

\begin{figure}[H]
\begin{center}

 
\tikzset{
pattern size/.store in=\mcSize, 
pattern size = 5pt,
pattern thickness/.store in=\mcThickness, 
pattern thickness = 0.3pt,
pattern radius/.store in=\mcRadius, 
pattern radius = 1pt}
\makeatletter
\pgfutil@ifundefined{pgf@pattern@name@_8j4fb5qhu}{
\pgfdeclarepatternformonly[\mcThickness,\mcSize]{_8j4fb5qhu}
{\pgfqpoint{0pt}{0pt}}
{\pgfpoint{\mcSize+\mcThickness}{\mcSize+\mcThickness}}
{\pgfpoint{\mcSize}{\mcSize}}
{
\pgfsetcolor{\tikz@pattern@color}
\pgfsetlinewidth{\mcThickness}
\pgfpathmoveto{\pgfqpoint{0pt}{0pt}}
\pgfpathlineto{\pgfpoint{\mcSize+\mcThickness}{\mcSize+\mcThickness}}
\pgfusepath{stroke}
}}
\makeatother
\tikzset{every picture/.style={line width=0.75pt}} 

\begin{tikzpicture}[x=0.75pt,y=0.75pt,yscale=-1,xscale=1]

\draw    (143.67,835.76) -- (99.33,794.32) ;
\draw [shift={(121.5,815.04)}, rotate = 43.07] [fill={rgb, 255:red, 0; green, 0; blue, 0 }  ][line width=0.08]  [draw opacity=0] (5.36,-2.57) -- (0,0) -- (5.36,2.57) -- cycle    ;
\draw    (99.33,794.32) -- (148.67,754.99) ;
\draw [shift={(124,774.66)}, rotate = 141.43] [fill={rgb, 255:red, 0; green, 0; blue, 0 }  ][line width=0.08]  [draw opacity=0] (5.36,-2.57) -- (0,0) -- (5.36,2.57) -- cycle    ;
\draw    (99.35,795.82) .. controls (97.7,797.51) and (96.03,797.52) .. (94.35,795.87) .. controls (92.67,794.22) and (91,794.23) .. (89.35,795.91) .. controls (87.7,797.59) and (86.03,797.61) .. (84.35,795.96) .. controls (82.67,794.31) and (81,794.32) .. (79.35,796) .. controls (77.7,797.68) and (76.03,797.7) .. (74.35,796.05) .. controls (72.67,794.4) and (71,794.41) .. (69.35,796.09) .. controls (67.7,797.77) and (66.03,797.79) .. (64.35,796.14) .. controls (62.67,794.49) and (61,794.5) .. (59.35,796.18) .. controls (57.7,797.86) and (56.03,797.88) .. (54.35,796.23) .. controls (52.67,794.58) and (51,794.59) .. (49.35,796.27) .. controls (47.7,797.95) and (46.03,797.97) .. (44.35,796.32) -- (42.68,796.33) -- (42.68,796.33)(99.32,792.82) .. controls (97.67,794.51) and (96,794.52) .. (94.32,792.87) .. controls (92.64,791.22) and (90.97,791.23) .. (89.32,792.91) .. controls (87.67,794.59) and (86,794.61) .. (84.32,792.96) .. controls (82.64,791.31) and (80.97,791.32) .. (79.32,793) .. controls (77.67,794.68) and (76,794.7) .. (74.32,793.05) .. controls (72.64,791.4) and (70.97,791.41) .. (69.32,793.09) .. controls (67.67,794.77) and (66,794.79) .. (64.32,793.14) .. controls (62.64,791.49) and (60.97,791.5) .. (59.32,793.18) .. controls (57.67,794.86) and (56,794.88) .. (54.32,793.23) .. controls (52.64,791.58) and (50.97,791.59) .. (49.32,793.27) .. controls (47.67,794.95) and (46,794.97) .. (44.32,793.32) -- (42.65,793.33) -- (42.65,793.33) ;
\draw    (58.67,809.07) -- (79.17,809.07) ;
\draw [shift={(82.17,809.07)}, rotate = 180] [fill={rgb, 255:red, 0; green, 0; blue, 0 }  ][line width=0.08]  [draw opacity=0] (5.36,-2.57) -- (0,0) -- (5.36,2.57) -- cycle    ;
\draw  [pattern=_8j4fb5qhu,pattern size=3.2249999999999996pt,pattern thickness=0.75pt,pattern radius=0pt, pattern color={rgb, 255:red, 0; green, 0; blue, 0}] (22.83,794.83) .. controls (22.83,789.36) and (27.27,784.92) .. (32.75,784.92) .. controls (38.23,784.92) and (42.67,789.36) .. (42.67,794.83) .. controls (42.67,800.31) and (38.23,804.75) .. (32.75,804.75) .. controls (27.27,804.75) and (22.83,800.31) .. (22.83,794.83) -- cycle ;
\draw [line width=1.5]    (32.67,753.76) -- (32.75,784.92) ;
\draw [line width=1.5]    (32.75,804.75) -- (33.17,838.76) ;

\draw (141,759.66) node [anchor=north west][inner sep=0.75pt]    {$p+q$};
\draw (136,815.66) node [anchor=north west][inner sep=0.75pt]    {$p$};
\draw (63.5,811.66) node [anchor=north west][inner sep=0.75pt]    {$q$};
\draw (194.5,780.51) node [anchor=north west][inner sep=0.75pt]    {$=\displaystyle{{i\kappa }\,\tilde{h}^{\mu\nu}(q)\,p_{\mu} (p+q)_{\nu}.}$};

\end{tikzpicture}

\end{center}

\end{figure}

Armed with this knowledge it is then possible to write, following the usual procedure, the classical leading order deflection. This is immediately found to be
\begin{equation}\label{oookkkk}
\Delta p^\mu_{\text{LO}}=\frac{i\kappa^2 m}{2} \int \hat{{d}}^4 \bq\, \hat{\delta}(u\cdot \bq)\hat{\delta}(V\cdot \bq)e^{-ib\cdot \bq} \, \bq^\mu \,\tilde{h}^{\alpha\beta} (\bq) u_\alpha u_\beta
\end{equation}
where, as before, we have factored  out both the source delta function and coupling $ \tilde{h}\to \kappa \hat{\delta}(q\cdot V) \tilde{h}$. 

\subsubsection{1-loop gravitational deflection}\label{mosoccazzi}

At the next order in $G_N$ we find again just one diagram to be relevant for the $I_1^\mu $ piece of  \eqref{i1i2}. This is the box diagram 
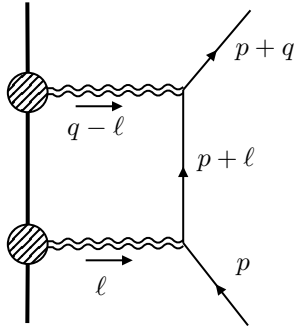
\begin{figure}[H]
\begin{center} 

 
\tikzset{
pattern size/.store in=\mcSize, 
pattern size = 5pt,
pattern thickness/.store in=\mcThickness, 
pattern thickness = 0.3pt,
pattern radius/.store in=\mcRadius, 
pattern radius = 1pt}
\makeatletter
\pgfutil@ifundefined{pgf@pattern@name@_6z7ltdjx8}{
\pgfdeclarepatternformonly[\mcThickness,\mcSize]{_6z7ltdjx8}
{\pgfqpoint{0pt}{0pt}}
{\pgfpoint{\mcSize+\mcThickness}{\mcSize+\mcThickness}}
{\pgfpoint{\mcSize}{\mcSize}}
{
\pgfsetcolor{\tikz@pattern@color}
\pgfsetlinewidth{\mcThickness}
\pgfpathmoveto{\pgfqpoint{0pt}{0pt}}
\pgfpathlineto{\pgfpoint{\mcSize+\mcThickness}{\mcSize+\mcThickness}}
\pgfusepath{stroke}
}}
\makeatother

 
\tikzset{
pattern size/.store in=\mcSize, 
pattern size = 5pt,
pattern thickness/.store in=\mcThickness, 
pattern thickness = 0.3pt,
pattern radius/.store in=\mcRadius, 
pattern radius = 1pt}
\makeatletter
\pgfutil@ifundefined{pgf@pattern@name@_v23xdhdlh}{
\pgfdeclarepatternformonly[\mcThickness,\mcSize]{_v23xdhdlh}
{\pgfqpoint{0pt}{0pt}}
{\pgfpoint{\mcSize+\mcThickness}{\mcSize+\mcThickness}}
{\pgfpoint{\mcSize}{\mcSize}}
{
\pgfsetcolor{\tikz@pattern@color}
\pgfsetlinewidth{\mcThickness}
\pgfpathmoveto{\pgfqpoint{0pt}{0pt}}
\pgfpathlineto{\pgfpoint{\mcSize+\mcThickness}{\mcSize+\mcThickness}}
\pgfusepath{stroke}
}}
\makeatother
\tikzset{every picture/.style={line width=0.75pt}} 

\begin{tikzpicture}[x=0.75pt,y=0.75pt,yscale=-1,xscale=1]

\draw    (410.33,622.48) -- (443.89,582.46) ;
\draw [shift={(427.11,602.47)}, rotate = 129.98] [fill={rgb, 255:red, 0; green, 0; blue, 0 }  ][line width=0.08]  [draw opacity=0] (5.36,-2.57) -- (0,0) -- (5.36,2.57) -- cycle    ;
\draw    (410.35,623.98) .. controls (408.7,625.66) and (407.03,625.68) .. (405.35,624.03) .. controls (403.67,622.38) and (402,622.4) .. (400.35,624.08) .. controls (398.7,625.76) and (397.03,625.78) .. (395.35,624.13) .. controls (393.67,622.48) and (392,622.5) .. (390.35,624.18) .. controls (388.7,625.86) and (387.03,625.88) .. (385.35,624.23) .. controls (383.67,622.58) and (382,622.59) .. (380.35,624.27) .. controls (378.7,625.95) and (377.03,625.97) .. (375.35,624.32) .. controls (373.67,622.67) and (372,622.69) .. (370.35,624.37) .. controls (368.7,626.05) and (367.03,626.07) .. (365.35,624.42) .. controls (363.67,622.77) and (362,622.79) .. (360.35,624.47) .. controls (358.7,626.15) and (357.03,626.17) .. (355.35,624.52) .. controls (353.67,622.87) and (352,622.89) .. (350.35,624.57) .. controls (348.7,626.25) and (347.03,626.27) .. (345.35,624.62) -- (342.68,624.65) -- (342.68,624.65)(410.32,620.98) .. controls (408.67,622.66) and (407,622.68) .. (405.32,621.03) .. controls (403.64,619.38) and (401.97,619.4) .. (400.32,621.08) .. controls (398.67,622.76) and (397,622.78) .. (395.32,621.13) .. controls (393.64,619.48) and (391.97,619.5) .. (390.32,621.18) .. controls (388.67,622.86) and (387,622.88) .. (385.32,621.23) .. controls (383.64,619.58) and (381.97,619.59) .. (380.32,621.27) .. controls (378.67,622.95) and (377,622.97) .. (375.32,621.32) .. controls (373.64,619.67) and (371.97,619.69) .. (370.32,621.37) .. controls (368.67,623.05) and (367,623.07) .. (365.32,621.42) .. controls (363.64,619.77) and (361.97,619.79) .. (360.32,621.47) .. controls (358.67,623.15) and (357,623.17) .. (355.32,621.52) .. controls (353.64,619.87) and (351.97,619.89) .. (350.32,621.57) .. controls (348.67,623.25) and (347,623.27) .. (345.32,621.62) -- (342.65,621.65) -- (342.65,621.65) ;
\draw    (356.67,630.73) -- (377.17,630.73) ;
\draw [shift={(380.17,630.73)}, rotate = 180] [fill={rgb, 255:red, 0; green, 0; blue, 0 }  ][line width=0.08]  [draw opacity=0] (5.36,-2.57) -- (0,0) -- (5.36,2.57) -- cycle    ;
\draw  [pattern=_6z7ltdjx8,pattern size=3.2249999999999996pt,pattern thickness=0.75pt,pattern radius=0pt, pattern color={rgb, 255:red, 0; green, 0; blue, 0}] (322.83,623.8) .. controls (322.83,618.32) and (327.27,613.88) .. (332.75,613.88) .. controls (338.23,613.88) and (342.67,618.32) .. (342.67,623.8) .. controls (342.67,629.27) and (338.23,633.71) .. (332.75,633.71) .. controls (327.27,633.71) and (322.83,629.27) .. (322.83,623.8) -- cycle ;
\draw    (442.67,740.99) -- (410.33,699.48) ;
\draw [shift={(426.5,720.23)}, rotate = 52.08] [fill={rgb, 255:red, 0; green, 0; blue, 0 }  ][line width=0.08]  [draw opacity=0] (5.36,-2.57) -- (0,0) -- (5.36,2.57) -- cycle    ;
\draw    (410.33,699.48) -- (410.33,622.48) ;
\draw [shift={(410.33,660.98)}, rotate = 90] [fill={rgb, 255:red, 0; green, 0; blue, 0 }  ][line width=0.08]  [draw opacity=0] (5.36,-2.57) -- (0,0) -- (5.36,2.57) -- cycle    ;
\draw    (410.35,700.98) .. controls (408.7,702.66) and (407.03,702.68) .. (405.35,701.03) .. controls (403.67,699.38) and (402,699.4) .. (400.35,701.08) .. controls (398.7,702.76) and (397.03,702.78) .. (395.35,701.13) .. controls (393.67,699.48) and (392,699.5) .. (390.35,701.18) .. controls (388.7,702.86) and (387.03,702.88) .. (385.35,701.23) .. controls (383.67,699.58) and (382,699.59) .. (380.35,701.27) .. controls (378.7,702.95) and (377.03,702.97) .. (375.35,701.32) .. controls (373.67,699.67) and (372,699.69) .. (370.35,701.37) .. controls (368.7,703.05) and (367.03,703.07) .. (365.35,701.42) .. controls (363.67,699.77) and (362,699.79) .. (360.35,701.47) .. controls (358.7,703.15) and (357.03,703.17) .. (355.35,701.52) .. controls (353.67,699.87) and (352,699.89) .. (350.35,701.57) .. controls (348.7,703.25) and (347.03,703.27) .. (345.35,701.62) -- (342.68,701.65) -- (342.68,701.65)(410.32,697.98) .. controls (408.67,699.66) and (407,699.68) .. (405.32,698.03) .. controls (403.64,696.38) and (401.97,696.4) .. (400.32,698.08) .. controls (398.67,699.76) and (397,699.78) .. (395.32,698.13) .. controls (393.64,696.48) and (391.97,696.5) .. (390.32,698.18) .. controls (388.67,699.86) and (387,699.88) .. (385.32,698.23) .. controls (383.64,696.58) and (381.97,696.59) .. (380.32,698.27) .. controls (378.67,699.95) and (377,699.97) .. (375.32,698.32) .. controls (373.64,696.67) and (371.97,696.69) .. (370.32,698.37) .. controls (368.67,700.05) and (367,700.07) .. (365.32,698.42) .. controls (363.64,696.77) and (361.97,696.79) .. (360.32,698.47) .. controls (358.67,700.15) and (357,700.17) .. (355.32,698.52) .. controls (353.64,696.87) and (351.97,696.89) .. (350.32,698.57) .. controls (348.67,700.25) and (347,700.27) .. (345.32,698.62) -- (342.65,698.65) -- (342.65,698.65) ;
\draw    (361.67,708.23) -- (382.17,708.23) ;
\draw [shift={(385.17,708.23)}, rotate = 180] [fill={rgb, 255:red, 0; green, 0; blue, 0 }  ][line width=0.08]  [draw opacity=0] (5.36,-2.57) -- (0,0) -- (5.36,2.57) -- cycle    ;
\draw  [pattern=_v23xdhdlh,pattern size=3.2249999999999996pt,pattern thickness=0.75pt,pattern radius=0pt, pattern color={rgb, 255:red, 0; green, 0; blue, 0}] (322.83,700.15) .. controls (322.83,694.67) and (327.27,690.23) .. (332.75,690.23) .. controls (338.23,690.23) and (342.67,694.67) .. (342.67,700.15) .. controls (342.67,705.62) and (338.23,710.06) .. (332.75,710.06) .. controls (327.27,710.06) and (322.83,705.62) .. (322.83,700.15) -- cycle ;
\draw [line width=1.5]    (332.56,578.46) -- (332.75,613.88) ;
\draw [line width=1.5]    (332.75,633.71) -- (332.75,690.23) ;
\draw [line width=1.5]    (332.75,710.06) -- (332.56,739.79) ;

\draw (365,714.81) node [anchor=north west][inner sep=0.75pt]    {$\ell$};
\draw (351,634.31) node [anchor=north west][inner sep=0.75pt]    {$q-\ell$};
\draw (435.5,705.81) node [anchor=north west][inner sep=0.75pt]    {$p$};
\draw (416,649.81) node [anchor=north west][inner sep=0.75pt]    {$p+\ell$};
\draw (435,595.81) node [anchor=north west][inner sep=0.75pt]    {$p+q$};
\draw (497,640.16) node [anchor=north west][inner sep=0.75pt]    { };

\end{tikzpicture}

\end{center}

\end{figure}

corresponding to the expression 

\begin{equation}
iq^\mu \mathcal{M}(q)=\left({i\kappa}\right)^2\kappa^2 i q^\mu \int \hat{d}^4 \ell\,\hat{\delta}(V\cdot \ell) \frac{\tilde{h}^{\alpha\beta}(\ell)p_{\alpha}(p+\ell)_{\beta} \,\,\tilde{h}^{\rho\sigma}(q-\ell)(p+\ell)_{\rho}(p+q)_{\sigma}}{2p\cdot \ell +\ell^2+i\epsilon}
\end{equation}
while the cut box is found by cutting the massive propagator
\begin{equation}
\mathcal{D}^\mu (q) = \kappa^4  \int \hat{d}^4 \ell\,\hat{\delta}(V\cdot \ell) \hat{\delta}(2p\cdot \ell+\ell^2)\,\ell\ell^\mu \, {\tilde{h}^{\alpha\beta}(\ell)p_{\alpha}(p+\ell)_{\beta} \,\,\tilde{h}^{\rho\sigma}(q-\ell)(p+\ell)_{\rho}(p+q)_{\sigma}}.
\end{equation}
Let us remark again the magic of the KS gauge: we are scattering two black holes at one loop and we have to consider one Feynman diagram!

Before computing the NLO deflection let us carry out a first sanity check, which is to show how the two contributions have a well defined classical limit.
The box singularity is, introducing wave numbers and averaging
\begin{equation}\label{boxsing}
\begin{split}
i \hbar^3 \mathcal{M}^\mu(\bq)&=- \frac{i \kappa^4 \bq^\mu}{2\hbar} \int \hat{d}^4 \bl \, \hat{\delta}(V\cdot \bl)\frac{\tilde{h}^{\alpha\beta}( \bl)p_{\alpha}p_{\beta} \,\,\tilde{h}^{\rho\sigma}(\bq-\bl)p_{\rho}p_{\sigma}}{p\cdot  \bl  +i\epsilon}+\mathcal{O}(\hbar^0) \\&
=- \frac{ \kappa^4 \bq^\mu}{4\hbar} \int \hat{d}^4 \bl\,\hat{\delta}(V\cdot \bl) {\tilde{h}^{\alpha\beta}( \bl)p_{\alpha}p_{\beta} \,\,\tilde{h}^{\rho\sigma}(\bq-\bl)p_{\rho}p_{\sigma}}
\hat{\delta}({p\cdot  \bl  })+\mathcal{O}(\hbar^0)
\end{split}
\end{equation}
while the leading $\hbar$ piece in the cross box reads
\begin{equation}
\begin{split}
 \hbar^3 \mathcal{D}^\mu(\bq)&=\frac{\kappa^4  }{2\hbar} \int \hat{d}^4 \bl \,\hat{\delta}(V\cdot \bl) \bl^\mu {\tilde{h}^{\alpha\beta}( \bl)p_{\alpha}p_{\beta} \,\,\tilde{h}^{\rho\sigma}(\bq-\bl)p_{\rho}p_{\sigma}}\hat{\delta}({p\cdot  \bl })+\mathcal{O}(\hbar^0)\\&
= \frac{ \kappa^4  \bq^\mu }{4\hbar} \int \hat{d}^4 \bl\,\hat{\delta}(V\cdot \bl)\bl^\mu {\tilde{h}^{\alpha\beta}( \bl)p_{\alpha}p_{\beta} \,\,\tilde{h}^{\rho\sigma}(\bq-\bl)p_{\rho}p_{\sigma}}
\hat{\delta}({p\cdot  \bl  }) +\mathcal{O}(\hbar^0)
\end{split}
\end{equation}
which is minus \eqref{boxsing} as expected.

The classical contributions from the box diagram can be obtained in 3 ways, just as before these come from picking a power of a null momenta from the numerator, expanding the propagator and shifting to take care of the implicit $\hbar$ in  $2p\cdot \bq=-\hbar \bq^2$. 
We have
\begin{equation}
\begin{split}
i \hbar^3 \mathcal{M}^\mu(\bq) & =-\frac{ {i\kappa}^4}{2 \hbar}\bq^\mu\int \hat{d}^4 \bl \,\hat{\delta}(V\cdot \bl)\frac{\tilde{h}^{\alpha\beta}(\bl)p_{\alpha}(p+\hbar \bl)_{\beta} \,\,\tilde{h}^{\rho\sigma}(\bq-\bl)(p+\hbar \bl)_{\rho}(p+\hbar \bq)_{\sigma}}{p\cdot \bl +{\hbar \bl^2}/{2}+i\epsilon}\\& =
-\frac{{i\kappa}^4}{2 } \bq^\mu \int \hat{d}^4 \bl\,\hat{\delta}(V\cdot \bl) \frac{\h{p}{\bl}{\bl} \,\h{p}{\bq-\bl}{p}+ \h{p}{\bl}{p}\,\h{p}{\bq-\bl}{(\bq+\bl)} }{p\cdot \bl +i\varepsilon}
\\&\,\,\,\,\,\,+\frac{{i\kappa}^4}{2 }\bq^\mu \int \hat{d}^4 \bl \,\hat{\delta}(V\cdot \bl)\frac{\h{p}{\bl}{p}\,\h{p}{\bq-\bl}{p} }{(p\cdot \bl+i\varepsilon)^2} \frac{\bl^2}{2}
\\&\,\,\,\,\,\,-\frac{{i\kappa}^4}{2 }\bq^\mu \int \hat{d}^4 \bl\,\hat{\delta}(V\cdot \bl) \frac{\h{p}{\bl}{p}\,\h{p}{\bq-\bl}{p} }{(-p\cdot \bl+i\varepsilon)^2} \frac{\bq^2}{4} +\mathcal{O}(\hbar)\\&=
\mathcal{M}^\mu_a(\bar{q})+\mathcal{M}^\mu_b(\bar{q})+\mathcal{M}^\mu_c(\bar{q})
 +\mathcal{O}(\hbar)
\end{split}
\end{equation}
having defined the 3 different classical terms and the notation $\h{p_1}{l}{p_2}\equiv \tilde{h}_{\alpha\beta}(l)p_1^\alpha p_2^\beta$. The cross box  expansion goes along in an equivalent manner, this time we find
\begin{equation}
\begin{split}
 \hbar^3 \mathcal{D}^\mu(\bq) & =\frac{ {\kappa}^4}{2 \hbar} \int \hat{d}^4 \bl \,\hat{\delta}(V\cdot \bl)\, \bl^\mu {\tilde{h}^{\alpha\beta}(\bl)p_{\alpha}(p+\hbar \bl)_{\beta} \,\,\tilde{h}^{\rho\sigma}(\bq-\bl)(p+\hbar \bl)_{\rho}(p+\hbar \bq)_{\sigma}}
\hat{\delta}(p\cdot \bl +{\hbar \bl^2}/{2}) 
 \\& =
 \frac{{\kappa}^4}{2 }   \int \hat{d}^4 \bl\,\hat{\delta}(V\cdot \bl) \bl^\mu \left( {\h{p}{\bl}{\bl} \,\h{p}{\bq-\bl}{p}+ \h{p}{\bl}{p}\,\h{p}{\bq-\bl}{(\bq+\bl)} }\right)
 \hat{\delta}(p\cdot \bl ) 
\\&\,\,\,\,\,\,+\frac{{\kappa}^4}{2 }  \int \hat{d}^4 \bl\, \hat{\delta}(V\cdot \bl)\bl^\mu  \,{\h{p}{\bl}{p}\,\h{p}{\bq-\bl}{p} }{ } \,\hat{\delta}'(p\cdot \bl )  \frac{\bl^2}{2}
\\&\,\,\,\,\,\,+\frac{{\kappa}^4}{2 }  \int \hat{d}^4 \bl\, \hat{\delta}(V\cdot \bl) (\bq-\bl)^\mu  \h{p}{\bl}{p}\,\h{p}{\bq-\bl}{p} \,  \hat{\delta}'(p\cdot \bl ) \frac{\bq^2}{4} +\mathcal{O}(\hbar)\\&=
\mathcal{D}^\mu_a(\bar{q})+\mathcal{D}^\mu_b(\bar{q})+\mathcal{D}^\mu_c(\bar{q})
 +\mathcal{O}(\hbar).
\end{split}
\end{equation}
Let us now simplify these classical expressions. As in the EM case a small calculation shows that
\begin{equation}
\begin{split}
(\mathcal{M}_a(\bq)+\mathcal{D}_a(\bq))^\mu & = -\frac{{i\kappa}^4}{2 } 
\int \hat{d}^4 \bl \, \hat{\delta}(V\cdot \bl)\left(\h{p}{\bl}{\bl} \,\h{p}{\bq-\bl}{p}+ \h{p}{\bl}{p}\,\h{p}{\bq-\bl}{(\bq+\bl)} \right) \\&
\qquad\qquad\qquad
\times\left(
\frac{\bq^\mu }{p\cdot \bl +i\varepsilon}+i\bl^\mu \hat{\delta}(p\cdot \bl)
\right)\\&=
-{{i\kappa}^4}{}   \int \hat{d}^4 \bl \,\hat{\delta}(V\cdot \bl)\h{p}{\bl}{p} \,\h{p}{\bq-\bl}{\bl}\, \left(
\frac{\bq^\mu }{p\cdot \bl +i\varepsilon}+i\bl^\mu \hat{\delta}(p\cdot \bl)
\right)
\end{split}
\end{equation}
as well as
\begin{equation}
\begin{split}
(\mathcal{D}_b(\bq)+\mathcal{D}_c(\bq))^\mu= 
\frac{{\kappa}^4}{2 }  \int \hat{d}^4 \bl\, \bl^\mu  \,\hat{\delta}(V\cdot \bl){\h{p}{\bl}{p}\,\h{p}{\bq-\bl}{p} }{ } \,\hat{\delta}'(p\cdot \bl ) \bl\cdot(\bl-\bq)+\mathcal{Z}^\mu
\end{split}
\end{equation}
where we defined the vector
\begin{equation}
\mathcal{Z}^\mu=-
\frac{{\kappa}^4}{4 }  \bq^\mu \int \hat{d}^4 \bl \,\hat{\delta}(V\cdot \bl){\h{p}{\bl}{p}\,\h{p}{\bq-\bl}{p} }{ } \,\hat{\delta}'(p\cdot \bl ) \bl\cdot(\bl-2\bq).
\end{equation}
When added to $\mathcal{M}_b(\bq)+\mathcal{M}_c(\bq)$, this allows us to nicely gather together the remaining parts
\begin{equation}
(\mathcal{M}_b(\bq)+\mathcal{M}_c(\bq)+\mathcal{Z} )^\mu=
\frac{{i\kappa}^4}{4 }\bq^\mu \int \hat{d}^4 \bl\,\hat{\delta}(V\cdot \bl) \frac{\h{p}{\bl}{p}\,\h{p}{\bq-\bl}{p} }{(p\cdot \bl+i\varepsilon)^2} \bl\cdot(\bl-\bq).
\end{equation}

Now we can finally write down the classical momentum deflection of a scalar body travelling on a KS background 
\begin{equation}\label{deltaplano}
\begin{split}
\Delta p^\mu_{\text{NLO}}=\frac{i\kappa^4 m}{8}&\int \hat{{d}}^4 \bq\,\hat{\delta}(u\cdot \bq)\hat{\delta}(V\cdot  \bq)e^{-ib\cdot \bq}\int \hat{d}^4\bl\, \hat{\delta}(V\cdot \bl)\\&\times \left[
\bq^\mu \left(
{ -4}\frac{\h{u}{\bl}{u}\,\h{u}{\bq-\bl}{\bl}}{u\cdot \bl + i\varepsilon}+ \frac{\h{u}{\bl}{u}\,\h{u}{\bq-\bl}{u} }{(u\cdot \bl+i\varepsilon)^2} \bl\cdot(\bl-\bq)
\right)\right.
\\&-\left.
i\bl^\mu \left(4 \hat{\delta} (u\cdot l) {\h{u}{\bl}{u}\,\h{u}{\bq-\bl}{\bl} }+
     \hat{\delta}'( \bar{\ell}\cdot u)\,  \bl\cdot (\bl-\bq) {\h{u}{\bl}{u}\,\h{u}{\bq-\bl}{u}}
\right)\frac{}{}\right].
\end{split}
\end{equation}
This also satisfies the on-shell check, as can be seen following the same steps outlined in \eqref{oskk}. A further confirmation on our results was obtained by working out $\Delta p^\mu$, both at LO and NLO, by perturbatively solving the classical equations followed by the probe. We will indeed sketch the proof of \eqref{oookkkk} and \eqref{deltaplano} using the geodesic equation in the appendix. We stress that, at this point, fourmulae \eqref{okok}, \eqref{deltaaaa}, \eqref{oookkkk} and \eqref{deltaplano} hold for a generic heavy source that can be put in  a Kerr-Schild form. 

The reader will appreciate how similar the electromagnetic and gravity expressions are, which is a consequence of working with Kerr-Schild expressions for the source. From inspection of our equations, it is clear that the gravity observables we obtained are equivalent to their gauge theory counterparts, up to a number,  by the simple replacement 
\begin{equation}
\tilde{A}^\mu (\bq)\leftrightarrow \tilde{h}^{\mu\nu}(\bq)u_\nu
\end{equation}
which is somehow very reminiscent of the classical double copy prescription. We believe this classical correspondence deserves further scrutiny.

\subsection{Further comments and checks}

Kerr-Schild field expressions are not very easy to deal with in practical computations. This is because both the momentum and position space dependence form of $\phi$ and $k^\mu$ can be tricky, especially for spinning sources. Luckily, sometimes, we can find a way around this impasse by switching to harmonic coordinates. One of the first works to adopt such a strategy was \cite{Vines:18}, where an explicit linear diffeomorphism between the Kerr metric in KS and de Donder/harmonic/Lorenz gauge was found. One has in position space, 
\begin{equation}\label{kerrrr}
h_{\mu\nu}^{\text{DD}}(x)= - \frac{\kappa M}{8\pi} {P_{\mu\nu}}^{\alpha\beta} \,V_{(\alpha} {\exp \left( a*\partial\right) _{\beta)}}^\rho \frac{V_\rho}{r}= h_{\mu\nu}^{\text{KS}}(x)+ \partial_{(\mu}\xi_{\nu)}.
\end{equation}
The exponential operator of \eqref{kerrrr} is defined as the power series of the 2-form 
\begin{equation}
(a*\partial)_{\mu\nu}=\epsilon_{\mu\nu }(a, \partial) .
\end{equation}
The explicit form of $\xi_\mu$ in the expression above will not interest us so we do not report it. \eqref{kerrrr} satisfies $P^{\mu\nu\alpha\beta}\partial_\nu h_{\alpha\beta}=0$ at the linear order in $G_N$, further corrections can in principle be obtained by solving Einstein's equations up to the desired order.

In the spirit of the Double copy we can define a Lorenz expression for a spinning source too
\begin{equation}\label{sqrtker}
A_\mu^\text{L}(x)=\frac{Q}{4\pi} {\exp \left( a*\partial\right) _{\mu}}^\rho \frac{V_\rho}{r}={A}_\mu^\text{KS}(x)+\partial_\mu \chi (x)
\end{equation}
which is often called $\sqrt{\text{Kerr}}$ in the literature.

equations \eqref{kerrrr} and \eqref{sqrtker} are more suitable for actual computation and their Fourier transforms can be readily computed. Let us consider EM first, we find
\begin{equation}\label{not}
\tilde{A}^{\text{L}}_{\mu}=-Q\hat{\delta }(q\cdot V) {\exp \left(-i a*q\right) _{\mu}}^\rho \frac{V_\rho}{q^2}\equiv -Q\hat{\delta }(q\cdot V) \frac{U_\mu (q)}{q^2}
\end{equation}
having hidden the exponential into the $q$-dependent vector $U_\mu$. But how to practically use this expression? One option  is to plug in the transformation \eqref{sqrtker} inside \eqref{okok} and \eqref{deltaaaa}, after the dust settles one is able to show that these become 
\begin{equation} 
{\Delta} p^\mu_\text{LO}=-ieQ \int \hat{d}^4 \bar{q}\, \hat{\delta}(u\cdot \bar{q})\hat{\delta} (V\cdot \bar{q})e^{-ib\cdot\bar{q}}{\bar{q}^\mu}\,\frac{ u\cdot U(\bar{q})}{q^2}.
\end{equation} 
and
\begin{equation} \label{lorenzgauge}
\begin{split}
\Delta p^\mu_{\text{NLO}}=\frac{ie^2Q^2}{2m}\int \hat{{d}}^4 \bq\,& \hat{\delta}(u\cdot \bq)\hat{\delta}(V\cdot \bq)e^{-ib\cdot \bq}\int \frac{\hat{d}^4 \bl}{\bar{\ell}^2(\bar{q}-\bar{\ell})^2} \, \hat{\delta}(V\cdot \bl)\\&\times \left[
\bq^\mu \left( U(\bl)\cdot U(\bq-\bl)
{ -2}\frac{U(\bl)\cdot u \,U(\bar{q}-\bar{\ell})\cdot \bl }{u\cdot \bl + i\varepsilon}+\frac{U(\bl)\cdot u\,U(\bar{q}-\bar{\ell})\cdot u}{(u\cdot \bl  +i \varepsilon)^2} \,\bl\cdot (\bl-\bq)
\right)\right.
\\&-\left.
i\bl^\mu \left(2 \hat{\delta} (u\cdot l) {U(\bl)\cdot u \,U(\bar{q}-\bar{\ell})\cdot \bl }+
     \hat{\delta}'( \bar{\ell}\cdot u)\,  \bl\cdot (\bl-\bq) {U(\bl)\cdot u\, U(\bar{q}-\bar{\ell})\cdot u}
\right)\frac{}{}\right].
\end{split}
\end{equation}
These equations, as already stressed before, are valid for generic spin orientations. While the leading order term is essentially unchanged in its form, it is interesting to see how switching to a Loernz gauge expression made the Feynman triangle contribution $\sim \bq^\mu U(\bl)\cdot U(\bq-\bl)$ appear.
Another way to prove this is to repeat the steps of \eqref{mosoccazzi} but in Lorenz gauge, with an $\Phi^2 A^2$ interaction term in the Lagrangian. It is also reassuring to see that  \eqref{lorenzgauge}  reproduces the results of \cite{Kosower:19}  when the mass ratio $M_1/M_2$ there is taken to be small and one takes the spinless limit
\begin{equation}
a\to 0, \,\,\,\,U_\mu(q)\to V_\mu,\,\,\,\, U(q)\cdot u\to \gamma.
\end{equation}

Let's carry some numerical checks now. We will  work in a frame where $u^\mu=\gamma (1, 0,0,\beta),\,\,\,V^\mu=(1, \vec{0})$. It is easy to show the agreement of the leading order deflection with our unitarity-based result of \eqref{ii1}. To realize this we only have to bring the exponential operator into a nicer form. On the support of the on shell measure we can write
\begin{equation} \label{algebbr}
\begin{split}
\bq^\mu \, u\cdot U(\bq)&= \bq^\mu\, \sum_{n\geq 0} \frac{(-ia*\bar{q})^n_{\alpha\beta}}{n!} u^\alpha V^\beta=\gamma  \bq^\mu\,\cosh a\cdot \bar{q}-i\bq^\mu \frac{\epsilon(u, V, a, \bq)}{ a\cdot\bar{q}} \sinh a\cdot\bar{q}
\\&=
\gamma  \bq^\mu\,\cosh a\cdot \bar{q}-i \epsilon^{\mu}(\bar{q}, u, V)\sinh a\cdot\bar{q}.
\end{split}
\end{equation}
 Observe that in the last equality above we made use of the useful property  \cite{Vines:18, Emond:2020lwi}
\begin{equation}
\bar{q}_\mu{\epsilon(u, V, a, \bq)}{}= a\cdot\bar{q}\,\epsilon_{\mu}(\bar{q}, u, V)
\end{equation}
which makes it clear that our two $\mathcal{O}(e Q)$ expressions equal each other.

Perhaps more interesting is to see what happens at one loop, since both the probe limit expressions  \eqref{deltaaaa} and \eqref{lorenzgauge} look very different from  \eqref{ii2}. To this end we will check that our formulae agree at $\mathcal{O}(e^2Q^2a^1)$ in the aligned spin case $a\cdot b=0$. Although going to an arbitrary high order in the spin poses no fundamental obstruction (one would just get higher rank 1-loop integrals to be reduced), computations become quickly messy, which is due to the presence of the exponential matrix that defines $U_\mu (q)$. So let us expand in $a^\mu$ equation \eqref{lorenzgauge}. We find the only term that contributes to the scattering angle comes from the second term inside the parenthesis of \eqref{lorenzgauge}. This is
\begin{equation}\label{aaaa3}
\Delta p_{\text{NLO}} ^\mu\to\frac{-ie^2Q^2 }{m} \int { \hat{d}^4 \bar{q} \, } e^{-{i}{b\cdot \bar{q}}}\hat{\delta} (u\cdot \bar{q})  \hat{\delta}(\bar{q}\cdot V) \bar{q}^\mu  \int \frac{\hat{d}^4 \bl}{\bar{\ell}^2(\bar{q}-\bar{\ell})^2}  \hat{\delta}(\bar{\ell}\cdot V) \frac{u\cdot V\, i \epsilon (\bar{q}, a, \bar{\ell}, V)}{u\cdot \bar{\ell}+i \varepsilon}+\mathcal{O}(a).
\end{equation}
Let's focus on the loop integral 
\begin{equation}
\int\frac{  \hat{d}^4 \bar{\ell}\,}{\bar{\ell}^2(\bar{q}-\bar{\ell})^2} \frac{\hat{\delta}(\bar{\ell}\cdot V)  }{p\cdot \bar{\ell}+i \varepsilon}\epsilon (\bar{q}, a, \bar{\ell}, V)=\epsilon (\bar{q}, a, I, V)
\end{equation}
with
\begin{equation}
I^\mu\equiv\int\frac{  \hat{d}^4 \bar{\ell}\,}{\bar{\ell}^2(\bar{q}-\bar{\ell})^2} \frac{\hat{\delta}(\bar{\ell}\cdot V)  }{p\cdot \bar{\ell}+i \varepsilon}\bar{\ell}^\mu.
\end{equation}
In general $I^\mu$ can be expanded as
\begin{equation}
I^\mu=A V^\mu+B p^\mu+C\bar{q}^\mu
\end{equation}
for some   $A, B, C$. However, when we plug this back into $\varepsilon$ the terms proportional to $V$ and $\bar{q}$ will be projected out, so we're only really interested in $B$. Contracting we have
\begin{equation}
I\cdot V=0 =A + m\gamma B, \,\,\, I\cdot p = m\gamma A+Bm^2=\int\frac{  \hat{d}^4 \bar{\ell}\,}{\bar{\ell}^2(\bar{q}-\bar{\ell})^2} \hat{\delta}(\bar{\ell}\cdot V)=\frac{1}{8}\frac{1}{\sqrt{-\bar{{q}}^2}}.
\end{equation}
Recall we are in a frame such that $V^\mu=(1, \vec{0})$ and $p^\mu=mu^\mu=m(\gamma, 0, 0, \gamma\beta)$. In the end
\begin{equation}
\epsilon (\bar{q}, a, I, V)=\frac{1}{8m^2}\frac{1}{1-\gamma^2}\frac{\epsilon (\bar{q}, a, p, V)}{\sqrt{-\bar{{q}}^2}}
\end{equation}
thus we can write \eqref{aaaa3} as
\begin{equation}
\begin{split}
\Delta p_{\text{NLO}} &\to{e^2Q^2 \gamma } \int \hat{d}^4 \bar{q} \,e^{-{i}{b\cdot \bar{q}}}\hat{\delta} (u\cdot \bar{q})\hat{\delta}(\bar{q}\cdot V) \bar{q}^\mu   \epsilon (\bar{q}, a, I, V)\\&
=\frac{e^2Q^2 \gamma }{8m^2 } \frac{\epsilon_{\rho \nu\alpha\beta}}{1-\gamma ^2}a^\nu p^\alpha V^\beta \int \hat{d}^4 \bar{q} \,e^{-{i}{b\cdot \bar{q}}}\hat{\delta} (u\cdot \bar{q})\hat{\delta}(\bar{q}\cdot V)   \frac{\bar{q}^\mu \bar{q}^\rho}{\sqrt{-{\bar{q}}^2}}
\end{split}
\end{equation} 
and use \cite{Maybee:19, Liu:2021zxr}
\begin{equation}
\begin{split}
\int \hat{d}^4 \bar{q}\,\hat{\delta} (\bar{q}\cdot u)\hat{\delta}(\bar{q}\cdot V)e^{-i\bar{q}\cdot b}\frac{\bar{q}^\mu\bar{q}^\rho}{\sqrt{-{\bar{q}}^2}}=\frac{1}{2\pi \sinh\phi \, b^3}\left( \Pi^{\mu\rho}-3\frac{b^\mu b^\rho}{b^2} \right)
\end{split}
\end{equation}
where the projector is the one initially defined in \eqref{proj1}. We thus obtain, for this piece of the deflection 
\begin{equation}
\Delta p_{\text{NLO}} ^\mu\to\frac{e^2Q^2 \gamma }{16\pi \gamma\beta m } \frac{\epsilon_{\rho \nu\alpha\beta}}{1-\gamma ^2}a^\nu u^\alpha V^\beta \frac{1}{ b^3}\left( \Pi^{\mu\rho}-3\frac{b^\mu b^\rho}{b^2} \right)+\mathcal{O}(a^2).
\end{equation}
Let us  get the scattering angle dotting with the impact parameter. We find, noting that ${\Pi^\mu}_\nu b^\nu=b^\mu$,
\begin{equation}\label{angolo}
\theta=\frac{b\cdot {\Delta} p_\text{NLO}}{b|\vec{p}|}\to-\frac{1}{|\vec{p}|}\frac{e^2Q^2 \gamma }{8\pi \beta
\gamma m b^4 } \frac{\epsilon(b, a, u, V)}{1-\gamma ^2}+\mathcal{O}(a^2)=-\frac{1}{8\pi}\frac{e^2Q^2 }{m^2 \gamma^3\beta^4 b^4 } \epsilon(b, a, u, V)+\mathcal{O}(a^2).
\end{equation}
The well known $\mathcal{O}(e^2Q^2a^0)$ term can also be obtained using standard integration techniques and reads
\begin{equation}
\theta\to -\frac{e^2Q^2}{32\pi (m\beta\gamma b)^2}
\end{equation}
so that
\begin{equation}\label{anggg}
\theta=- \frac{e^2Q^2}{32\pi m^2\beta^2\gamma^2}\frac{1}{b^2}\left(1+
\frac{2}{  \gamma\beta^2}\frac{ \epsilon(b, a, u, V)}{b^2}
\right)+\mathcal{O}(a^2)
\end{equation}
which, in the aligned case,  matches  the Taylor expansion of \eqref{angolettino} up to the respective spin order and equals the projection on the scattering plane contribution of our non aligned spin result \eqref{rootkerrdp}. A simple calculation also shows that the longitudinal vector components of  \eqref{lorenzgauge} and  \eqref{ii2}
match, as expected. 

We believe it is fascinating to see at work two different frameworks (unitarity/probe limit) and how, even if the calculations organise themselves so differently, they agree in the end. The leading singularity has the advantage that (in the aligned case) all the gauge invariant scattering information is nicely encoded in the unitarity triangle cut, while in \eqref{lorenzgauge} the bit that actually contributed to \eqref{anggg} came from a box (Feynman) diagram. This is even less unclear if one had used the KS expression \eqref{deltaaaa}, where the source Fourier transform has an even less trivial momentum dependence, thus isolating the unitarity triangle terms may be even more cumbersome.

The gravity side of the story streamlines as in the electromagnetic case at leading order. The probe limit deflection  \eqref{oookkkk} can be related to the Fourier transform of the harmonic metric by a linear diffeomorphism. We find, on shell of the delta functions
\begin{equation}\label{ppgr}
\Delta p^\mu_{\text{LO}}=\frac{i\kappa^2 mM}{4} \int \hat{{d}}^4 \bq\, \hat{\delta}(u\cdot \bq)\hat{\delta}(V\cdot \bq)e^{-ib\cdot \bq} \, \bq^\mu \,\frac{P^{\alpha\beta \rho\sigma} V_{(\rho} {\exp \left( -ia*q\right) _{\sigma)}} ^\lambda {V_\lambda}(\bq) u_\alpha u_\beta}{q^2}
\end{equation}
we observe that, inside the integral, following the same steps outlined in   \eqref{firstcosh} and \eqref{algebbr}  
\begin{equation}
 u_\alpha u_\beta P^{\alpha\beta \rho\sigma} V_{(\rho} {\exp \left( -ia*q\right) _{\sigma)}} ^\lambda {V_\lambda}(\bq)= \frac{1}{2}\left(
\bq^\mu (2\gamma^2-1)\cosh a\cdot \bq -2i \gamma\, \epsilon^\mu (\bq, u, V) \sinh a \cdot \bq
\right)
\end{equation}
which is enough to see that \eqref{ppgr} agrees with   \eqref{gravprob}.
Unfortunately the NLO harmonic expression of the Kerr metric is not known, or at least to our knowledge, so carrying the one loop integral becomes more complicated since we would have to work with the momentum space expression of the metric in KS coordinates. A preliminary check suggests that \eqref{deltaplano} yields the right answer for the case of a spinless heavy source but a proper treatment of the spinning case is still lacking.

 Another reason why momentum KS expressions are worth further investigation is related to higher loops. For instance, it is easy to write a generic $L$-loop amplitude. Due to the linearity of the exact KS vertex we only really have 1 ladder diagram at every loop order,
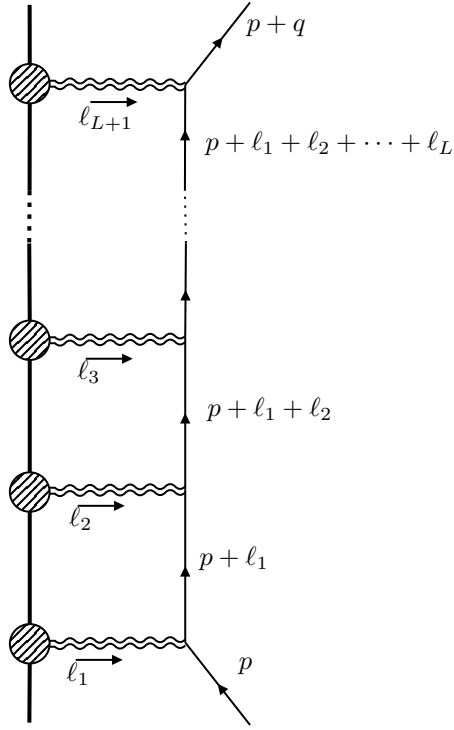
\begin{figure}[H]
\begin{center}

 
\tikzset{
pattern size/.store in=\mcSize, 
pattern size = 5pt,
pattern thickness/.store in=\mcThickness, 
pattern thickness = 0.3pt,
pattern radius/.store in=\mcRadius, 
pattern radius = 1pt}
\makeatletter
\pgfutil@ifundefined{pgf@pattern@name@_7k1ray68r}{
\pgfdeclarepatternformonly[\mcThickness,\mcSize]{_7k1ray68r}
{\pgfqpoint{0pt}{0pt}}
{\pgfpoint{\mcSize+\mcThickness}{\mcSize+\mcThickness}}
{\pgfpoint{\mcSize}{\mcSize}}
{
\pgfsetcolor{\tikz@pattern@color}
\pgfsetlinewidth{\mcThickness}
\pgfpathmoveto{\pgfqpoint{0pt}{0pt}}
\pgfpathlineto{\pgfpoint{\mcSize+\mcThickness}{\mcSize+\mcThickness}}
\pgfusepath{stroke}
}}
\makeatother

 
\tikzset{
pattern size/.store in=\mcSize, 
pattern size = 5pt,
pattern thickness/.store in=\mcThickness, 
pattern thickness = 0.3pt,
pattern radius/.store in=\mcRadius, 
pattern radius = 1pt}
\makeatletter
\pgfutil@ifundefined{pgf@pattern@name@_ybw3p0k1u}{
\pgfdeclarepatternformonly[\mcThickness,\mcSize]{_ybw3p0k1u}
{\pgfqpoint{0pt}{0pt}}
{\pgfpoint{\mcSize+\mcThickness}{\mcSize+\mcThickness}}
{\pgfpoint{\mcSize}{\mcSize}}
{
\pgfsetcolor{\tikz@pattern@color}
\pgfsetlinewidth{\mcThickness}
\pgfpathmoveto{\pgfqpoint{0pt}{0pt}}
\pgfpathlineto{\pgfpoint{\mcSize+\mcThickness}{\mcSize+\mcThickness}}
\pgfusepath{stroke}
}}
\makeatother

 
\tikzset{
pattern size/.store in=\mcSize, 
pattern size = 5pt,
pattern thickness/.store in=\mcThickness, 
pattern thickness = 0.3pt,
pattern radius/.store in=\mcRadius, 
pattern radius = 1pt}
\makeatletter
\pgfutil@ifundefined{pgf@pattern@name@_eyqz8mlsc}{
\pgfdeclarepatternformonly[\mcThickness,\mcSize]{_eyqz8mlsc}
{\pgfqpoint{0pt}{0pt}}
{\pgfpoint{\mcSize+\mcThickness}{\mcSize+\mcThickness}}
{\pgfpoint{\mcSize}{\mcSize}}
{
\pgfsetcolor{\tikz@pattern@color}
\pgfsetlinewidth{\mcThickness}
\pgfpathmoveto{\pgfqpoint{0pt}{0pt}}
\pgfpathlineto{\pgfpoint{\mcSize+\mcThickness}{\mcSize+\mcThickness}}
\pgfusepath{stroke}
}}
\makeatother

 
\tikzset{
pattern size/.store in=\mcSize, 
pattern size = 5pt,
pattern thickness/.store in=\mcThickness, 
pattern thickness = 0.3pt,
pattern radius/.store in=\mcRadius, 
pattern radius = 1pt}
\makeatletter
\pgfutil@ifundefined{pgf@pattern@name@_3jx9amvb1}{
\pgfdeclarepatternformonly[\mcThickness,\mcSize]{_3jx9amvb1}
{\pgfqpoint{0pt}{0pt}}
{\pgfpoint{\mcSize+\mcThickness}{\mcSize+\mcThickness}}
{\pgfpoint{\mcSize}{\mcSize}}
{
\pgfsetcolor{\tikz@pattern@color}
\pgfsetlinewidth{\mcThickness}
\pgfpathmoveto{\pgfqpoint{0pt}{0pt}}
\pgfpathlineto{\pgfpoint{\mcSize+\mcThickness}{\mcSize+\mcThickness}}
\pgfusepath{stroke}
}}
\makeatother
\tikzset{every picture/.style={line width=0.75pt}} 

\begin{tikzpicture}[x=0.75pt,y=0.75pt,yscale=-1,xscale=1]

\draw    (133.35,1525.17) .. controls (131.7,1526.85) and (130.04,1526.87) .. (128.35,1525.22) .. controls (126.67,1523.57) and (125,1523.59) .. (123.35,1525.27) .. controls (121.7,1526.95) and (120.03,1526.96) .. (118.35,1525.31) .. controls (116.67,1523.66) and (115,1523.68) .. (113.35,1525.36) .. controls (111.7,1527.04) and (110.03,1527.06) .. (108.35,1525.41) .. controls (106.67,1523.76) and (105,1523.78) .. (103.35,1525.46) .. controls (101.7,1527.14) and (100.03,1527.16) .. (98.35,1525.51) .. controls (96.67,1523.86) and (95,1523.88) .. (93.35,1525.56) .. controls (91.7,1527.24) and (90.03,1527.26) .. (88.35,1525.61) .. controls (86.67,1523.96) and (85,1523.98) .. (83.35,1525.66) .. controls (81.7,1527.34) and (80.03,1527.36) .. (78.35,1525.71) .. controls (76.67,1524.06) and (75,1524.08) .. (73.35,1525.76) .. controls (71.7,1527.44) and (70.03,1527.46) .. (68.35,1525.81) -- (65.68,1525.83) -- (65.68,1525.83)(133.32,1522.17) .. controls (131.67,1523.85) and (130.01,1523.87) .. (128.32,1522.22) .. controls (126.64,1520.57) and (124.97,1520.59) .. (123.32,1522.27) .. controls (121.67,1523.95) and (120,1523.96) .. (118.32,1522.31) .. controls (116.64,1520.66) and (114.97,1520.68) .. (113.32,1522.36) .. controls (111.67,1524.04) and (110,1524.06) .. (108.32,1522.41) .. controls (106.64,1520.76) and (104.97,1520.78) .. (103.32,1522.46) .. controls (101.67,1524.14) and (100,1524.16) .. (98.32,1522.51) .. controls (96.64,1520.86) and (94.97,1520.88) .. (93.32,1522.56) .. controls (91.67,1524.24) and (90,1524.26) .. (88.32,1522.61) .. controls (86.64,1520.96) and (84.97,1520.98) .. (83.32,1522.66) .. controls (81.67,1524.34) and (80,1524.36) .. (78.32,1522.71) .. controls (76.64,1521.06) and (74.97,1521.08) .. (73.32,1522.76) .. controls (71.67,1524.44) and (70,1524.46) .. (68.32,1522.81) -- (65.65,1522.83) -- (65.65,1522.83) ;
\draw    (79.67,1531.92) -- (100.17,1531.92) ;
\draw [shift={(103.17,1531.92)}, rotate = 180] [fill={rgb, 255:red, 0; green, 0; blue, 0 }  ][line width=0.08]  [draw opacity=0] (5.36,-2.57) -- (0,0) -- (5.36,2.57) -- cycle    ;
\draw  [pattern=_7k1ray68r,pattern size=3.2249999999999996pt,pattern thickness=0.75pt,pattern radius=0pt, pattern color={rgb, 255:red, 0; green, 0; blue, 0}] (45.83,1524.98) .. controls (45.83,1519.51) and (50.27,1515.07) .. (55.75,1515.07) .. controls (61.23,1515.07) and (65.67,1519.51) .. (65.67,1524.98) .. controls (65.67,1530.46) and (61.23,1534.9) .. (55.75,1534.9) .. controls (50.27,1534.9) and (45.83,1530.46) .. (45.83,1524.98) -- cycle ;
\draw    (165.67,1642.18) -- (133.33,1600.67) ;
\draw [shift={(149.5,1621.42)}, rotate = 52.08] [fill={rgb, 255:red, 0; green, 0; blue, 0 }  ][line width=0.08]  [draw opacity=0] (5.36,-2.57) -- (0,0) -- (5.36,2.57) -- cycle    ;
\draw    (133.33,1600.67) -- (133.33,1523.67) ;
\draw [shift={(133.33,1562.17)}, rotate = 90] [fill={rgb, 255:red, 0; green, 0; blue, 0 }  ][line width=0.08]  [draw opacity=0] (5.36,-2.57) -- (0,0) -- (5.36,2.57) -- cycle    ;
\draw    (133.35,1602.17) .. controls (131.7,1603.85) and (130.04,1603.87) .. (128.35,1602.22) .. controls (126.67,1600.57) and (125,1600.59) .. (123.35,1602.27) .. controls (121.7,1603.95) and (120.03,1603.96) .. (118.35,1602.31) .. controls (116.67,1600.66) and (115,1600.68) .. (113.35,1602.36) .. controls (111.7,1604.04) and (110.03,1604.06) .. (108.35,1602.41) .. controls (106.67,1600.76) and (105,1600.78) .. (103.35,1602.46) .. controls (101.7,1604.14) and (100.03,1604.16) .. (98.35,1602.51) .. controls (96.67,1600.86) and (95,1600.88) .. (93.35,1602.56) .. controls (91.7,1604.24) and (90.03,1604.26) .. (88.35,1602.61) .. controls (86.67,1600.96) and (85,1600.98) .. (83.35,1602.66) .. controls (81.7,1604.34) and (80.03,1604.36) .. (78.35,1602.71) .. controls (76.67,1601.06) and (75,1601.08) .. (73.35,1602.76) .. controls (71.7,1604.44) and (70.03,1604.46) .. (68.35,1602.81) -- (65.68,1602.83) -- (65.68,1602.83)(133.32,1599.17) .. controls (131.67,1600.85) and (130.01,1600.87) .. (128.32,1599.22) .. controls (126.64,1597.57) and (124.97,1597.59) .. (123.32,1599.27) .. controls (121.67,1600.95) and (120,1600.96) .. (118.32,1599.31) .. controls (116.64,1597.66) and (114.97,1597.68) .. (113.32,1599.36) .. controls (111.67,1601.04) and (110,1601.06) .. (108.32,1599.41) .. controls (106.64,1597.76) and (104.97,1597.78) .. (103.32,1599.46) .. controls (101.67,1601.14) and (100,1601.16) .. (98.32,1599.51) .. controls (96.64,1597.86) and (94.97,1597.88) .. (93.32,1599.56) .. controls (91.67,1601.24) and (90,1601.26) .. (88.32,1599.61) .. controls (86.64,1597.96) and (84.97,1597.98) .. (83.32,1599.66) .. controls (81.67,1601.34) and (80,1601.36) .. (78.32,1599.71) .. controls (76.64,1598.06) and (74.97,1598.08) .. (73.32,1599.76) .. controls (71.67,1601.44) and (70,1601.46) .. (68.32,1599.81) -- (65.65,1599.83) -- (65.65,1599.83) ;
\draw    (78.67,1609.42) -- (99.17,1609.42) ;
\draw [shift={(102.17,1609.42)}, rotate = 180] [fill={rgb, 255:red, 0; green, 0; blue, 0 }  ][line width=0.08]  [draw opacity=0] (5.36,-2.57) -- (0,0) -- (5.36,2.57) -- cycle    ;
\draw  [pattern=_ybw3p0k1u,pattern size=3.2249999999999996pt,pattern thickness=0.75pt,pattern radius=0pt, pattern color={rgb, 255:red, 0; green, 0; blue, 0}] (45.83,1601.33) .. controls (45.83,1595.86) and (50.27,1591.42) .. (55.75,1591.42) .. controls (61.23,1591.42) and (65.67,1595.86) .. (65.67,1601.33) .. controls (65.67,1606.81) and (61.23,1611.25) .. (55.75,1611.25) .. controls (50.27,1611.25) and (45.83,1606.81) .. (45.83,1601.33) -- cycle ;
\draw [line width=1.5]    (55.75,1534.9) -- (55.75,1591.42) ;
\draw [line width=1.5]    (55.75,1611.25) -- (55.56,1640.98) ;
\draw  [pattern=_eyqz8mlsc,pattern size=3.2249999999999996pt,pattern thickness=0.75pt,pattern radius=0pt, pattern color={rgb, 255:red, 0; green, 0; blue, 0}] (45.83,1448.64) .. controls (45.83,1443.16) and (50.27,1438.72) .. (55.75,1438.72) .. controls (61.23,1438.72) and (65.67,1443.16) .. (65.67,1448.64) .. controls (65.67,1454.11) and (61.23,1458.55) .. (55.75,1458.55) .. controls (50.27,1458.55) and (45.83,1454.11) .. (45.83,1448.64) -- cycle ;
\draw [line width=1.5]    (55.75,1458.55) -- (55.75,1515.07) ;
\draw    (133.33,1523.67) -- (133.33,1446.67) ;
\draw [shift={(133.33,1485.17)}, rotate = 90] [fill={rgb, 255:red, 0; green, 0; blue, 0 }  ][line width=0.08]  [draw opacity=0] (5.36,-2.57) -- (0,0) -- (5.36,2.57) -- cycle    ;
\draw    (133.35,1449.47) .. controls (131.7,1451.15) and (130.03,1451.17) .. (128.35,1449.52) .. controls (126.67,1447.87) and (125,1447.89) .. (123.35,1449.57) .. controls (121.7,1451.25) and (120.03,1451.27) .. (118.35,1449.62) .. controls (116.67,1447.97) and (115,1447.99) .. (113.35,1449.67) .. controls (111.7,1451.35) and (110.03,1451.37) .. (108.35,1449.72) .. controls (106.67,1448.07) and (105,1448.08) .. (103.35,1449.76) .. controls (101.7,1451.44) and (100.03,1451.46) .. (98.35,1449.81) .. controls (96.67,1448.16) and (95,1448.18) .. (93.35,1449.86) .. controls (91.7,1451.54) and (90.03,1451.56) .. (88.35,1449.91) .. controls (86.67,1448.26) and (85,1448.28) .. (83.35,1449.96) .. controls (81.7,1451.64) and (80.03,1451.66) .. (78.35,1450.01) .. controls (76.67,1448.36) and (75,1448.38) .. (73.35,1450.06) .. controls (71.7,1451.74) and (70.03,1451.76) .. (68.35,1450.11) -- (65.68,1450.14) -- (65.68,1450.14)(133.32,1446.47) .. controls (131.67,1448.15) and (130,1448.17) .. (128.32,1446.52) .. controls (126.64,1444.87) and (124.97,1444.89) .. (123.32,1446.57) .. controls (121.67,1448.25) and (120,1448.27) .. (118.32,1446.62) .. controls (116.64,1444.97) and (114.97,1444.99) .. (113.32,1446.67) .. controls (111.67,1448.35) and (110,1448.37) .. (108.32,1446.72) .. controls (106.64,1445.07) and (104.97,1445.09) .. (103.32,1446.77) .. controls (101.67,1448.45) and (100,1448.46) .. (98.32,1446.81) .. controls (96.64,1445.16) and (94.97,1445.18) .. (93.32,1446.86) .. controls (91.67,1448.54) and (90,1448.56) .. (88.32,1446.91) .. controls (86.64,1445.26) and (84.97,1445.28) .. (83.32,1446.96) .. controls (81.67,1448.64) and (80,1448.66) .. (78.32,1447.01) .. controls (76.64,1445.36) and (74.97,1445.38) .. (73.32,1447.06) .. controls (71.67,1448.74) and (70,1448.76) .. (68.32,1447.11) -- (65.65,1447.14) -- (65.65,1447.14) ;
\draw    (133.33,1446.67) -- (133.67,1399.84) ;
\draw [shift={(133.5,1423.26)}, rotate = 90.41] [fill={rgb, 255:red, 0; green, 0; blue, 0 }  ][line width=0.08]  [draw opacity=0] (5.36,-2.57) -- (0,0) -- (5.36,2.57) -- cycle    ;
\draw [line width=1.5]    (55.5,1399.93) -- (55.75,1438.72) ;
\draw    (83.67,1457.92) -- (104.17,1457.92) ;
\draw [shift={(107.17,1457.92)}, rotate = 180] [fill={rgb, 255:red, 0; green, 0; blue, 0 }  ][line width=0.08]  [draw opacity=0] (5.36,-2.57) -- (0,0) -- (5.36,2.57) -- cycle    ;
\draw    (165.67,1278.7) -- (133.33,1320.21) ;
\draw [shift={(152.02,1296.22)}, rotate = 127.92] [fill={rgb, 255:red, 0; green, 0; blue, 0 }  ][line width=0.08]  [draw opacity=0] (5.36,-2.57) -- (0,0) -- (5.36,2.57) -- cycle    ;
\draw    (133.33,1320.21) -- (133.17,1373.43) ;
\draw [shift={(133.26,1342.72)}, rotate = 90.18] [fill={rgb, 255:red, 0; green, 0; blue, 0 }  ][line width=0.08]  [draw opacity=0] (5.36,-2.57) -- (0,0) -- (5.36,2.57) -- cycle    ;
\draw    (133.32,1321.71) .. controls (131.64,1323.36) and (129.97,1323.34) .. (128.32,1321.66) .. controls (126.67,1319.98) and (125,1319.96) .. (123.32,1321.61) .. controls (121.64,1323.26) and (119.97,1323.24) .. (118.32,1321.56) .. controls (116.67,1319.88) and (115,1319.86) .. (113.32,1321.51) .. controls (111.64,1323.16) and (109.97,1323.14) .. (108.32,1321.46) .. controls (106.67,1319.78) and (105,1319.76) .. (103.32,1321.41) .. controls (101.64,1323.06) and (99.97,1323.04) .. (98.32,1321.36) .. controls (96.67,1319.68) and (95,1319.66) .. (93.32,1321.31) .. controls (91.64,1322.96) and (89.97,1322.95) .. (88.32,1321.27) .. controls (86.67,1319.59) and (85,1319.57) .. (83.32,1321.22) .. controls (81.64,1322.87) and (79.97,1322.85) .. (78.32,1321.17) .. controls (76.67,1319.49) and (75,1319.47) .. (73.32,1321.12) .. controls (71.64,1322.77) and (69.97,1322.75) .. (68.32,1321.07) -- (65.65,1321.04) -- (65.65,1321.04)(133.35,1318.71) .. controls (131.66,1320.36) and (130,1320.34) .. (128.35,1318.66) .. controls (126.7,1316.98) and (125.03,1316.96) .. (123.35,1318.61) .. controls (121.67,1320.26) and (120,1320.24) .. (118.35,1318.56) .. controls (116.7,1316.88) and (115.03,1316.86) .. (113.35,1318.51) .. controls (111.67,1320.16) and (110,1320.14) .. (108.35,1318.46) .. controls (106.7,1316.78) and (105.03,1316.76) .. (103.35,1318.41) .. controls (101.67,1320.06) and (100,1320.04) .. (98.35,1318.36) .. controls (96.7,1316.68) and (95.03,1316.66) .. (93.35,1318.31) .. controls (91.67,1319.96) and (90,1319.95) .. (88.35,1318.27) .. controls (86.7,1316.59) and (85.03,1316.57) .. (83.35,1318.22) .. controls (81.67,1319.87) and (80,1319.85) .. (78.35,1318.17) .. controls (76.7,1316.49) and (75.03,1316.47) .. (73.35,1318.12) .. controls (71.67,1319.77) and (70,1319.75) .. (68.35,1318.07) -- (65.68,1318.04) -- (65.68,1318.04) ;
\draw    (86,1328.79) -- (106.5,1328.79) ;
\draw [shift={(109.5,1328.79)}, rotate = 180] [fill={rgb, 255:red, 0; green, 0; blue, 0 }  ][line width=0.08]  [draw opacity=0] (5.36,-2.57) -- (0,0) -- (5.36,2.57) -- cycle    ;
\draw  [pattern=_3jx9amvb1,pattern size=3.2249999999999996pt,pattern thickness=0.75pt,pattern radius=0pt, pattern color={rgb, 255:red, 0; green, 0; blue, 0}] (45.83,1319.54) .. controls (45.83,1325.02) and (50.27,1329.46) .. (55.75,1329.46) .. controls (61.23,1329.46) and (65.67,1325.02) .. (65.67,1319.54) .. controls (65.67,1314.07) and (61.23,1309.63) .. (55.75,1309.63) .. controls (50.27,1309.63) and (45.83,1314.07) .. (45.83,1319.54) -- cycle ;
\draw [line width=1.5]    (55.67,1373.43) -- (55.75,1329.46) ;
\draw [line width=1.5]    (55.75,1309.63) -- (55.7,1279.9) ;
\draw [line width=1.5]  [dash pattern={on 1.69pt off 2.76pt}]  (55.67,1373.43) -- (55.5,1399.93) ;
\draw  [dash pattern={on 0.84pt off 2.51pt}]  (133.67,1399.84) -- (133.17,1373.43) ;

\draw (72.67,1609.42) node [anchor=north west][inner sep=0.75pt]    {$\ell _{1}$};
\draw (74,1531.5) node [anchor=north west][inner sep=0.75pt]    {$\ell _{2}$};
\draw (158.5,1607) node [anchor=north west][inner sep=0.75pt]    {$p$};
\draw (139,1551) node [anchor=north west][inner sep=0.75pt]    {$p+\ell _{1}$};
\draw (78,1457.5) node [anchor=north west][inner sep=0.75pt]    {$\ell _{3}$};
\draw (143,1476) node [anchor=north west][inner sep=0.75pt]    {$p+\ell _{1} +\ell _{2}$};
\draw (162.5,1283.92) node [anchor=north west][inner sep=0.75pt]    {$p+q$};
\draw (78.67,1329.5) node [anchor=north west][inner sep=0.75pt]    {$\ell _{L+1}$};
\draw (142,1342) node [anchor=north west][inner sep=0.75pt]    {$p+\ell _{1} +\ell _{2} +\cdots +\ell _{L}$};

\end{tikzpicture}

\end{center}
\end{figure}
giving us a box diagram which reads
\begin{equation}\label{allloop}
\begin{split}
i\mathcal{M}(q)&=(i\kappa)^{L+1}i^L\int \hat{d}^4 \ell_1\cdots  \hat{d}^4 \ell_{L+1}\,   \hat{\delta}^4\left(q-\sum_{i=1}^{L+1}\ell_i \right)
\\&\times
  \mathcal{V}_1\frac{\mathcal{V}_2}{(p+\ell_1)-m^2+i\varepsilon} \frac{\mathcal{V}_3 }{(p+\ell_1+\ell_2)-m^2+i\varepsilon} \cdots \frac{\mathcal{V}_{L+1}}{\left( p+  \sum_{i=1}^L  \ell_i \right)^2-m^2+i\varepsilon}.
\end{split}
\end{equation}
Here
\begin{equation}
\mathcal{V}_k\equiv \tilde{h}_{\mu\nu} (\ell_k )\left(p^\mu +\sum_{i=1}^{k-1}\ell_i^\mu  \right)     \left(p^\nu +\sum_{i=1}^{k}\ell_i^\nu  \right) 
\end{equation}
is the $k$-th KS vertex. This  seems to suggest that higher order observables can be computed in an analogous way to the one presented above, since the amplitude structure extends itself in a straightforward fashion. That is take the classical limit of \eqref{allloop} and subtract all cut contributions to make sure all superclassical terms add up to zero. 
Therefore, we leave the study of loop integration of KS metrics in momentum space to a future work.  

\section{Summary}\label{fin}

In this paper  we  focused on the understanding of the classical impulse up to NLO associated with scattering of spinning particles for electromagnetic and gravitational interactions. We have considered both aligned and non-aligned spin configurations, thereby extending the results presented in Refs.~\cite{Kosower:19,Arkani-Hamed:20}. Our results were achieved using two frameworks, a unitarity based one and a more traditional  one involving background amplitudes.
In the  case of gravity, and within the point-particle approximation, our results describe the scattering of two Kerr black holes. However, as our equations reveal, studying the EM case first is extremely valuable, since it leads the way to the analogous gravitational calculation through fascinating double copy correspondences.
As our results clearly demonstrate, the Newman-Janis complex shift $b \to b + i a$ does not seem to be so straightforward at NLO, in comparison with the tree-level case. To this respect we believe it would be interesting to   investigate whether the recently proposed worldsheet interpretation of the NJ shift \cite{Guevara:2020xjx} holds at NLO as well, we believe our calculations could help to this respect.

Our work is also preparatory and we believe it can be extended in multiple directions.
An outstanding  issue is to improve our comprehension on the Compton amplitude at higher spin orders, which is one of the  main limitations of the present work. In a historical perspective, perhaps one way to deepen our understanding would be to investigate subleading soft theorems for external particles with higher spin. This is certainly an issue that deserves further studies, and indeed different and fascinating works are gaining grounds on this matter \cite{Aoude:22,camilascompton, paolocompton, yutinspin, Bern:22, Aoude:2022thd, Chen:2022yxw}. An easier task would be to build up on the results of section \ref{probesss}  to incorporate higher spin orders in the light body. This can be done  by studying fermion or a general massive spin-$S$ Lagrangian, as was already argued in \cite{Maybee:19}, in the presence of classic sources.   
Other direct extensions would involve applying our tools to the calculation of the classical NLO angular impulse, following a similar recipe as given in ~\cite{Maybee:19,Guevara:19}, or to turn to the  investigation of the double copy relation between dyons and the Taub-NUT solution~\cite{Luna:2015paa,Moynihan:2019bor, Emond:2020lwi, donaleikonal, Huang:2019cja}.
Finally,  it would be useful to link our results to recently rediscovered eikonal methods \cite{carloeikonal,Ciafaloni:2018uwe, Bern:2020gjj,Parra-Martinez:2020dzs,DiVecchia:2020ymx,Mogull:2020sak,DiVecchia:2021bdo,DiVecchia:2021ndb,Jakobsen:2021smu,Shi:2021qsb,Herrmann:2021tct,Damgaard:2021ipf,Emond:2021lfy, Cristofoli:2021jas, DiVecchia:2022nna, AccettulliHuber:2020oou} where one is able to obtain observables by differentiation of a scalar function (or a coherent state parameter) which entails the conservative (and radiative) dynamics.

These subjects would indeed be interesting to explore, and we hope to return to such calculations in the future.

\section*{Acknowledgements} 

We thank Donal O'Connell, Nathan Moynihan, Alasdair Ross, Andrea Cristofoli, Riccardo Gonzo,  Justin Vines and David Peinador Veiga  for interesting and stimulating discussions. The work of GM has been partially supported by Conselho Nacional de Desenvolvimento Cient\'ifico e Tecnol\'ogico - CNPq under grant 317548/2021-2 and Funda\c{c}\~ao Carlos Chagas Filho de Amparo \`a Pesquisa do Estado do Rio de Janeiro - FAPERJ under grants E-26/202.725/2018 and E-26/201.142/2022.
MS is supported by a Principal's Career Development Scholarship from the University
of Edinburgh and the School of Physics and Astronomy.
The work of MS was also supported in part by the National Science Foundation under Grant No. NSF PHY-1748958.

\vspace{1.25cm}
 
\section*{Conventions}\label{sec:Conventions}
We work in a Minkowski spacetime with signature 
\begin{equation}
\eta_{\mu\nu}=\text{diag}(+1, -1, -1, -1)
\end{equation}
  and the Riemann curvature tensor given by 
  \begin{equation}
  R^{\lambda}_{\ \mu\nu\kappa} = \partial_{\kappa}\Gamma^{\lambda}_{\mu\nu}-\partial_{\nu}\Gamma^{\lambda}_{\mu\kappa} + \Gamma^{\eta}_{\mu\nu}\Gamma^{\lambda}_{\kappa\eta} - \Gamma^{\eta}_{\mu\kappa}\Gamma^{\lambda}_{\nu\eta} .
  \end{equation}

Our Fourier transforms conventions are 
\[
f(x) &= \int \hat{d} ^4 k \, e^{-i k \cdot x} \, \tilde f(k) \\
\tilde f(k) &= \int d^4 x \, e^{ik\cdot x} \, f(x) \,
\]
where to tidy up factors of $2\pi$ we write
\[\hat{d}
 ^n k =\frac{d^n k }{(2\pi)^n}\,,\qquad  \hat{\delta} ^n(k)=(2\pi)^n \delta^n(k) \,.
\]
We also define  
\[
v^{(\mu}w^{\nu)}=v^\mu w^\nu+v^\nu w^\mu, \qquad v^{[\mu}w^{\nu]}=v^\mu w^\nu-v^\nu w^\mu.
\label{eq:antisymmDef}
\]

\section*{Appendix A -- Classical calculation of $\Delta p^\mu$ in the probe limit.} 
Here we provide a purely classical proof of \eqref{deltaplano}, which is integrating up to second order in Newton's constant the geodesic equation of a light scalar travelling on a Kerr-Schild Background.

 The geodesic equation for the probe on the Kerr background is 
\begin{equation}
\frac{d p ^\mu (\tau)}{d\tau}=-m{\Gamma^{\mu}}_{\alpha\beta}(x(\tau))u^\alpha(\tau)u^\beta(\tau)
\end{equation}
where
\begin{equation}
{\Gamma^{\mu}}_{\alpha\beta}=\frac{1}{2}g^{\mu\rho}\left(
\partial_\alpha g_{\rho\beta}+\partial_\beta g_{\rho\alpha}-\partial_{\rho}g_{\alpha\beta}
\right),\,\,\,\,\,g_{\mu\nu}=\eta_{\mu\nu}+\kappa h_{\mu\nu}\,\,\,g^{\mu\nu}=\eta^{\mu\nu}-\kappa h^{\mu\nu}.
\end{equation}
We will define the graviton Fourier transform by 
\begin{equation}
h_{\mu\nu}(x)=\int \hat{d}^4 \bar{q}\, e^{-i\bar{q}\cdot x} \tilde{h}_{\mu\nu}(\bar{q})=\int \hat{d}^4 \bar{q}\, e^{-i\bar{q}\cdot x} \left(
{\kappa\,\hat{\delta}(\bar{q}\cdot V)}  H_{\mu\nu}(\bar{q})
\right).
\end{equation} 
The  specific form of $H$  will not really matter so  we will leave it implicit for now.
 
 At LO we  evaluate the metric on a straight-line trajectory $x^\mu(\tau)=b^\mu+u^\mu \tau$ where we choose $\tau$ to be the curved proper time. One finds immediately
\begin{equation}
\begin{split}
\frac{d p^\mu }{d\tau}&=-\kappa m u^\alpha u^\beta\left(  \partial_\alpha {h^\mu}_\beta-\frac{1}{2}\partial^\mu h_{\alpha\beta}
\right)\\&=
-\kappa^2 m  u^\alpha u^\beta \int \hat{d}^4 \bar{q}\, e^{-i\bar{q}\cdot (b+u\tau)}{\hat{\delta}(\bar{q}\cdot V)}  i \left( 
\bar{q}_\alpha {H^\mu}_\beta(\bar{q})+\frac{\bar{q}^\mu}{2}H_{\alpha\beta}(\bar{q})
\right)
\end{split}
\end{equation}
which, integrating over all proper times, gives us the LO deflection
\begin{equation}
\Delta p^\mu_{\text{LO}}=-\frac{\kappa^2 m }{2}  \int \hat{d}^4 \bar{q}\, e^{-i\bar{q}\cdot b}{\hat{\delta}(\bar{q}\cdot V)} \hat{\delta}(\bar{q}\cdot u) i {\bar{q}^\mu} u^\alpha u^\beta H_{\alpha\beta}(\bar{q}).
\end{equation}

From now on we will define and use the vector
\begin{equation}
E^\mu(\bar{q}):= H^{\mu\nu}(\bar{q})u_\nu
\end{equation}
so, for instance 
\begin{equation}
\Delta p^\mu_{\text{LO}}=-\frac{\kappa m N}{2}  \int \hat{d}^4 \bar{q}\, e^{-i\bar{q}\cdot b}{\hat{\delta}(\bar{q}\cdot V)} \hat{\delta}(\bar{q}\cdot u) i {\bar{q}^\mu} E(\bar{q})\cdot u
\end{equation}
this will help us avoiding long strings of indices.

\subsection{NLO}

Now we compute the correction to the velocity/momentum and to the trajectory. We have
\begin{equation}\label{deltap}
\begin{split}
p^\mu_{(1)} (\tau)(\tau)=m u^\mu_{(1)}(\tau)&=\int_{-\infty}^\tau d\tau \frac{d p^\mu }{d\tau}
\\&=
\kappa^2 m   \int \hat{d}^4 \bar{q}\, e^{-i\bar{q}\cdot (b+u\tau)}\frac{\hat{\delta}(\bar{q}\cdot V)}{ \bar{q}\cdot u +i\varepsilon}  \left( 
-\bar{q}\cdot u \,{E^\mu}(\bar{q})+\frac{\bar{q}^\mu}{2}E(\bar{q})\cdot u
\right)
\end{split}
\end{equation}
where we have regulated \cite{Jackson} the lower bound divergence of the integral by 
\begin{equation}
\int_{-\infty}^\tau d\tau\,e^{-i\tau u\cdot \bar{q}}\to \int_{-\infty}^\tau d\tau\, e^{-i\tau (u\cdot \bar{q}+i\varepsilon) }=\frac{i}{u\cdot \bar{q}+i\varepsilon}e^{-i\tau (u\cdot \bar{q}+i\varepsilon) }.
\end{equation}
We can integrate again to get an expression for the trajectory
\begin{equation}\label{deltax}
\begin{split}
 x^\mu_{(1)} (\tau)&=\frac{1}{m}\int_{-\infty}^\tau d\tau\,p^\mu_{(1)}
 \\&=
i\kappa^2 \int \hat{d}^4 \bar{q}\, e^{-i\bar{q}\cdot (b+u\tau)}\frac{\hat{\delta}(\bar{q}\cdot V)}{ (\bar{q}\cdot u +i\varepsilon)^2}  \left( 
-\bar{q}\cdot u \,{E^\mu}(\bar{q})+\frac{\bar{q}^\mu}{2}E(\bar{q})\cdot u
\right).
\end{split}
\end{equation}

Now we identify all possible contributions at NLO. These are 
\begin{equation}
\frac{d p^\mu_{(2)}(\tau)}{d \tau} =-m\left( 2\, {\Gamma^{\mu}}_{\alpha\beta}(b+u \tau)u^\alpha u^\beta _{(1)}(\tau)+ \left( {\Gamma^{\mu}_{(1)}}_{\alpha\beta}(\tau)+{x_{(1)}(\tau)\cdot \partial\, \Gamma^{\mu}}_{\alpha\beta}(b+u \tau)\right)u^\alpha u^\beta
\right)
\end{equation}
where the first one corresponds to considering a product ok $\kappa$ in the connection and another one in the velocity correction. The last two have two powers of $\kappa$ from expanding the Christoffel symbols only instead (expanding the trajectory inside and expanding $\Gamma\sim h\partial h$).  We will also write the corresponding deflections as $\Delta p^\mu _{1,2,3}$ with $\Delta p^\mu_{\text{NLO}}=\sum_i \Delta p^\mu _{i} $.

The first term above yields, using \eqref{deltap}
\begin{equation}
\begin{split}
-2m {\Gamma^{\mu}}_{\alpha\beta}(b+u \tau)u^\alpha u^\beta _{(1)}(\tau)=im& \kappa^4\int  {\hat{d}^4 \bar{q}}\,{\hat{d}^4 \bar{\ell}}\,e^{-ib\cdot (\bar{q}+\bar{\ell})}e^{-i\tau u\cdot (\bar{q}+\bar{\ell})}\frac{\hat{\delta}(\bar{q}\cdot V)\hat{\delta}(\bar{\ell}\cdot V)}{\bar{\ell}\cdot u +i \varepsilon}  \\&
\times
\left(
\bar{q}\cdot u \,{H^\mu}_\beta (\bar{q})+\bar{q}_\beta E^\mu (\bar{q})-\bar{q}^\mu E_{\beta}(\bar{q})
\right)\left(
-\bar{\ell}\cdot u\, E^\beta (\bar{\ell})+\frac{\bar{\ell}^\beta}{2}E(\bar{\ell})\cdot u
\right).
\end{split}
\end{equation}
Thus
\begin{equation}
\begin{split}
\Delta p^\mu _1=im& \kappa^4\int  {\hat{d}^4 \bar{q}}\,{\hat{d}^4 \bar{\ell}}\,e^{-ib\cdot (\bar{q}+\bar{\ell})}\hat{\delta}(u\cdot (\bar{q}+\bar{\ell}))\frac{\hat{\delta}(\bar{q}\cdot V)\hat{\delta}(\bar{\ell}\cdot V)}{\bar{\ell}\cdot u +i \varepsilon}  \\&
\times
\left(
\bar{q}\cdot u \,{H^\mu}_\beta (\bar{q})+\bar{q}_\beta E^\mu (\bar{q})-\bar{q}^\mu E_{\beta}(\bar{q})
\right)\left(
-\bar{\ell}\cdot u\, E^\beta (\bar{\ell})+\frac{\bar{\ell}^\beta}{2}E(\bar{\ell})\cdot u
\right).
\end{split}
\end{equation}

Similarly we find
\begin{equation}
\begin{split}
\Delta p^\mu _2&=-m\int d\tau\,  {\Gamma^{\mu}_{(1)}}_{\alpha\beta}(\tau)u^\alpha u^\beta \\&=
im\kappa^4\int  {\hat{d}^4 \bar{q}}\,{\hat{d}^4 \bar{\ell}}\,e^{-ib\cdot (\bar{q}+\bar{\ell})}\hat{\delta}(u\cdot (\bar{q}+\bar{\ell})){\hat{\delta}(\bar{q}\cdot V)\hat{\delta}(\bar{\ell}\cdot V)}  
\,{H^\mu}_\beta (\bar{q})\left(
-\bar{\ell}\cdot u\, E^\beta (\bar{\ell})+\frac{\bar{\ell}^\beta}{2}E(\bar{\ell})\cdot u
\right).
\end{split}
\end{equation}
Note already that, because on shell $\frac{u\cdot \bar{q}}{u\cdot \bar{\ell}+i\varepsilon}=-1$, the sum of these two first deflection contributions simplifies to 

\begin{equation}
\begin{split}
\Delta p^\mu _1+\Delta p^\mu _2=im& \kappa^4\int  {\hat{d}^4 \bar{q}}\,{\hat{d}^4 \bar{\ell}}\,e^{-ib\cdot (\bar{q}+\bar{\ell})}\hat{\delta}(u\cdot (\bar{q}+\bar{\ell}))\frac{\hat{\delta}(\bar{q}\cdot V)\hat{\delta}(\bar{\ell}\cdot V)}{\bar{\ell}\cdot u +i \varepsilon}  
\bar{q}^{[\beta} E^{\mu]}(\bar{q})\left(
-\bar{\ell}\cdot u\, E_\beta (\bar{\ell})+\frac{\bar{\ell}_\beta}{2}E(\bar{\ell})\cdot u
\right).
\end{split}
\end{equation}

Finally, using \eqref{deltax} we get to 

\begin{equation}
\begin{split}
\Delta p^\mu _3=im& \kappa^4\int  {\hat{d}^4 \bar{q}}\,{\hat{d}^4 \bar{\ell}}\,e^{-ib\cdot (\bar{q}+\bar{\ell})}\hat{\delta}(u\cdot (\bar{q}+\bar{\ell}))\frac{\hat{\delta}(\bar{q}\cdot V)\hat{\delta}(\bar{\ell}\cdot V)}{(\bar{\ell}\cdot u +i \varepsilon)^2}  \\&
\times
\left(
-\bar{\ell}\cdot u \,E(\bar{\ell})\cdot \bar{q}+\frac{\bar{q}\cdot \bar{\ell}}{2} E(\bar{\ell})\cdot u
\right)\left(
\bar{q}\cdot u\, E^\mu (\bar{q})-\frac{\bar{q}^\mu}{2}E(\bar{q})\cdot u
\right).
\end{split}
\end{equation}

In the end the complete NLO deflection looks like, simplifying a bit using the constraints imposed by the delta functions 
\begin{equation}\label{finalu}
\begin{split}
\Delta p^\mu _{\text{NLO}}&= im \kappa^4\int  {\hat{d}^4 \bar{q}}\,{\hat{d}^4 \bar{\ell}}\,e^{-ib\cdot (\bar{q}+\bar{\ell})}\hat{\delta}(u\cdot (\bar{q}+\bar{\ell})){\hat{\delta}(\bar{q}\cdot V)\hat{\delta}(\bar{\ell}\cdot V)}  \\&
\times \bar{q}^\mu \left(
E(\bar{q})\cdot E(\bar{\ell})-\frac{1}{2}
\frac{E(\bar{q})\cdot \bar{\ell}\, E(\bar{\ell})\cdot u}{\bar{\ell}\cdot u +i\varepsilon}+\frac{1}{2}
\frac{E(\bar{\ell})\cdot \bar{q}\, E(\bar{q})\cdot u}{\bar{\ell}\cdot u +i\varepsilon}-\frac{\bar{q}\cdot \bar{\ell}}{4} \frac{E(\bar{q})\cdot u\, E(\bar{\ell})\cdot u}{(\bar{\ell}\cdot u+i\varepsilon)^2}
\right).
\end{split}
\end{equation}
This satisfies the on-shell check $-2p\cdot \Delta p_{\text{NLO}}=(\Delta p_{\text{LO}})^2$ after a short computation. It can be seen that  \eqref{finalu} is indeed \eqref{deltaplano}. To begin with, we realise that the first term inside the parenthesis is zero in KS coordinates, this is because it is a pure convolution in momentum space or product in position space, as can be seen by changing $\bar{q}+\bar{\ell}\to \bar{q}$. Let's then look at
\begin{equation}
\begin{split}
&im \kappa^4\int  {\hat{d}^4 \bar{q}}\,{\hat{d}^4 \bar{\ell}}\,e^{-ib\cdot (\bar{q}+\bar{\ell})}\hat{\delta}(u\cdot (\bar{q}+\bar{\ell})){\hat{\delta}(\bar{q}\cdot V)\hat{\delta}(\bar{\ell}\cdot V)}   \bar{q}^\mu \left( -\frac{1}{2}
\frac{E(\bar{q})\cdot \bar{\ell}\, E(\bar{\ell})\cdot u}{\bar{\ell}\cdot u +i\varepsilon}+\frac{1}{2}
\frac{E(\bar{\ell})\cdot \bar{q}\, E(\bar{q})\cdot u}{\bar{\ell}\cdot u +i\varepsilon}
\right)
\end{split}
\end{equation}
we rewrite this, flipping labels $\bar{\ell}\leftrightarrow \bar{q}$, as
\begin{equation}
\begin{split}
 & im \kappa^4\int  {\hat{d}^4 \bar{q}}\,{\hat{d}^4 \bar{\ell}}\,e^{-ib\cdot (\bar{q}+\bar{\ell})}\hat{\delta}(u\cdot (\bar{q}+\bar{\ell})){\hat{\delta}(\bar{q}\cdot V)\hat{\delta}(\bar{\ell}\cdot V)}  \bar{\ell}^\mu \left( \frac{1}{2}
\frac{E(\bar{q})\cdot \bar{\ell}\, E(\bar{\ell})\cdot u}{-\bar{\ell}\cdot u +i\varepsilon}-\frac{1}{2}
\frac{E(\bar{\ell})\cdot \bar{q}\, E(\bar{q})\cdot u}{-\bar{\ell}\cdot u +i\varepsilon}
\right)
\end{split}
\end{equation}
then we switch to a new $\bar{q}$ variable defined by $\bar{q}+\bar{\ell}\to \bar{q}$ to obtain 
\begin{equation}
\begin{split}
 & im \kappa^4\int  {\hat{d}^4 \bar{q}}\,e^{-ib\cdot \bar{q}}\hat{\delta}(u\cdot \bar{q}){\hat{\delta}(\bar{q}\cdot V) \int {\hat{d}^4 \bar{\ell}}\, \hat{\delta}(\bar{\ell}\cdot V)}  \bar{\ell}^\mu \left( \frac{1}{2}
\frac{E(\bar{q}-\bar{\ell})\cdot \bar{\ell}\, E(\bar{\ell})\cdot u}{-\bar{\ell}\cdot u +i\varepsilon}-\frac{1}{2}
\frac{E(\bar{\ell})\cdot (\bar{q}-\bar{\ell})\, E(\bar{q}-\bar{\ell})\cdot u}{-\bar{\ell}\cdot u +i\varepsilon}
\right).
\end{split}
\end{equation}
Finally, we change the classical loop momentum $\bar{\ell}\to \bar{q}-\bar{\ell}$ in the second term only, obtaining
\begin{equation}
\begin{split}
  im \kappa^4\int  {\hat{d}^4 \bar{q}}\,e^{-ib\cdot \bar{q}}\hat{\delta}(u\cdot \bar{q})&{\hat{\delta}(\bar{q}\cdot V) \int {\hat{d}^4 \bar{\ell}}\, \hat{\delta}(\bar{\ell}\cdot V)} 
 \\&
\times 
 \left( \frac{ \bar{\ell}^\mu }{2}
\frac{E(\bar{q}-\bar{\ell})\cdot \bar{\ell}\, E(\bar{\ell})\cdot u}{-\bar{\ell}\cdot u +i\varepsilon}-\frac{\bar{q}^\mu-\bar{\ell}^\mu }{2}
\frac{E(\bar{q}-\bar{\ell})\cdot \bar{\ell}\, E(\bar{\ell})\cdot u}{\bar{\ell}\cdot u +i\varepsilon}
\right)
\end{split}
\end{equation}
which, using the delta-function trick, can be brought into the following form 
\begin{equation}
\begin{split}
  im \kappa^4\int  {\hat{d}^4 \bar{q}}\,e^{-ib\cdot \bar{q}}\hat{\delta}(u\cdot \bar{q})&{\hat{\delta}(\bar{q}\cdot V) \int {\hat{d}^4 \bar{\ell}}\, \hat{\delta}(\bar{\ell}\cdot V)} 
 \\&
\times 
 \left( -i \frac{ \bar{\ell}^\mu }{2} \hat{\delta}(\bar{\ell}\cdot u) 
{E(\bar{q}-\bar{\ell})\cdot \bar{\ell}\, E(\bar{\ell})\cdot u}-\frac{\bar{q}^\mu }{2}
\frac{E(\bar{q}-\bar{\ell})\cdot \bar{\ell}\, E(\bar{\ell})\cdot u}{\bar{\ell}\cdot u +i\varepsilon}
\right).
\end{split}
\end{equation}
Similar manipulations  and use of the formula 
\begin{equation}
i\hat{\delta}'(x)=\frac{1}{(x+i\varepsilon)^2}-\frac{1}{(-x+i\varepsilon)^2}
\end{equation}
reveal that  
\begin{equation} 
\begin{split}
& - \frac{im \kappa^4}{4} \int  {\hat{d}^4 \bar{q}}\,{\hat{d}^4 \bar{\ell}}\,e^{-ib\cdot (\bar{q}+\bar{\ell})}\hat{\delta}(u\cdot (\bar{q}+\bar{\ell})){\hat{\delta}(\bar{q}\cdot V)\hat{\delta}(\bar{\ell}\cdot V)}   \bar{q}^\mu \,{\bar{q}\cdot \bar{\ell}} \frac{E(\bar{q})\cdot u\, E(\bar{\ell})\cdot u}{(\bar{\ell}\cdot u+i\varepsilon)^2}\\&=
 \frac{im \kappa^4}{8}  \int  {\hat{d}^4 \bar{q}}\,e^{-ib\cdot \bar{q}}\hat{\delta}(u\cdot \bar{q}){\hat{\delta}(\bar{q}\cdot V) \int {\hat{d}^4 \bar{\ell}}\, \hat{\delta}(\bar{\ell}\cdot V)}  \bar{\ell}\cdot (\bar{\ell}-\bar{q})
 \\&\qquad\qquad\qquad
\times 
 \left(  {\bar{q}^\mu }
\frac{E(\bar{q}-\bar{\ell})\cdot u\, E(\bar{\ell})\cdot u}{(\bar{\ell}\cdot u +i\varepsilon)^2}
-i { \bar{\ell}^\mu } \hat{\delta}'(\bar{\ell}\cdot u) 
{E(\bar{q}-\bar{\ell})\cdot u\, E(\bar{\ell})\cdot u}\right)
\end{split}
\end{equation}
proving the equivalence of the two derivations.

We avoid repeating the computation for a generic $\sqrt{\text{Kerr}}$ electromagnetic source, since this is essentially the one found in \cite{Kosower:19}. We only report the final deflection  which one finds integrating  Maxwell's equations 
\begin{equation} 
\begin{split}
{\Delta} p^\mu &  
= \frac{ie^2Q^2}{m}\int { \hat{d}^4 \bar{\ell}\,\hat{d}^4 \bar{q}} \, e^{-i(\bar{q}+\bar{\ell})\cdot b}{\hat{\delta}(\bar{\ell}\cdot V)}\hat{\delta}((\bar{q}+\bar{\ell})\cdot u)\hat{\delta}(\bar{q}\cdot V) \\& \times \bar{q}^\mu   \left(   -\frac{\tilde{A} (\bar{q})\cdot\bar{\ell}\,\tilde{A}  (\bar{\ell})\cdot u}{\bar{\ell}\cdot u+i\varepsilon}  + \frac{   \tilde{A}  (\bar{\ell})\cdot\bar{q}\,\tilde{A} (\bar{q})\cdot u }{\bar{\ell}\cdot u+i\varepsilon}-\frac{\bar{\ell}\cdot\bar{q}\, \tilde{A} (\bar{\ell})\cdot u\,  \tilde{A}  (\bar{q})\cdot  u}{(\bar{q}\cdot u+i\varepsilon)^2}\right)
 \end{split}
\end{equation} 
which can be shown, tracing the same steps above, to equal what we had found using amplitudes.

\bibliographystyle{utphys.bst}
\bibliography{bibliography_nlo}
\end{document}